\pdfoutput=1
\newcommand*{\ATLASLATEXPATH}{latex/}
\documentclass[cernpreprint=true, atlasdraft=false, texlive=2016, UKenglish, LANGEDIT=false]{\ATLASLATEXPATH atlasdoc}
\usepackage[subfigure]{\ATLASLATEXPATH atlaspackage}
\usepackage{\ATLASLATEXPATH atlasbiblatex}
\usepackage{\ATLASLATEXPATH atlasphysics}
\usepackage{rotating}
\graphicspath{{logos/}{figures/}}
\usepackage{EXOT-2016-38-PAPER-defs}
\AtlasTitle{%
   A strategy for a general search for new phenomena \\
   using data-derived signal regions and \\
   its application within the ATLAS experiment}
\AtlasAbstract{%
   This paper describes a strategy for a general search used by the ATLAS Collaboration 
   to find potential indications of new physics.
   Events are classified according to their final state into many event classes.
   For each event class an automated search algorithm tests whether the 
   data are compatible with the Monte Carlo simulated expectation in  several
   distributions sensitive to the effects of new physics.
   The significance of a deviation is quantified using pseudo-experiments.
   A data selection with a significant deviation defines a signal region for 
   a dedicated follow-up analysis
   with an improved background expectation. 
   The analysis of the data-derived signal regions on a new dataset 
   allows a statistical interpretation without the large look-elsewhere effect.
   The sensitivity of the approach is discussed using Standard Model processes
   and benchmark signals of new physics.
   As an example, results are shown for 3.2~fb$^{-1}$ of proton--proton collision
   data at a centre-of-mass energy of 13~\TeV{} collected with the ATLAS
   detector at the LHC in 2015, in which more than 700 event classes and more than $10^5$ regions
   have been analysed.
   No significant deviations are found and consequently 
   no data-derived signal regions for a follow-up analysis 
   have been defined.}
\AtlasRefCode{EXOT-2016-38}
\PreprintIdNumber{CERN-EP-2018-070}
\AtlasJournal{\EPJC}
\AtlasJournalRef{Eur.\ Phys.\ J.\ C 79 (2019) 120}
\AtlasDOI{10.1140/epjc/s10052-019-6540-y}

\begin{document}

\maketitle

\section{Introduction}
   \label{sec:introduction}
   Direct searches for unknown particles and interactions are one of the primary objectives of the physics programme at the Large Hadron Collider (LHC). 
The ATLAS experiment at the LHC has thoroughly analysed the \RunOne \pp collision dataset (recorded in 2010--2012) and 
roughly a quarter 
of the expected \RunTwo dataset (2015--2018). No evidence of physics beyond the Standard Model (SM) has been found in any of the searches performed so far.

Searches that 
have been performed 
to date do not fully cover the enormous parameter space of masses, cross-sections and decay 
channels
of possible new particles. Signals might be hidden in kinematic regimes and final states that have remained unexplored. 
This motivates a model-independent\footnote{`Model-independent' refers to the absence of a beyond the Standard Model signal assumption. 
The analysis depends on the Standard Model prediction.}
analysis to search for physics beyond the Standard Model (BSM) in a structured, global and automated way, where many of the final states not yet covered can be probed.

General searches without an explicit BSM signal assumption have been 
been performed 
by the D\O\ Collaboration~\cite{Abbott:2000fb,Abbott:2000gx,Abbott:2001ke, Abazov:2011ma} at the Tevatron,
by the H1 Collaboration~\cite{Aktas:2004pz,Aaron:2008aa} at HERA, 
and by the CDF Collaboration~\cite{Aaltonen:2007dg, Aaltonen:2008vt} at the Tevatron. 
At the LHC, preliminary versions of such searches have been performed 
by the ATLAS Collaboration at $\sqs=7$, $8$ and $13$~\TeV{},
and by the CMS Collaboration at $\sqs=7$ and $8$ \TeV{}.

This paper outlines a strategy employed by the ATLAS Collaboration to search
in a systematic and (quasi\nobreakdash-)model-independent way for deviations of the data from the SM prediction.
This approach assumes only generic features of the potential BSM signals. 
Signal events are expected to have reconstructed objects with relatively large momentum transverse to the beam axis. 
The main objective of this strategy is not to 
finally assess 
the exact level of significance of a deviation with all available data, but rather to 
identify with a first dataset 
those phase-space regions 
where significant deviations of the data from SM prediction are 
present for a further dedicated analysis.
The observation of one or more significant deviations in some phase-space region(s) serves as a trigger to perform 
dedicated and model-dependent analyses where 
these `data-derived' 
phase-space region(s) can be used as 
signal regions. 
Such an analysis can then determine the level of significance using a second dataset.
The main advantage of this procedure is that it allows a large number of phase-space regions to be tested with the available resources, 
thereby minimizing the possibility 
of missing a signal 
for new physics, 
while simultaneously maintaining a low false discovery 
rate 
by testing the data-derived signal region(s) on an independent dataset in a dedicated analysis. The dedicated analysis with data-derived signal regions also allows
an improved background prediction.

In this approach, events are first classified into different 
(exclusive) categories,
labelled with the multiplicity of final-state objects (e.g.\ muons, electrons, jets, missing transverse momentum, etc.) in an event. 
These final-state categories are then automatically analysed for deviations of the data from the SM prediction in several BSM-sensitive distributions 
using an algorithm that locates the region of largest excess or deficit. 
Sensitivity tests for specific signal models are performed to demonstrate the effectiveness of this 
approach. 
The methodology has been applied to a subset of the $\sqs=13$~\TeV{} proton--proton 
collision data as reported in this paper. 
The data were collected with the ATLAS detector in 2015, and correspond to an integrated luminosity of $3.2$~\ifb.

The paper is organized as follows: 
the general analysis strategy is outlined in \Sect{\ref{sec:strategy}}, 
while \Sect{\ref{sec:application_strategy}} provides specific details about its application to the ATLAS 2015 \pp collision dataset. 
Conclusions are given in \Sect{\ref{sec:Conclusions}}.

\section{Strategy}
   \label{sec:strategy}
   The analysis strategy assumes that 
a signal 
of unknown origin
can be revealed as a statistically significant deviation of the event counts in the data from the expectation in a specific data selection. 
A data selection can be any set of requirements on objects or variables needed to define a signal region (e.g.\ an event class or a specific range in one or multiple observables).
In order to search for these signals a large variety of data selections 
need 
to be tested.
This requires a high degree of automation and a 
categorization of the data events 
according to their main features.
The main 
objective of this analysis
is 
to identify selections 
for which the data deviates
significantly 
from the SM expectation. 
These selections can then be
applied 
as data-derived signal regions
in a dedicated analysis
to determine the level of significance
using a new dataset. 
This has
the advantage of a more reliable background
expectation, which should allow
an increase in signal sensitivity
compared to a strategy that only relies on Monte
Carlo expectations with a typically conservative evaluation
of uncertainties.
The 
strategy is divided into the seven steps described below.

\subsection{Step 1: Data selection and Monte Carlo simulation}
The recorded data are reconstructed via the ATLAS software chain. 
Events are selected by applying event-quality and trigger criteria, 
and are classified according to the type and multiplicity of reconstructed objects with high transverse momentum (\pt). 
Objects that can be considered in the classification are those typically used to characterize hadron collisions such as 
electrons, muons, $\tau$-leptons, photons, jets, \btagged jets and missing transverse momentum. 
More complex objects, which were not implemented in the 
example described in \Sect{\ref{sec:application_strategy}}, 
could also be considered.
Examples are resonances reconstructed by a specific decay 
(e.g.\ \Zboson or Higgs bosons decaying into two or four isolated leptons respectively,
or decaying hadronically and giving rise to large radius jets with substructure) and displaced vertices. 
Event classes (or channels) are then defined
as the set of events with a given number of reconstructed objects for each type, e.g.\ two muons and a jet.

Monte Carlo (MC) simulations are used 
to estimate the expected event counts from SM processes.
To allow the investigation of signal regions with a low number of expected events it is important that 
the equivalent integrated luminosity of the MC 
samples significantly exceeds that of the data, 
and that 
all relevant background processes are included, in particular rare processes which might dominate certain multi-object event classes.

\subsection{Step 2: Systematic uncertainties and validation}
\label{sec:step2}
The 
particular nature of this analysis, 
in which a large number of final states are explored,
makes the definition of control and validation regions difficult. 
In searches for BSM physics at the LHC, control regions 
are used to constrain MC-based 
background predictions
with auxiliary measurements. Validation
regions are used to test the validity of the background model prediction with data.

The simplest 
way to construct a background model is to obtain the background expectation from the MC prediction including the corresponding theoretical and experimental uncertainties. 
This approach, which is applied in the example in \Sect{\ref{sec:application_strategy}}, 
has the advantage that it prevents the absorption of BSM signal contributions into a rescaling of the SM processes.
Another possible approach is to automatically define, for each data selection  and  algorithmic hypothesis test, statistically independent control selections. 
The data in the control selections can be used to rescale the MC background predictions and to constrain the systematic uncertainties. 
This comes at the price of reduced sensitivity for the case in which a BSM model predicts a simultaneous effect in the signal region and control region, 
which would be absorbed in the rescaling.

To verify the proper modelling of the SM background processes, several validation 
distributions are 
defined 
using inclusive selections 
for which observable signals for new physics are excluded.
If these validation distributions show problems in the MC modelling, 
either corrections to the MC backgrounds are applied or the affected event class is excluded.

Uncertainties in the background estimate arise from  experimental effects,
and the theoretical accuracy of the prediction of the 
(differential) cross-section and acceptance of the MC simulation.
Their effect is evaluated for all 
contributing background processes as well as for  benchmark signals.

\subsection{Step 3: Sensitive variables and search algorithm}
Distributions of observables in the form of histograms are investigated for all 
event classes considered in the analysis. 
Observables are included if they have a high sensitivity to a wide range of BSM signals. 
The total number of observables considered is, however, restricted to a few to avoid  
a large increase in the number of
hypothesis tests, as the latter also increases 
the rate of deviations from background fluctuations.
In high-energy physics this 
effect is commonly known as
the 
`trial factor' or `look-elsewhere effect'.
Examples of such observables are the effective mass \meff{} (defined as the sum of the scalar transverse momenta of all objects plus the scalar missing transverse momentum), 
the total invariant mass \minv{} (defined as the invariant mass of all visible objects), 
the invariant mass  of any combination of objects (such as the dielectron invariant mass in events
with two electrons and two muons),
event shape variables such as thrust~\cite{Brandt:1964sa,Farhi:1977sg}
or even more complicated variables such as the output of a machine-learning algorithm.

A statistical algorithm is used to scan these distributions for each event class and quantify the deviations of the data from the SM expectation. 
The algorithm identifies the data selection 
that has the largest deviation in the distribution of the investigated observable 
by testing many data selections to 
minimize a test statistic. 
An example of a possible test-statistic which has also been used in the analysis described in \Sect{\ref{sec:application_strategy}}, is the 
local \pvalue-value, which gives the expected probability of observing a fluctuation that is at least as far from the SM expectation 
as the  observed number of data events in a given region, if the experiment were to be repeated: 
\begin{linenomath}
\begin{align}
\pvalue				& = 2 \cdot \min \Bigr[ P(n \leq \Nobs),\ P(n \geq \Nobs) \Bigr]  \label{eq:pvalue1} \\
P(n \leq \Nobs) 	& = \int_{0}^{\infty} \mathrm{d}x \
						\gauss \left( x; \Nsm, \delta \Nsm \right)
                        \cdot \sum_{n=0}^{\Nobs} \frac{\text{e}^{-x}x^n}{n!}
                  	  +	\int_{-\infty}^{0} \mathrm{d}x \ 
                    	\gauss \left( x; \Nsm, \delta \Nsm \right) \label{eq:pvalue2} \\
P(n \geq \Nobs)		& = \int_{0}^{\infty} \mathrm{d}x \ 
						\gauss \left(x; \Nsm, \delta \Nsm \right)
                        \cdot \sum_{n=\Nobs}^{\infty} \frac{\text{e}^{-x}x^n}{n!}	\label{eq:pvalue3}
\end{align}
\end{linenomath}
where $n$ is the independent variable of the Poisson probability mass function (pmf), \Nobs is the observed number of data events for a given selection,
$P(n \leq \Nobs)$ is the probability of observing no more than the number of events observed in the data and
$P(n \geq \Nobs)$ is the probability of observing at least the number of events observed in the data.
The quantity \Nsm is the expectation for the number of events with its total uncertainty 
$\delta \Nsm$ for a given selection. The convolution of the Poisson pmf 
(with mean $x$) with a Gaussian probability density function (pdf), $\gauss(x;\Nsm,\delta \Nsm)$ with mean \Nsm 
and width $\delta \Nsm$, takes the effect of both non-negligible systematic uncertainties and statistical uncertainties into account.\footnote{
  The second term in \Eqn{(\ref{eq:pvalue2})} gives the probability of observing no events 
  given a negative expectation from downward variations of the systematic uncertainties. 
  It can be derived as follows:
  \[
  \int_{-\infty}^{0} \mathrm{d}x \ \gauss \left( x; \Nsm, \delta \Nsm \right) 
  \cdot \sum_{n=0}^{\Nobs} \ \lim_{\mu \to 0} \left(\frac{\text{e}^{-\mu}\mu^n}{n!}\right)
  = \int_{-\infty}^{0} \mathrm{d}x \ \gauss \left( x; \Nsm, \delta \Nsm \right) 
  \cdot \sum_{n=0}^{\Nobs}
  \delta_{n0}
  = \int_{-\infty}^{0} \mathrm{d}x \ \gauss \left( x; \Nsm, \delta \Nsm \right)
  \]
  where 
  $\mu$ is the mean of the Poisson pmf
  and 
  $\delta_{n0} = \left\{ 1 \text{ if } n=0 \text{, \ }0 \text{ if } n\neq0 \right\}$ is the Kronecker delta.
  In \Eqn{(\ref{eq:pvalue3})} this term vanishes for $\Nobs>0$.
}
If the Gaussian pdf \gauss is replaced by a Dirac delta function $\delta(x-\Nsm)$ the estimator \pvalue results in the usual Poisson probability. 
The selection with the largest deviation identified by the algorithm is defined as the selection giving the smallest \pvalue-value.
The smallest \pvalue for a given
channel
is defined as \pchannel, which therefore corresponds to the local \pvalue-value of the largest deviation in that channel.

Data selections are not considered  in the scan 
if large uncertainties in the expectation  arise due to a lack of MC events, or from large systematic uncertainties.  
To avoid overlooking potential excesses in these selections the \pvalue-values of selections with more than three data events 
are  monitored separately. Single outstanding events with atypical object multiplicities (e.g.\ events with 12 muons) are visible as an event class. 
Single outstanding events in the scanned distributions are  monitored separately.

The result of scanning the distributions for all event classes is a list of
data selections, one per event class containing the largest deviation in that class, and their
local statistical significance. 
Details of the procedure and the statistical algorithm used for the 2015 dataset are explained in \Sect{\ref{sec:Interpretation}}.

\subsection{Step 4: Generation of pseudo-experiments}
\label{sec:pseudoexperiments}
The probability that 
for a given observable
one or more deviations of a certain size occur somewhere in the event classes 
considered
is modelled by pseudo-experiments. 
Each pseudo-experiment consists of exactly the same event classes 
as those considered when applying the search algorithm to data. 
However, the data counts are replaced by pseudo-data counts which are generated from the SM expectation using an MC technique. 
Pseudo-data distributions are produced taking into account both statistical and systematic uncertainties by drawing pseudo-random data counts for each bin from the convolved 
pmf used
in \Eqnrange{(\ref{eq:pvalue1})}{(\ref{eq:pvalue3})} to compute a \pvalue-value.

Correlations in the uncertainties of the SM expectation affect the 
chance of observing one or more deviations of a given size.
The effect of correlations 
between bins of the same distribution 
or between distributions of different event classes
are therefore taken into account when generating pseudo-data for pseudo-experiments. 
Correlations between distributions of different observables are not taken into account,
since the results obtained for different observables are not combined in the interpretation.

The search algorithm is then applied to each of the distributions, resulting in a \pchannel-value for each event class.
The \pchannel distributions of many pseudo-experiments and their statistical properties can be compared with the \pchannel distribution obtained from data 
to interpret the test statistics in a frequentist manner.
The fraction of 
pseudo-experiments 
having one of the \pchannel-values smaller than a given value $p_{\text{min}}$ indicates the probability of observing such a deviation by chance, taking into account the number of selections and event classes tested.

\begin{figure}[htbp]
  \begin{center}
    \includegraphics[page=1,width=0.495\textwidth]{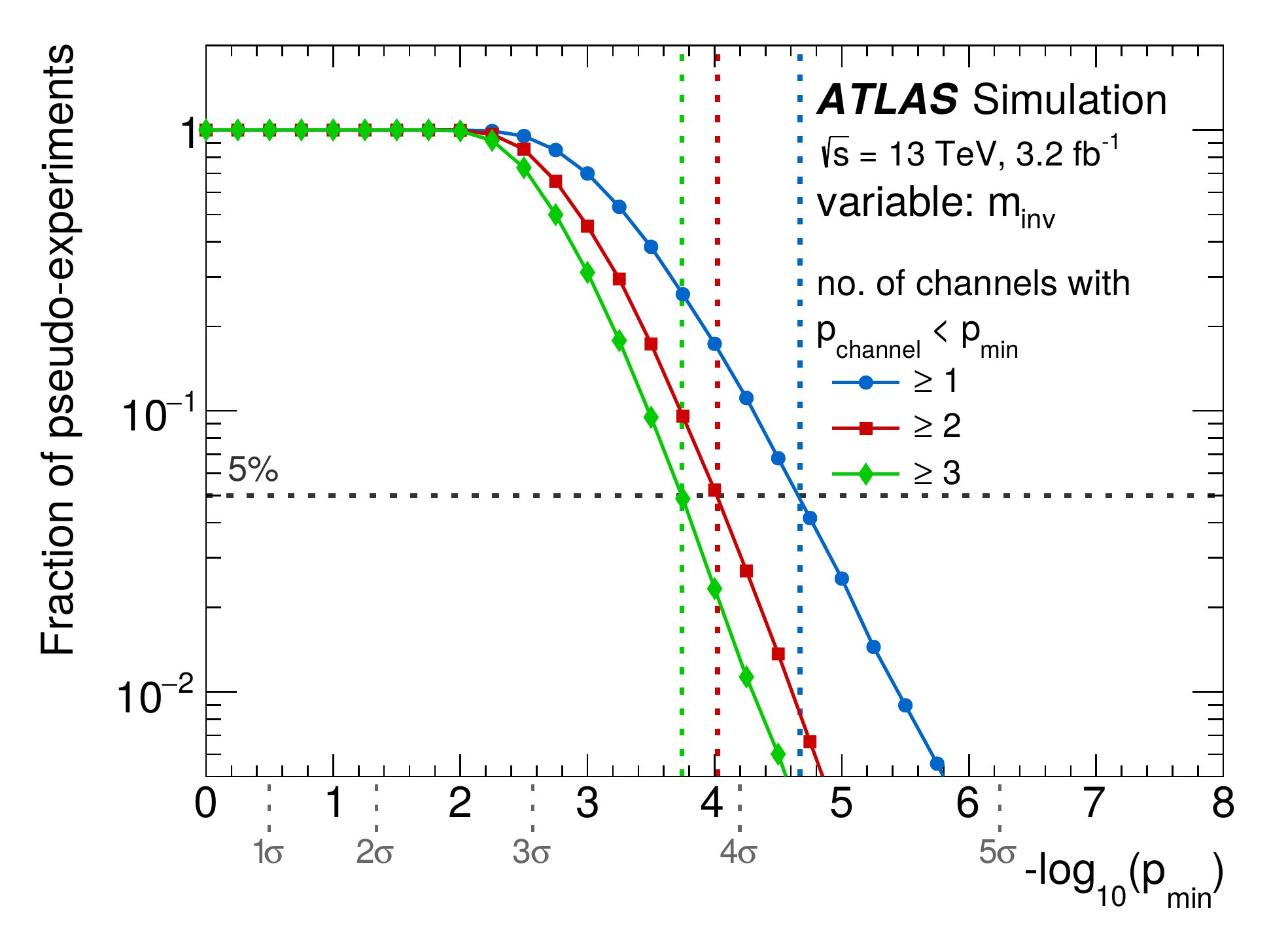}
  \end{center}
  \vspace*{-0.5cm}
  \caption{\label{fig:nominal_smonly} 
The fractions of pseudo-experiments ($\pexp(p_{\text{min}})$) in the \minv{} scan, 
which have at least one, two or three \pchannel-values 
smaller than a given threshold ($p_{\text{min}}$).
Pseudo-datasets are generated from the SM expectation.
Dotted lines are drawn at $\pexp=5\%$ and at the corresponding $-\log_{10}(p_{\text{min}})$-values.}
\end{figure}

To illustrate this, \Fig{\ref{fig:nominal_smonly}} shows three cumulative distributions of \pchannel-values from pseudo-experiments. 
The number of event classes (686) and the \minv{} distributions used to generate these pseudo-experiments coincide 
with the example application in \Sect{\ref{sec:application_strategy}}.
The distribution in \Fig{\ref{fig:nominal_smonly}} 
with circular markers 
is the fraction of pseudo-experiments with at least one
\pchannel-value smaller than $p_{\text{min}}$.
For example, about 15\% of the pseudo-experiments have at least one \pchannel-value smaller than $p_{\text{min}}=10^{-4}$.
Therefore, the estimated probability (\pexp) of obtaining at least one \pchannel-value ($i=1$) smaller than $10^{-4}$ from data in the absence of a signal is about 15\%, 
or $P_{\text{exp},1}(10^{-4}) = 0.15$. To estimate the probability of observing deviations of a given size in at least two or three different event classes, the second or third smallest \pchannel-value of a pseudo-experiment is compared with a given $p_{\textrm{min}}$ threshold.
From \Fig{\ref{fig:nominal_smonly}} it follows for instance that 2\% of the pseudo-experiments have at least three \pchannel-values smaller than $10^{-4}$.
Consequently, the probability of obtaining a third smallest \pchannel-value smaller than $10^{-4}$ from data in the absence of a signal is about 2\%,
or $P_{\text{exp},3}(10^{-4}) = 0.02$

In \Fig{\ref{fig:nominal_smonly}} a horizontal dotted line is drawn at a fraction of pseudo-experiments of 5\%
and corresponding vertical dotted lines are drawn at the three $p_{\text{min}}$ thresholds.
The observation of one, two or three \pchannel-values in data below the corresponding $p_{\text{min}}$ threshold,
i.e.\ an observation with a $\pexp < 0.05$,
promotes the selections that yielded these deviations to signal regions that can be 
tested in a new dataset.

\subsection{Step 5: Evaluation of the sensitivity}
The sensitivity of the procedure 
to a priori unspecified BSM signals 
can be evaluated with two different methods
that 
either use a modified background estimation through the removal of SM processes 
or 
in which signal contributions are added to the pseudo-data sample.

In the first method, a rare
SM process
(with either a low cross-section or a low reconstruction efficiency)
is removed from the background model. 
The search algorithm is applied again to test the data 
or `signal' pseudo-experiments generated from the unmodified SM expectation,
against the modified background expectation. 
The data samples would be expected to reveal excesses 
relative to the modified background prediction.

In the second method,
pseudo-experiments  are used  to test the sensitivity of the analysis to benchmark signal models of new physics.
The prediction of a model is added to the SM prediction,
and this modified expectation is 
used to generate `signal' pseudo-experiments.
The search algorithm is 
applied to the pseudo-experiments 
and the distribution of 
\pchannel-values is derived.

To provide a figure of merit for the sensitivity of the analysis, the fraction of 
`signal' pseudo-experiments 
with $\pexp < 5\%$ for $i=1,2,3$ is computed.

\subsection{Step 6: Results}
Finding one or more deviations in the data with $\pexp<5\%$
triggers a dedicated analysis that uses 
the data selection in which the deviation is observed as a signal region (step 7). 
If no significant deviations are found, the outcome of the analysis technique includes information such as: the number of events and expectation per event class, 
a comparison of the data with the SM expectation in the distributions of observables considered,
the scan results (i.e.\ the location and the local \pvalue-value of the largest deviation per event class) 
and the comparison with the expectation from pseudo-experiments.

\subsection{Step 7 (only in the case of $\pexp<5\%$): Dedicated analysis of deviation}
\paragraph{Dedicated analysis on original dataset}
Deviations are investigated 
using methods similar to those of 
a conventional analysis. 
In particular, the background 
prediction is determined using control selections to control and validate the background modelling. 
Such a procedure further constrains the background expectation and uncertainty, and reduces the dependence on simulation.
If such a re-analysis of the region results in an insignificant deviation, it can be inferred that the deviation seen before was due to mismodellings or not well-enough understood backgrounds.

\paragraph{Dedicated analysis on an independent dataset}
If a deviation 
persists in
a dedicated analysis
using the original dataset, 
the 
data selection 
in which the deviation is observed 
defines a 
data-derived signal region 
that is tested 
in an independent new dataset with a similar or larger integrated luminosity. 
At this point, 
a particular model of new physics can be
used to interpret the result of testing the
data-derived signal region.
Since 
the signal region is known, the corresponding  data 
can be excluded (`blinded') from the analysis until the very end to minimize any possible bias in the analysis. 
Additionally, 
since only a few optimized hypothesis tests 
are performed on the independent dataset,
the large look-elsewhere effect due to the large number of hypothesis tests performed in step 3
is not present in the dedicated analysis of the signal region(s). The assumptions of Gaussian uncertainties for the background models can also be tested in the dedicated analysis.
If the full LHC data yields a significant deviation, the LHC 
running time may need to be increased, or the excess may have to be followed up at a future collider.

\subsection{Advantages and disadvantages}
The features of 
this 
strategy lead to several advantages and disadvantages that are outlined below.   

Advantages:
\begin{itemize}
\item It can find unexpected signals 
for new physics due to the large number of event classes and phase-space regions probed, which may otherwise remain uninvestigated. 
\item A relatively small excess in two or three independent data selections, each of which is not big enough to trigger 
a dedicated analysis by itself ($P_{\text{exp},1} > 5\%$), can trigger one in combination  ($P_{\text{exp},2,3} < 5\%$).
\item The approach is broad, and the scanned distributions can be used to probe the overall description of the data by the event generators for many SM processes.
\item The probability of a deviation 
occurring in \emph{any} of the many different event classes under study 
can be determined with pseudo-experiments, resulting in a 
truly global interpretation of the probability 
of finding a deviation within an experiment
such as ATLAS. 
\end{itemize}

Disadvantages:
\begin{itemize}
\item
The outcome depends on the MC-based description of physics processes and simulations of the detector response. 
Event classes in which the majority of the events contain misreconstructed objects are typically poorly modelled by MC simulation and might need to be excluded from the analysis.
Although step 2  validates  the  description  of the data by the MC simulation, there is still a
possibility 
of triggering 
false 
positives due to an MC mismodelling in a corner of phase space. Step 7 aims to minimize this by reducing the 
dependence
on MC simulations in a dedicated analysis performed for each significant deviation.
In future implementations a better background model could be constructed with the help of control regions or data-derived fitting functions. 
This might allow the detection of excesses that are small compared to the uncertainties in the MC-based description of the SM processes.
\item
Since this analysis is not optimized for a specific class of BSM signals, a dedicated analysis optimized for a given BSM signal achieves a larger sensitivity to that signal. 
The enormous parameter space of possible signals makes an optimized search for each of them impossible.
\item
The large  number of data selections introduce a large look-elsewhere effect, which reduces the significance of a real signal. 
Step 7 circumvents this problem since the final discovery significance is determined with a dedicated analysis of one or a few data selection(s) 
and a statistically independent dataset. 
This can yield an improved signal sensitivity if the background uncertainty can be constrained in the dedicated analysis.
\item
Despite being broad, the procedure might miss a certain signal because it does not 
show a localized excess in one of the studied distributions. 
This might be overcome with better observables, better event classification or modified algorithms, which may then be sensitive to such signals.
\end{itemize}

\section{Application of the strategy to ATLAS data}
   \label{sec:application_strategy}
   This section describes the application of the strategy
   outlined in the previous section to the 13~\TeV{} $pp$ collision data recorded by  the ATLAS experiment in 2015.

   \subsection{Step 1: Data selection and Monte Carlo simulation}
      \subsubsection{ATLAS detector and dataset}
\label{sec:ATLAS}
The ATLAS detector~\cite{PERF-2007-01} is a multipurpose particle physics detector with a forward-backward 
symmetric cylindrical geometry and a coverage of nearly $4\pi$ in solid angle.\footnote{
  ATLAS uses
  a right-handed coordinate system with its origin at the nominal interaction point in the centre of the detector.
  The positive $x$-axis is defined by the direction from the interaction point to the centre of the LHC ring,
  with the positive $y$-axis pointing upwards, while the beam direction defines the $z$-axis.
  Cylindrical coordinates ($r$, $\phi$) are used in the transverse plane, $\phi$ being the azimuthal angle around the $z$-axis.
  The pseudorapidity $\eta$ is defined in terms of the polar angle $\theta$ by $\eta = - \ln\tan(\theta/2)$.
  The angular distance is defined as $\Delta R = \sqrt{(\Delta \eta)^2 + (\Delta\phi)^2}$.
  Rapidity is defined as $y = 0.5\cdot\ln[(E + p_z)/(E - p_z)]$ where $E$ denotes the energy and $p_z$ is the component of the momentum along the beam direction.} 
The inner tracking detector (ID) consists of 
silicon pixel and 
microstrip detectors covering the pseudorapidity region $|\eta| < 2.5$,
surrounded by a 
straw-tube 
transition radiation tracker which enhances electron identification in the region $|\eta| < 2.0$.
Between \RunOne and \RunTwo, a new inner pixel layer, the insertable B-layer~\cite{Capeans:1291633}, was inserted at a mean sensor radius of 3.3 cm.
The inner detector is surrounded by a thin superconducting solenoid providing an axial 2~T magnetic field 
and by a fine-granularity lead/liquid-argon (LAr) electromagnetic calorimeter covering $|\eta|<3.2$.
A steel/scintillator-tile calorimeter provides hadronic coverage in the central pseudorapidity range ($|\eta| < 1.7$).
The endcap and forward calorimeter coverage ($1.5 < |\eta| < 4.9$) is completed by
LAr active layers with either copper or tungsten as the absorber material. 
An extensive muon spectrometer with an air-core toroid magnet system surrounds the calorimeters.
Three layers of high-precision tracking chambers provide coverage in the range $|\eta| < 2.7$,
while dedicated fast chambers 
provide a muon trigger 
in the region $|\eta| < 2.4$.
The ATLAS trigger system consists of a hardware-based level-1 trigger followed by a software-based high-level trigger~\cite{TRIG-2016-01}.

The data used in this analysis 
were 
collected by the ATLAS detector during 2015 in \pp collisions at the LHC with a centre-of-mass energy of 13~\TeV{} and a 25~ns bunch crossing interval.
After applying quality criteria for the beam, data and detector, the available dataset corresponds to an integrated luminosity of 3.2~\ifb.
In this dataset, each event includes an average of approximately 14 additional inelastic \pp collisions in the same bunch crossing (pile-up).

Candidate events are 
required to have a reconstructed vertex~\cite{ATL-PHYS-PUB-2015-026}, 
with at least two associated tracks with $\pt> 400$~\MeV.
The vertex with the highest sum of squared transverse momenta of the tracks is considered 
to be 
the primary vertex.

      \subsubsection{Monte Carlo samples}
\label{sec:BGSamples}
Monte Carlo simulated event samples~\cite{SOFT-2010-01} are used 
to describe SM background processes and to model 
possible signals. 
The ATLAS detector is simulated  either by a software system based on \textsc{Geant4}~\cite{Agostinelli:2002hh}
or by a faster simulation based on a parameterization 
of the calorimeter response and \textsc{Geant4} for the other detector systems.
The impact of detector conditions on the simulation is typically corrected for as part of the calibrations and scale factors applied to the reconstructed objects.

To account for additional \pp interactions from the same or nearby bunch crossings,
a set of minimum-bias interactions generated using \PYTHIAV{8.186}~\cite{Sjostrand:2007gs}, 
the MSTW2008LO~\cite{Martin:2009iq} parton distribution function (PDF) set 
and the A2 set of tuned parameters (tune)~\cite{ATL-PHYS-PUB-2012-003} was superimposed onto the hard-scattering events
to reproduce the observed distribution of the average number of interactions per bunch crossing.

Any further study of time-dependent detector variations would be part of the dedicated search following any interesting deviation. 

In all MC samples, except those produced by \textsc{Sherpa}~\cite{Gleisberg:2008ta}, the \textsc{EvtGen}~v1.2.0 program~\cite{EvtGen}
was used to model the properties of the bottom and charm hadron decays. 
The SM MC 
programs 
are listed in \Tab{\ref{tab:MC}} and a detailed explanation
can be found in \App{\ref{appendixA1}}.

\begin{sidewaystable}
  \centering\footnotesize
  \caption{A summary of the MC samples used in the analysis to model  SM background processes.
    For each sample the corresponding generator, 
    matrix element (ME) accuracy, 
    parton shower, cross-section normalization accuracy, PDF set and  tune are indicated. Details are given in \App{\ref{appendixA1}}. Samples with `data' in the `cross-section normalization' column are scaled to data as described in \Sect{\ref{sec:reweighting}}. \Zboson refers to $\gamma^*$/~$Z$.}
  \label{tab:MC}
  \begin{tabular}{l l c c c c c}
    \toprule
    Physics process						& Generator	&ME accuracy		& Parton shower 			& Cross-section 		& PDF set 			& Tune  			\\
    							        &		&					& 							& normalization 		& 					& 	 			\\
    \midrule
    \Wboson (\ra\lnu) + jets			& \SHERPAV{2.1.1}&	0,1,2j@NLO + 3,4j@LO	& \SHERPAV{2.1.1}			& NNLO  				& NLO CT10    		& \SHERPA default \\
    
    \Zboson(\ra\leplep) + jets 			& \SHERPAV{2.1.1} &	0,1,2j@NLO + 3,4j@LO	& \SHERPAV{2.1.1}  			& NNLO 					& NLO CT10         	& \SHERPA default \\
    
    \Zboson~/~\Wboson(\ra\qqbar) + jets 	& \SHERPAV{2.1.1} &	1,2,3,4j@LO				& \SHERPAV{2.1.1}  			& NNLO 					& NLO CT10         	& \SHERPA default \\

    \Zboson~/~\Wboson + $\gamma$	& \SHERPAV{2.1.1} & 0,1,2,3j@LO	& \SHERPAV{2.1.1} 			& NLO  					& NLO CT10         	& \SHERPA default	\\

    \Zboson~/~\Wboson + $\gamma\gamma$	& \SHERPAV{2.1.1} 	& 0,1,2,3j@LO& \SHERPAV{2.1.1} 			& NLO  					& NLO CT10         	& \SHERPA default	\\
    
    $\gamma$ + jets 			& \SHERPAV{2.1.1} & 0,1,2,3,4j@LO	& \SHERPAV{2.1.1}       	& data  				& NLO CT10         	& \SHERPA default	\\
        
    $\gamma\gamma$ + jets 			& \SHERPAV{2.1.1}& 0,1,2j@LO 	& \SHERPAV{2.1.1}       	& data  				& NLO CT10         	& \SHERPA default	\\
        
    $\gamma\gamma\gamma$ + jets 		& MG5\_aMC@NLO~2.3.3	& 0,1j@LO 	&     \PYTHIAV{8.212}   	& 	LO			&  NNPDF23LO         	& A14	\\
    
 	$\ttbar$              				& \POWHEGBOX~v2 & NLO	& \PYTHIAV{6.428}  			& NNLO+NNLL				& NLO CT10  		& \Perugia~2012	\\
        
    $\ttbar+\Wboson$       				& MG5\_aMC@NLO~2.2.2 &	0,1,2j@LO& \PYTHIAV{8.186}   		& NLO  					& NNPDF2.3LO 		& A14     		\\
    $\ttbar+\Zboson$       				& MG5\_aMC@NLO~2.2.2 &	0,1j@LO& \PYTHIAV{8.186}   		& NLO  					& NNPDF2.3LO 		& A14     		\\
    $\ttbar+\Wboson\Wboson$       				& MG5\_aMC@NLO~2.2.2&	LO & \PYTHIAV{8.186}   		& NLO  					& NNPDF2.3LO 		& A14     	\\
    $t\bar{t}+\gamma$       			& MG5\_aMC@NLO~2.2.2 &	LO& \PYTHIAV{8.186}   		& LO  					& NNPDF2.3LO 		& A14    \\
    \ttbar + \bbbar              		& \SHERPAV{2.2.0} & NLO 	& \SHERPAV{2.2.0}  			& NLO 			& NLO CT10f4  		& \SHERPA default		\\
    
    Single-top (t-channel)           		    & \POWHEGBOX~v1 & NLO	& \PYTHIAV{6.428} 			& app. NNLO  					& NLO CT10f4   		& \Perugia~2012	 \\
    Single-top (s- and $\Wboson t$-channel) 	& \POWHEGBOX~v2 & NLO	& \PYTHIAV{6.428}  			& app. NNLO  					& NLO CT10   		& \Perugia~2012	 \\ 
    $t\Zboson$       					& MG5\_aMC@NLO~2.2.2&	LO & \PYTHIAV{8.186}   		& LO  					& NNPDF2.3LO 		& A14     \\
    3-top      					& MG5\_aMC@NLO~2.2.2 &	LO& \PYTHIAV{8.186}   		& LO  					& NNPDF2.3LO 		& A14     		 \\
    4-top       				& MG5\_aMC@NLO~2.2.2&	LO & \PYTHIAV{8.186}   		& NLO  					& NNPDF2.3LO 		& A14     	\\
    
	$\Wboson\Wboson$    		& \SHERPAV{2.1.1}&	0j@NLO + 1,2,3j@LO 	& \SHERPAV{2.1.1}     		& NLO  					& NLO CT10   		& \SHERPA default \\
	$\Wboson\Zboson$     		& \SHERPAV{2.1.1}  &	0j@NLO + 1,2,3j@LO	& \SHERPAV{2.1.1}     		& NLO  					& NLO CT10   		& \SHERPA default \\
    $\Zboson\Zboson$    		& \SHERPAV{2.1.1} & 0,1j@NLO + 2,3j@LO	& \SHERPAV{2.1.1}     		& NLO  					& NLO CT10   		& \SHERPA default 	\\
    Multijets							& \PYTHIAV{8.186} 	&	LO& \PYTHIAV{8.186}   		& data  				& NNPDF2.3LO 		& A14     	\\
    
    Higgs (ggF/VBF)           			& \POWHEGBOX~v2 & NLO	& \PYTHIAV{8.186} 			& NNLO  				& NLO CT10   		& \textsc{AZNLO}\\
    Higgs ($t\bar{t}H$)           		& MG5\_aMC@NLO~2.2.2 & NLO	& \HERWIGpp 			& NNLO 					& NLO CT10   		& \textsc{UEEE5}	\\
	Higgs ($\Wboson/\Zboson H$)           			& \PYTHIAV{8.186} 	&	LO& \PYTHIAV{8.186} 			& NNLO  				& NNPDF2.3LO 		& A14 		\\
    
    Tribosons     						& \SHERPAV{2.1.1} & 0,1,2j@LO	& \SHERPAV{2.1.1}    		& NLO  					& NLO CT10   		& \SHERPA default 	\\
    
    \bottomrule
  \end{tabular}
\end{sidewaystable}

In addition to the SM background processes, two possible signals are considered as benchmarks.
The first benchmark model considered is the production of a new heavy neutral gauge boson of spin 1 ($Z'$), as predicted by many extensions of the SM. 
Here, the specific case of the 
sequential extension of the SM gauge group (SSM)~\cite{London:1986dk,Langacker:2008yv}
is considered, for which the couplings 
are the same as for the SM $Z$ boson.  
This process was generated at leading order (LO) using \PYTHIAV{8.212} with the NNPDF23LO~\cite{Ball:2012cx} PDF set and the A14 tune~\cite{ATL-PHYS-PUB-2014-021}, 
as a Drell--Yan process, 
for five different resonant masses,
covering the range from 2~\TeV{} to 4~\TeV, in steps of 0.5~\TeV.  The considered decays of $Z'$ bosons are inclusive, covering the full range
of lepton and quark pairs. 
Interference effects with SM Drell--Yan production are not included, 
and the $Z'$ boson is required to decay into fermions only.

The second signal considered is the supersymmetric~\cite{Golfand:1971iw,Volkov:1973ix,Wess:1974tw,Wess:1974jb,Ferrara:1974pu,Salam:1974ig} 
production of gluino pairs through strong interactions.
The gluinos are assumed to decay promptly into a pair of top quarks and an almost massless neutralino via an off-shell top squark $\tilde{g}\to tt\tilde{\chi}^0_1$. 
Samples for this process were generated at LO with up to two additional partons 
using MG5\_aMC@NLO~2.2.2~\cite{Alwall:2014hca} 
with the CTEQ6L1~\cite{Pumplin:2002vw} PDF set, interfaced to \PYTHIAV{8.186} with the A14 tune.
The matching with the parton shower was done using the CKKW-L~\cite{Lonnblad:2011xx} prescription,
with a matching scale set to one quarter of the pair-produced resonance mass.
The signal cross-sections were calculated at next-to-leading order (NLO) in the strong coupling constant,
adding the resummation of soft gluon emission at next-to-leading-logarithm (NLL) accuracy~\cite{Beenakker:1997ut,Beenakker:2010nq,Beenakker:2011fu}.

      \subsubsection{Object reconstruction}
\label{sec:Objects}
Reconstructed physics objects considered in the analysis are:
prompt and isolated electrons ($e$), 
muons ($\mu$)  
and photons ($\gamma$),
as well as
$b$-jets ($b$) and light (non-$b$-tagged) jets ($j$)
reconstructed with the anti-$k_t$ algorithm~\cite{Cacciari:2008gp} with radius parameter $R = 0.4$,
and large missing transverse momentum ($\met$). 
\Tab{\ref{tab:Classification}} lists the reconstructed physics objects along with their 
\pt and pseudorapidity requirements.
Jets and electrons misidentified as hadronically decaying $\tau$-leptons are difficult to model with the MC-based approach used in this analysis. Therefore, the identification 
of hadronically decaying $\tau$-leptons is not considered; they are 
mostly reconstructed as light
jets.
Details of the object reconstruction can be found in \App{\ref{appendixB}}.

\begin{table}[htbp]
\centering
\footnotesize
\caption{\label{tab:Classification} The physics objects used for classifying the events, 
  with their corresponding label, minimum \pt{} requirement, and pseudorapidity requirement.}
  \begin{tabular}{lc|cc}
    \toprule
    Object             		& Label 	& \pt{} (min) [\GeV] 		& Pseudorapidity \\
    \midrule
    Isolated electron      	& $e$ 		& 25 	& $|\eta| < 1.37$ or $1.52 < |\eta| < 2.47$		\\ 
    Isolated muon			& $\mu$		& 25	& $|\eta| < 2.7$								\\
    Isolated photon			& $\gamma$	& 50	& $|\eta| < 1.37$ or $1.52 < |\eta| < 2.37$		\\
    $b$-tagged jet          & $b$ 		& 60    & $|\eta| < 2.5$  								\\ 
    Light (non-$b$-tagged) jet & $j$  		& 60  	& $|\eta| < 2.8$ 								\\
    Missing transverse momentum& $\met$ & 200   & 												\\
    \bottomrule
  \end{tabular}
\end{table}

After object identification, overlaps between object candidates are resolved using the distance variable 
$\Delta R_y = \sqrt{(\Delta y)^2 + (\Delta\phi)^2}$.
If an electron and a muon share the same ID track, the electron is removed.
Any jet within a distance $\Delta R_y=0.2$ 
of an electron candidate is discarded, 
unless the jet has a value of the 
$b$-tagging 
\textsc{MV2c20} discriminant~\cite{PERF-2012-04,ATL-PHYS-PUB-2015-022}
larger than that corresponding to approximately 85\% $b$-tagging efficiency,
in which case the electron is discarded since it probably originated from a semileptonic $b$-hadron decay.
Any remaining electron within $\Delta R_y = 0.4$ of a jet is discarded.
Muons within $\Delta R_y =0.4$ of a jet are also removed. 
However, if the jet has fewer than three associated tracks, 
the muon is kept and the jet is discarded instead to avoid inefficiencies
for high-energy muons undergoing significant energy loss in the calorimeter.
If a photon candidate is found within $\Delta R_y=0.4$ of a jet,
the jet is discarded. 
Photons within a cone of size $\Delta R_y=0.4$ around an electron or muon candidate are discarded.

The missing transverse momentum (with magnitude \met) is defined as the negative vector sum of the transverse momenta
of all selected and calibrated physics objects (electrons, photons, muons and jets) in the event, with an additional soft-term~\cite{PERF-2016-07}.
The soft-term is constructed from all tracks that are not associated with any physics object, 
but are associated with the primary vertex.
The missing transverse momentum is reconstructed for all events; however, separate analysis channels are constructed for events with $\met > 200$~\GeV.
These events are taken exclusively from the \met{} trigger.

\subsubsection{Event selection and classification}
\label{sec:Classification}
The events are 
divided into mutually exclusive classes 
that are labelled with
the number and type of reconstructed objects 
listed in \Tab{\ref{tab:Classification}}.
The 
division can be regarded as a classification
according to the most important features of the event.
The classification includes all possible final-state configurations
and object multiplicities, e.g.\ if a data event with seven reconstructed muons
and no other objects is found,
it is classified in a `$7$-muon' event class ($7\mu$).
Similarly an event with missing transverse momentum, two muons,
one photon and four jets is classified and considered in the corresponding event class denoted $\met 2\mu1\gamma4j$.

\begin{figure}[htbp]
  \begin{center}
    \includegraphics{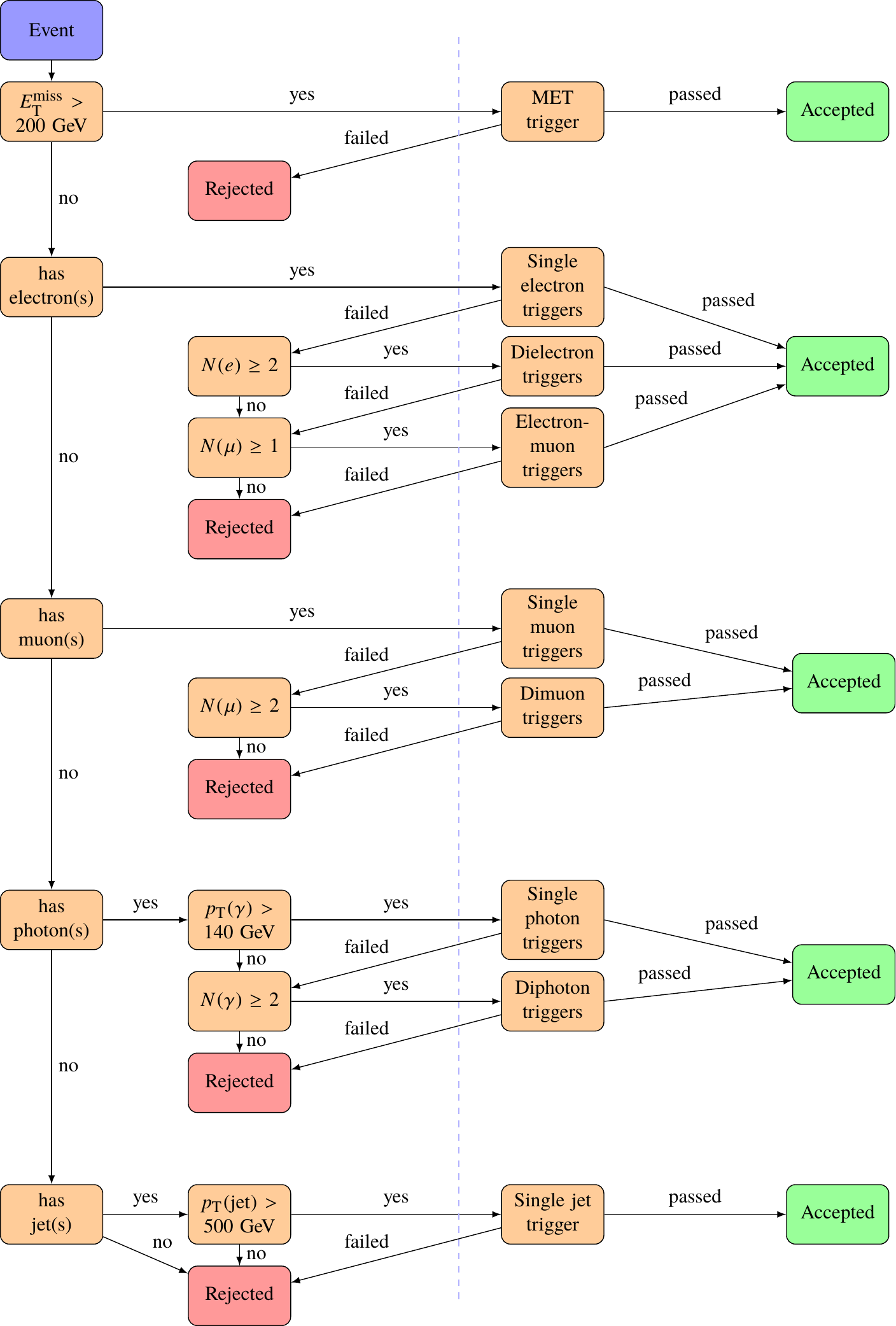}
  \end{center}
\caption{\label{fig:triggers} Flow diagram for the trigger and offline event selection strategy. The offline requirements are shown on the left of the dashed line and the trigger requirements are shown on the right of the dashed line.}   
\end{figure}

All events contributing to a
particular event class are 
also 
required to be selected by a 
trigger from a
corresponding 
class of 
triggers by
imposing a hierarchy in the event selection. 
This avoids ambiguities in the application of trigger efficiency corrections to MC simulations and avoids variations in the acceptance within an event class.
The flow diagram in \Fig{\ref{fig:triggers}} gives a graphical representation 
of the trigger and offline event selection, based on the class of the event. 
Since the thresholds for the single-photon and single-jet triggers are 
higher than the \pt requirements in the photon and jet 
object selection, an additional 
reconstruction-level \pt cut is imposed to avoid trigger inefficiencies.
For the other triggers, the \pt requirements in the object definitions exceed the trigger thresholds 
by a sufficient margin 
to avoid additional trigger inefficiencies.
Electrons are considered before muons in the event selection hierarchy because the electron trigger efficiency is considerably higher compared to the muon trigger efficiency.

Events with 
$\met > 200$~\GeV{} are required to pass the \met{} trigger which becomes fully efficient at 200~\GeV{}, otherwise they are rejected and not considered for further event selection.
If the event has $\met < 200$~\GeV{} but contains an electron with $\pt > 25$~\GeV{} it is required to pass the single-electron trigger.
However, events with more than one electron with $\pt > 25$~\GeV{} or with an additional muon with $\pt > 25$~\GeV{} can be selected by the dielectron trigger 
or electron-muon trigger respectively if the event fails to pass the single-electron trigger.
Events with a muon with $\pt > 25$~\GeV{} but no reconstructed electrons or large \met{} 
are required to pass the single-muon trigger.
If the event has more than one muon with $\pt > 25$~\GeV{} and fails to pass the single-muon trigger, it can additionally be selected by the dimuon trigger.
Remaining events with a photon with $\pt > 140$~\GeV{} or two photons with $\pt > 50$~\GeV{}
are required to pass the single-photon or diphoton trigger, respectively.
Finally, any remaining event with no large \met{}, leptons, or photons, but containing a jet with $\pt > 500$~\GeV{} is required to pass the single-jet trigger.

In addition to the thresholds imposed by the trigger,
a further selection is applied 
to event classes with $\met<200$~\GeV{} containing one lepton or one electron and one muon 
and 
possibly additional photons or jets ($1\mu+X$, $1e+X$ and $1\mu1e+X$), to reduce the overall data volume.
In these 
event classes, one lepton is required to have $\pt > 100$~\GeV{} if the event has less than three jets with $\pt > 60$~\GeV{}.

To suppress sources of fake $\met$, additional requirements are imposed on events
to be classified in $\met$ categories.
The ratio of \met{} 
to \meff{} is required to be greater than 0.2,
and the minimum azimuthal separation between the \met{} direction
and the three leading reconstructed jets (if present)
has to be greater than 0.4, otherwise the event is rejected.

   \subsection{Step 2: Systematic uncertainties and validation}
      \subsubsection{Systematic uncertainties}
\label{sec:Systematics}
\paragraph{Experimental uncertainties}
The dominant experimental systematic uncertainties
in the SM expectation for the different event classes typically
are 
the jet energy scale (JES) and resolution (JER)~\cite{PERF-2016-04}
and the scale and resolution of the \met{} soft-term.
The uncertainty related to the modelling of \met{}
in the simulation is estimated by propagating
the uncertainties in the energy and momentum scale
of each of the objects entering the calculation, 
with an additional uncertainty in the resolution and scale of the soft-term~\cite{PERF-2016-07}.
The uncertainties in 
correcting the efficiency of identifying jets containing $b$-hadrons in MC simulations
are determined 
in data samples enriched in top quark decays, and in simulated events~\cite{PERF-2012-04}.
Leptonic decays of $J/\psi$ mesons and $Z$ bosons in data and simulation 
are exploited to estimate the uncertainties in lepton reconstruction, 
identification, momentum/energy scale and resolution, 
and isolation criteria \cite{PERF-2015-10,ATLAS-CONF-2016-024,ATL-PHYS-PUB-2016-015}.
Photon reconstruction and identification efficiencies are evaluated from samples of $Z\to ee$ and $Z+\gamma$ events~\cite{ATL-PHYS-PUB-2016-014,ATL-PHYS-PUB-2016-015}.
The luminosity measurement was calibrated during dedicated beam-separation scans, using the same methodology 
as that described in \Ref{\cite{DAPR-2013-01}}.
The uncertainty of this measurement is found to be 2.1\%.

In total, 35 sources of experimental uncertainties are identified pertaining to one or more physics objects considered.
For each source the one-standard-deviation $(1\sigma)$ confidence interval (CI) is propagated to a $1\sigma$ CI around the nominal SM expectation. 
The total experimental uncertainty of the SM expectation is obtained from the sum in quadrature of these 35 $1\sigma$ CIs and the uncertainty of the luminosity measurement.

\paragraph{Theoretical modelling uncertainties}
Two different sources of uncertainty in the theoretical modelling of the SM production processes are considered.
A first uncertainty
is assigned to account for our knowledge of the cross-sections for the inclusive processes.
A second uncertainty
is used to cover the modelling 
of the shape of the differential cross-sections.
In order to derive the modelling uncertainties, either variations of the QCD 
factorization, renormalization, resummation and merging 
scales are used
or comparisons of the nominal MC samples with alternative ones are used.
For some SM processes additional modelling uncertainties are included.
\App{\ref{appendixA2}} describes all theoretical uncertainties considered
for the  various SM processes.
The total uncertainty is taken as the sum in quadrature of the two components and the statistical uncertainty of the MC prediction.

\subsubsection{Validation procedures}
\label{sec:Validation_procedures}
The evaluated SM processes, together with their standard selection cuts and the studied validation distributions, are detailed in \Tab{\ref{tab:valid}}. 
These validation distributions rely on inclusive selections to probe the general agreement between data and simulation and are evaluated in restricted ranges where large new-physics contributions have been excluded by previous direct searches.

\begin{table}[htbp]
  \centering
  \footnotesize
  \caption{A summary of the SM processes and their inclusive selections used to validate the background modelling. 
    For each selection the $\pt$, $\eta$, and $\phi$ distributions of the objects used in the selection and of additional jets are included as validation distributions by default.
    Additional validation distributions are listed per selection.
    In all cases, `jet(s)' refers to both $b$-tagged and non-$b$-tagged jets, except where `$b$-jet' is mentioned explicitly.
  	$\HT(\text{jets})$ is defined as the scalar $\pt$ sum of all the jets in the event. 
    Some selections rely on the transverse mass ($\mt(\ell,\met)$) which is defined as $[2\ \pt(\ell)\ \met\ (1 - \cos \Delta\phi(\ell,\met))]^{1/2}$.
    $N((b\text{-})\text{jets})$ is the number of ($b$-)jets in an event.
    For the distance variables $\Delta R$ and $\Delta \phi$, the two instances of the objects with the minimum distance between them are used.
    The $\HT(\text{jets})$, $\mt(\ell,\met)$, $\pt(\ell\ell)$ and $\minv$ validation distributions are evaluated in restricted ranges where large new-physics contributions have been excluded by previous direct searches.
    Same-flavour opposite-charge sign lepton pairs are referred to as SFOS\ pairs.}      
  \label{tab:valid}
  \begin{tabular}{l l l}
    \toprule
    Physics process & Event selection & Additional validation distributions \\
    \midrule
    $W(\rightarrow \ell\nu) + \text{jets}$   & 1 lepton, $\met>25$ \GeV{}						& $		N(\text{jets}) 
                                                                                                \quad 	N(b\text{-jets})
                                                                                                \quad	\mt(\ell,\met)
                                                                                                \quad	\HT(\text{jets})
                                                                                                $ \\
                                      & and $\mt(\ell,\met)>50$ \GeV{} 							& $		
                                      																	\Delta 	R({\ell,\text{jet}}) 
                                                                                                \quad 	\Delta \phi({\ell,\met})
                                                                                                \quad 	\Delta \phi({\text{jet},\met})
                                                                                                $ \\ 
    								  \cmidrule(l{13pt}){2-3}
									  & \quad \& \quad $N(\text{jets}) \geq 3$ 					& $		{\HT(\text{jets})}
                                                                                                $ \\
                                      \cmidrule(l{13pt}){2-3}
									  & \quad \& \quad $N(b\text{-jets}) \geq 1$				& $		{\mt(\ell\text{, }\met)}
                                                                                                $ \\
    								  \cmidrule(l{13pt}){2-3}
									  & \quad \& \quad $N(b\text{-jets}) \geq 2$				& $		{\mt(\ell\text{, }\met)}
                                                                                                $ \\
    \midrule
    $Z(\rightarrow \ell \ell) + \text{jets}$ & 1 SFOS\ pair									    & $		N(\text{jets}) 
                                                                                                \quad 	N(b\text{-jets}) 
                                                                                                \quad	\minv(\ell\ell) 
                                                                                                \quad 	\pt(\ell\ell)
                                                                                                $ \\
                                      & $66<\minv(\ell\ell)<116$ \GeV{}							& $		\HT(\text{jets})
                                      															\quad 	\Delta R(\ell,\ell) 
                                                                                                \quad 	\Delta R(\ell,\text{jet}) 
                                                                                                $ \\
    								  \cmidrule(l{13pt}){2-3}
									  & \quad \& \quad $N(\text{jets}) \geq 2$ 					& $		{\HT(\text{jets})}
                                                                                                $ \\
									  \cmidrule(l{13pt}){2-3}
									  & \quad \& \quad $N(b\text{-jets}) \geq 1$ 				& $		\minv(\ell\ell) 
                                                                                                \quad 	\pt(\ell\ell)
                                                                                                $ \\
									  \cmidrule(l{13pt}){2-3}
									  & \quad \& \quad $N(b\text{-jets}) \geq 2$ 				& $		\minv(\ell\ell) 
                                                                                                \quad 	\pt(\ell\ell)
                                                                                                $ \\
    \midrule
    $W + \gamma(\gamma)$ 			  & same selection as $W(\rightarrow \ell\nu) + \text{jets}$& same distributions as $W(\rightarrow \ell\nu) + \text{jets}$ and \\    
									  & and 1(2) additional photon(s) 							& $		\Delta R(\ell,\gamma) 
                                                                                                $ \\    
    \midrule
	$Z + \gamma(\gamma)$ 			  & same selection as $Z(\rightarrow \ell \ell)$ + jets 	&  same distributions as $Z(\rightarrow \ell\ell) + \text{jets}$ and \\    
									  & and 1(2) additional photon(s) 							& $		\Delta R(\ell,\gamma) 
                                      															$ \\   
    \midrule
    $\gamma(\gamma) + \text{jets}$ 		  & 1(2) photon(s), no leptons							& $		N(\text{jets}) 
    																							\quad 	N(b\text{-jets}) 
                                                                                                \quad 	\HT(\text{jets})
                                                                                                \quad 	\minv(\gamma\gamma)
                                                                                                $ \\
                            		  & and at least 1(0) jet(s) 								& $ 	\Delta R(\gamma,\gamma) 
                                      															\quad 	\Delta R(\gamma,\text{jet}) 
                                                                                                $ \\
    \midrule
    $t\bar{t} \to $              	  & 1 lepton, at least 2 $b$-jets and at least 2 light jets & $		N(\text{jets}) 
                                                                                                \quad 	N(b\text{-jets})
                                                                                                \quad 	\minv(jj) 
                                                                                                $ \\
$\quad W (\to jj)+ {}$      		  & $50< \minv(jj) <110$ \GeV{} 			                & $		\mt(\ell,\met)
																								\quad	\Delta R(\text{jet},\text{jet}) 
                                                                                                \quad 	\Delta R(\ell,\text{jet})
                                                                                                $ \\
$\quad W (\to \ell\nu)+ {}$           & electron channel: $\mt(e\text{, }\met)>50$ \GeV{} 		& $ 	 
                                                                                                $ \\ 
$\quad bb$                     		  & muon channel: $\mt(\mu\text{, }\met)>60$ \GeV{} 		& $ 	
                                                                                                $ \\     
                            		  & $\met>40$ \GeV{}  										& $		
                                                                                                $ \\
    \midrule
Diboson & & \\
	\cmidrule(l{13pt}r{0pt}){1-2}
    $\quad WW$						  & 1 electron and 1 muon of opposite charge and no  jets  \\
            						  & $\met>50$ \GeV{} \\
	\cmidrule(l{13pt}r{0pt}){1-2}
    $\quad WZ$    					  & 1 SFOS\ pair $(\ell\ell)$	                            & $		\minv(\ell \ell)\text{ (SFOS pair(s))}
                                                                                                $ \\
        					  		  & and 1 lepton of different flavour $(\ell')$				& $
                                      																	\pt(\ell \ell)\text{ (SFOS pair(s))}
                                                                                                $ \\
            						  & $66<\minv(\ell\ell)<116$ \GeV{} 						& $		\mt(\ell',\met)
                                                                                                $ \\
            						  & $\met>50$ \GeV{} and $\mt(\ell',\met)>50$ \GeV{}		& $ 	\Delta R(\ell,\ell)
                                                                                                \quad 	\Delta \phi(\ell,\met)
                                                                                                $ \\ 
	\cmidrule(l{13pt}r{0pt}){1-2}
    $\quad ZZ$    					  & 2 SFOS\ pairs   \\
            						  & $66<\minv(\ell\ell)<116$ \GeV{} (both SFOS pairs)       \\
    \midrule
    Multijets   					  & at least 2 jets 										& $		N(\text{jets}) 
                                                                                                \quad 	N(b\text{-jets})
                                                                                                \quad 	\Delta R(\text{jet},\text{jet})
                                                                                                $ \\
    								  & no leptons or photons 									& $		\met/\meff 
                                                                                                $ \\
    \bottomrule
  \end{tabular}
\end{table}

There are some cases in which the validation procedure finds modelling problems and MC background corrections are needed (multijets, $\gamma(\gamma)$ + jets). 
In other cases, 
the affected event classes are excluded from the analysis as their SM expectation dominantly 
arises from object misidentification
(e.g.\ jets reconstructed as electrons) which is poorly modelled in MC simulation. The excluded classes are: $1e1j$, $1e2j$, $1e3j$, $1e4j$, $1e1b$, $1e1b1j$, $1e1b2j$, $1e1b3j$.
Event classes containing a single  object, as well as those containing only $\met$ and a lepton
are 
also discarded from the analysis
due to difficulties in modelling final states with one high energy object recoiling against many soft (non-reconstructed) ones.

      \subsubsection{Corrections to the MC background}
\label{sec:reweighting}
The MC samples for multijet and $\gamma+\text{jets}$ production, 
while giving a good description of kinematic variables, 
predict an overall cross-section and a jet multiplicity distribution 
that disagrees with data. Following step 2, correction
procedures were applied.

In classes containing only $j$ and $b$ the multijet MC samples are scaled to data with normalization factors
ranging between approximately 0.8 and 1.2. 
The normalization factors are derived separately in each exclusive jet multiplicity class
by equating the expected total number of events to the observed number of events. 
Multijet production in other channels are not rescaled and found to be described by the MC samples within the theoretical uncertainties. 
If a channel contains less than four data events, no modifications are made.

For $\gamma+\text{jets}$ event classes the same rescaling procedure is applied
to classes with exactly one photon, no leptons or \met, and any number of jets.

The \SHERPAV{2.1.1} MC generator has a known deficiency in the modelling of \met{}
due to too large forward jet activity.
This results in a visible mismodelling of the \met{} distribution in event classes with two photons,
which also affects the \meff{} distribution.
To correct for this mismodelling a reweighting~\cite{SUSY-2016-04} is applied
to the background 
events containing two real photons 
($\gamma\gamma+\text{jets}$).
The diphoton MC events are reweighted as a function of \met{} and of the number of selected jets
to match the respective distributions in the data for the inclusive diphoton sample in the range $\met < 100$~\GeV.
In no other event classes was the mismodelling large enough to warrant such a procedure.

The application of scale factors also outside the region where data to Monte Carlo comparisons are made would be cross-checked in the dedicated reanalysis of any deviation.

\subsubsection{Comparison of the event yields with the MC prediction}
After classification,  704  event classes are found with at least one data event or an SM expectation greater than 0.1 events.
The data and the background  predictions from MC simulation for these classes  are shown in 
\Fig{\ref{fig:global1}} 
and \App{\ref{appendixA3}}.
Agreement is observed between data and the  prediction in most of the event classes.
In events classes having more than two $b$-jets and where the SM expectation is dominated by $t\bar{t}$ production, the nominal SM expectation is systematically slightly below the data.
Data events are found in 528 out of 704 event classes.
These include events with up to 
four leptons (muons and/or electrons),
three photons, twelve jets and eight $b$-jets.
There are 18 event classes 
with an SM expectation of less than 0.1 events;
no more than two data events are observed 
in any of these,
and they are not considered further in the analysis.
No outstanding event was found in those channels.
The remaining 686 classes are retained for statistical analysis.

\begin{figure}[htbp]
  \begin{center}
 \includegraphics[height=16cm,width=16cm,keepaspectratio]{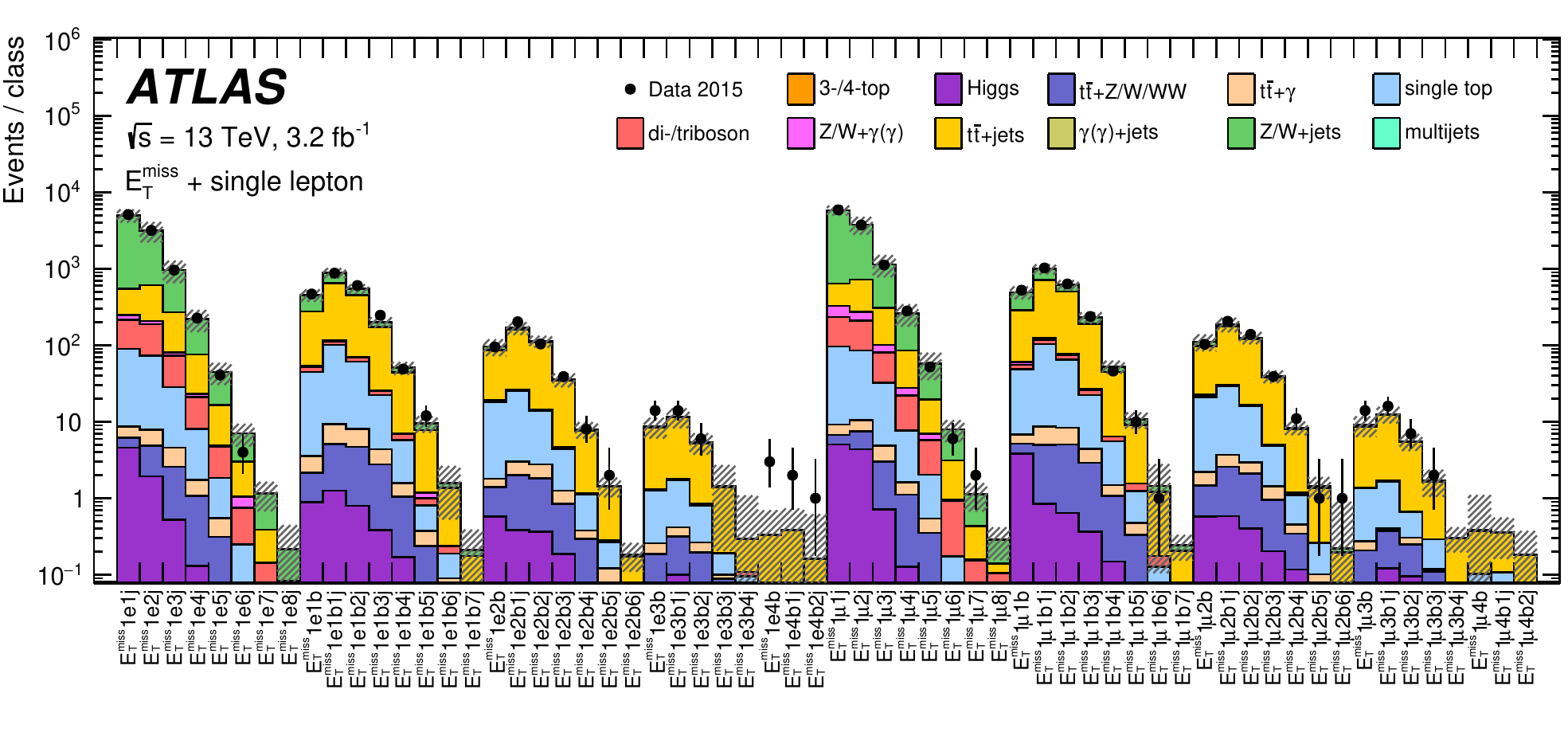}
  \end{center}
 \vspace*{-0.5cm}
  \caption{\label{fig:global1} The number of events in data, and for the different SM background predictions considered,
  	for classes with large 
    \met{}, one lepton and ($b$-)jets (no photons).
    The classes are labelled according to the multiplicity and type
    ($e$, $\mu$, $\gamma$, $j$, $b$, $\met$) of the reconstructed objects for the given event class.
    The hatched bands indicate the total uncertainty of the SM prediction. 
    This figure shows 60 out of 704 event classes, the remaining event classes can be found in \Figrange{\ref{fig:global2}}{\ref{fig:global14}} of \App{\ref{appendixA3}}. 
    }
\end{figure}

   \subsection{Step 3: Sensitive variables and search algorithm}
      \label{sec:Interpretation}
      In order to quantitatively determine the level of agreement between the
data and the SM expectation, and to identify regions of possible 
deviations, this analysis uses
an algorithm 
for multiple hypothesis testing.
The algorithm locates a single
region of largest deviation for
specific observables in each event class. 

In the following, 
an algorithm
derived from the algorithm used in~\Ref{\cite{Aktas:2004pz}} is
applied to the 2015 dataset.

\subsubsection{Choice of variables}
For each event class, the \meff{} 
and 
\minv{} distributions are considered in the form of histograms.
The invariant mass is computed from all visible objects in the event,
with no attempt to use the \met{} information. 
These variables have been widely used in  searches for new physics, 
and are sensitive to a large range of 
possible signals, manifesting either as bumps, deficits or wide excesses.
Several other commonly used kinematic variables 
have also been studied for various 
models, but were not found to significantly increase  the 
sensitivity. The approach is however not limited to these variables, as discussed in  
\Sect{\ref{sec:strategy}}.

For each histogram, the bin widths $h(x)$ as a function of the abscissa $x$ are determined using:
\begin{linenomath}
\[
h(x) = \sqrt{ \sum_{i=1}^{N_{\text{objects}}} k^2 \sigma_i^2(x/2) }
\]
\end{linenomath}
where $N_{\text{objects}}$ is the number of objects in the event class, 
$k$ is 
the width of the bin in standard deviations,
and 
$\sigma_i(x/2)$ is the expected detector resolution 
in the central region 
for the \pt of object $i$ evaluated at $\pt=x/2$ to roughly approximate the largest \pt-scale in the event.
An exception to this is the missing transverse momentum resolution (\sigmaMET), which is a function of $\sum \et$, 
where $\sum \et$ is approximated by the effective mass minus the \met object requirement: $\sigmaMET\left( \sum \et = x-200~\GeV\right)$.
The \met object is 
only 
considered in the binning of the 
effective 
mass histograms.
A $\pm 1 \sigma$ interval is used for the bin width ($k=2$) for all objects except for photons and electrons, 
for which a $\pm 3 \sigma$ interval is used ($k=6$) to avoid having too finely binned histograms with few MC events.
This results in variable bin widths with values ranging from 20~\GeV{} to about 2000~\GeV.
For a given event class, the scan starts at a value of the scanned observable larger than two times  the sum of the minimum \pt{} requirement
of each contributing object  considered (e.g. 100~\GeV{} for a $2\mu$ class).
This minimises spurious deviations which might arise from insufficiently well modelled threshold regions.

\subsubsection{Algorithm to search for deviations of the data from the expectation}
The algorithm identifies the single region with the largest upward or downward deviation
in a distribution, provided in the form of a histogram,
as the region of interest (ROI).
The total number of independent bins
is $36936$, leading to $518320$ combinations of contiguous bins (regions\footnote{
A histogram of $n$ bins has 1 region of $n$ contiguous bins, 
2 regions of $n-1$ contiguous bins, etc.\ down to $n$ regions of single bins.
Therefore, it has $\sum_{i=1}^n i = n(n+1)/2$ regions. When combining bins, background uncertainties are conservatively
treated as correlated among the bins with the exception of MC statistical uncertainties.
})
with an SM expectation larger than 0.01 events.
For each region 
with an SM expectation larger than 0.01,
the statistical 
estimator \pvalue is calculated 
as defined in \Eqnrange{(\ref{eq:pvalue1})}{(\ref{eq:pvalue3})}.
Here, \pvalue is to be interpreted as a local \pvalue-value.
The region of largest deviation found by the algorithm is the region with the
smallest \pvalue-value. Such a method is able to find narrow resonances
and single outstanding bins, as well as signals spread over
large regions of phase space in distributions of any shape.

\begin{figure}[htbp] 
\centering
\subfigure[]{\includegraphics[width=0.48\textwidth]{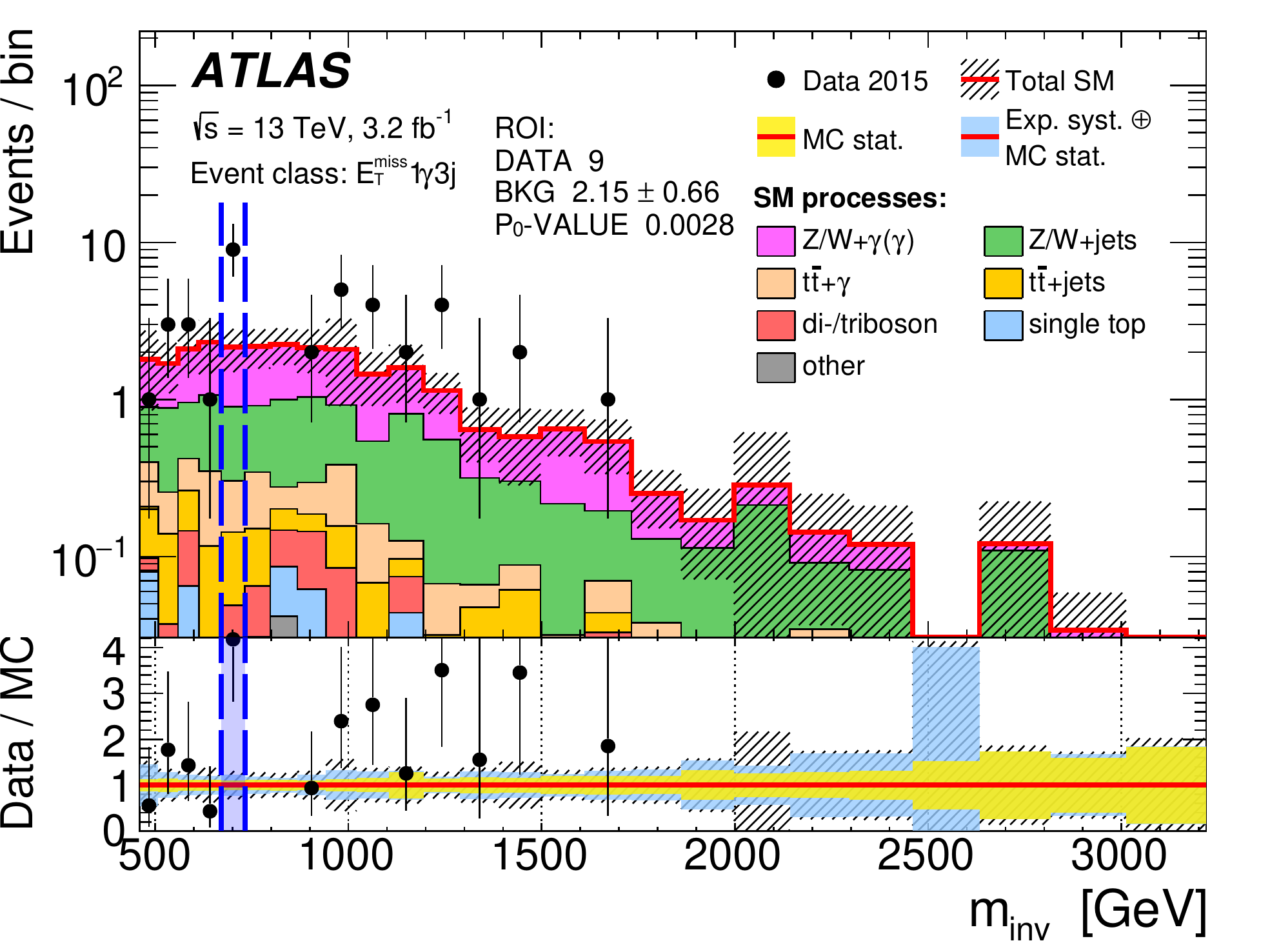}\label{fig:scan_example_a}}
\subfigure[]{\includegraphics[width=0.48\textwidth]{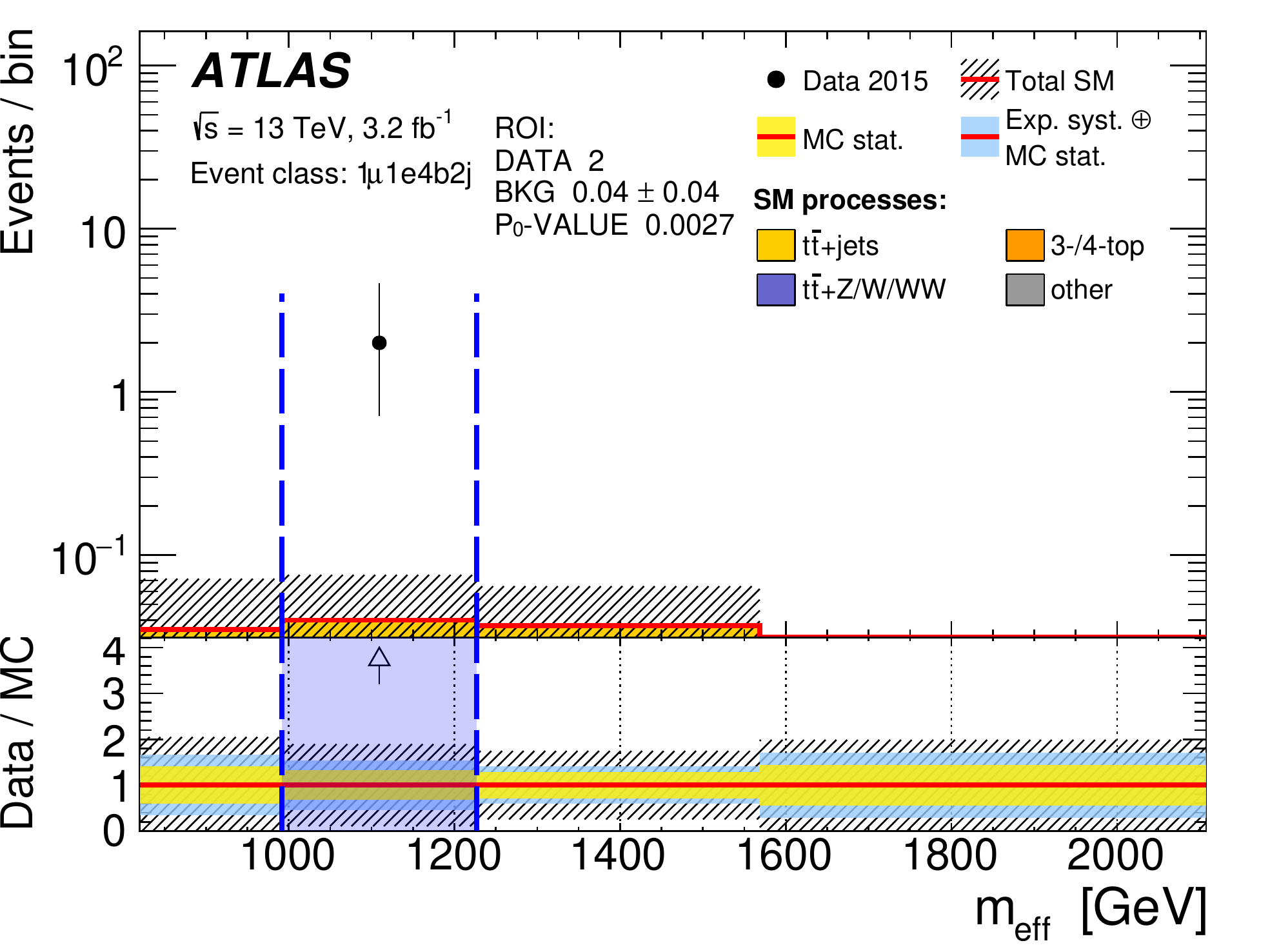}\label{fig:scan_example_b}} \\
\vspace*{-0.4cm} 
\subfigure[]{\includegraphics[width=0.48\textwidth]{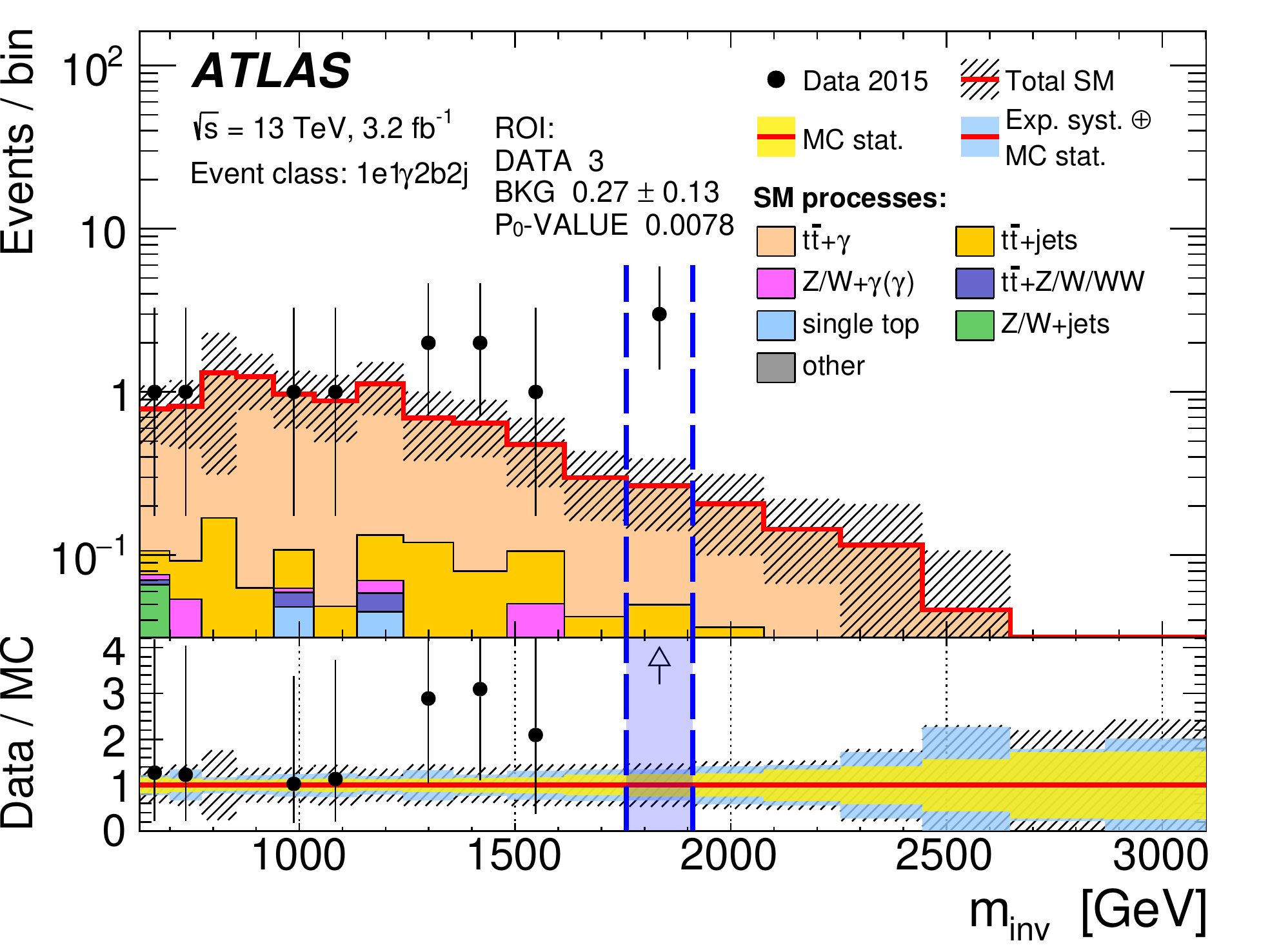}\label{fig:scan_example_c}}
\subfigure[]{\includegraphics[width=0.48\textwidth]{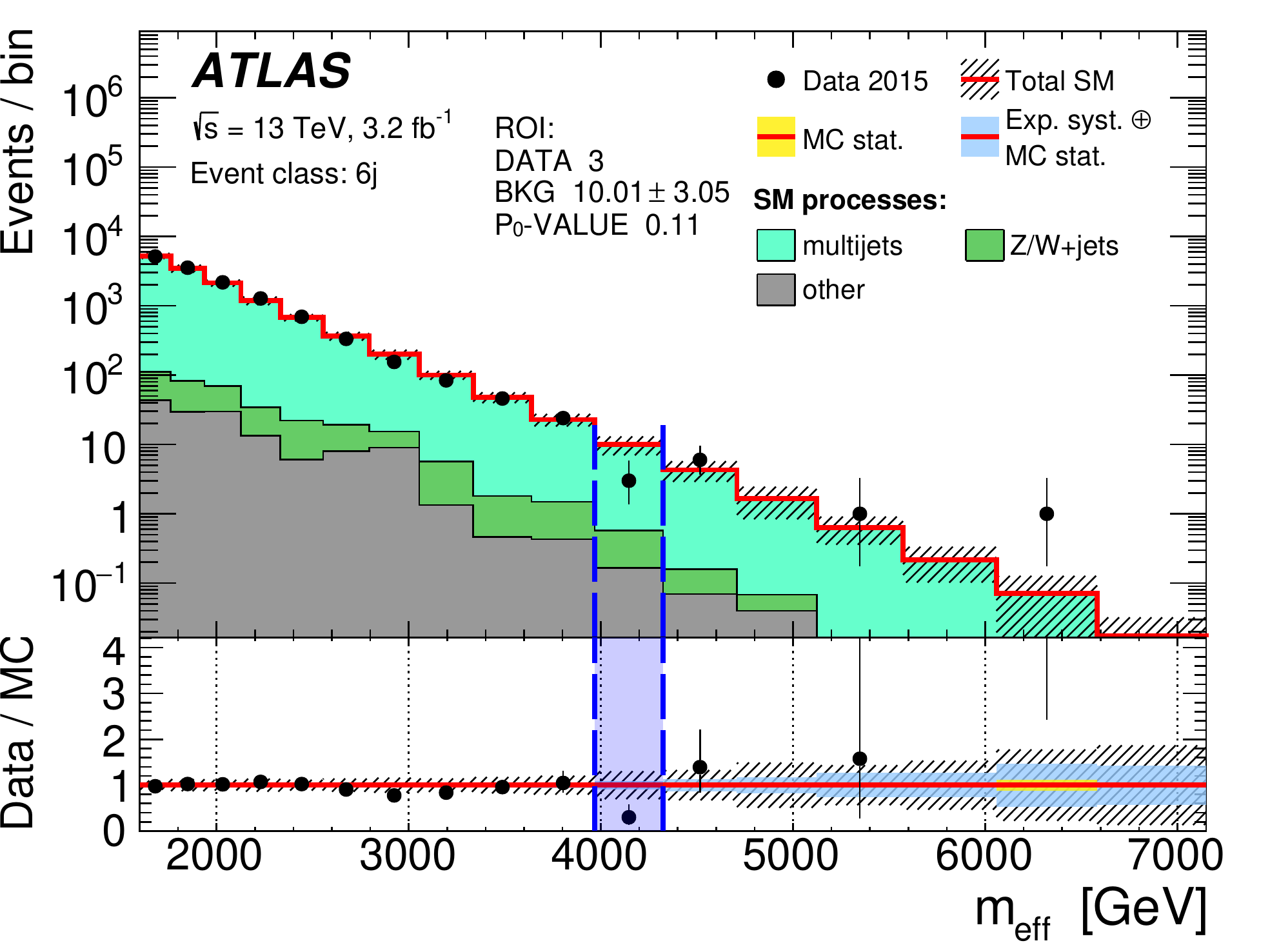}\label{fig:scan_example_d}} \\
\vspace*{-0.4cm} 
\subfigure[]{\includegraphics[width=0.48\textwidth]{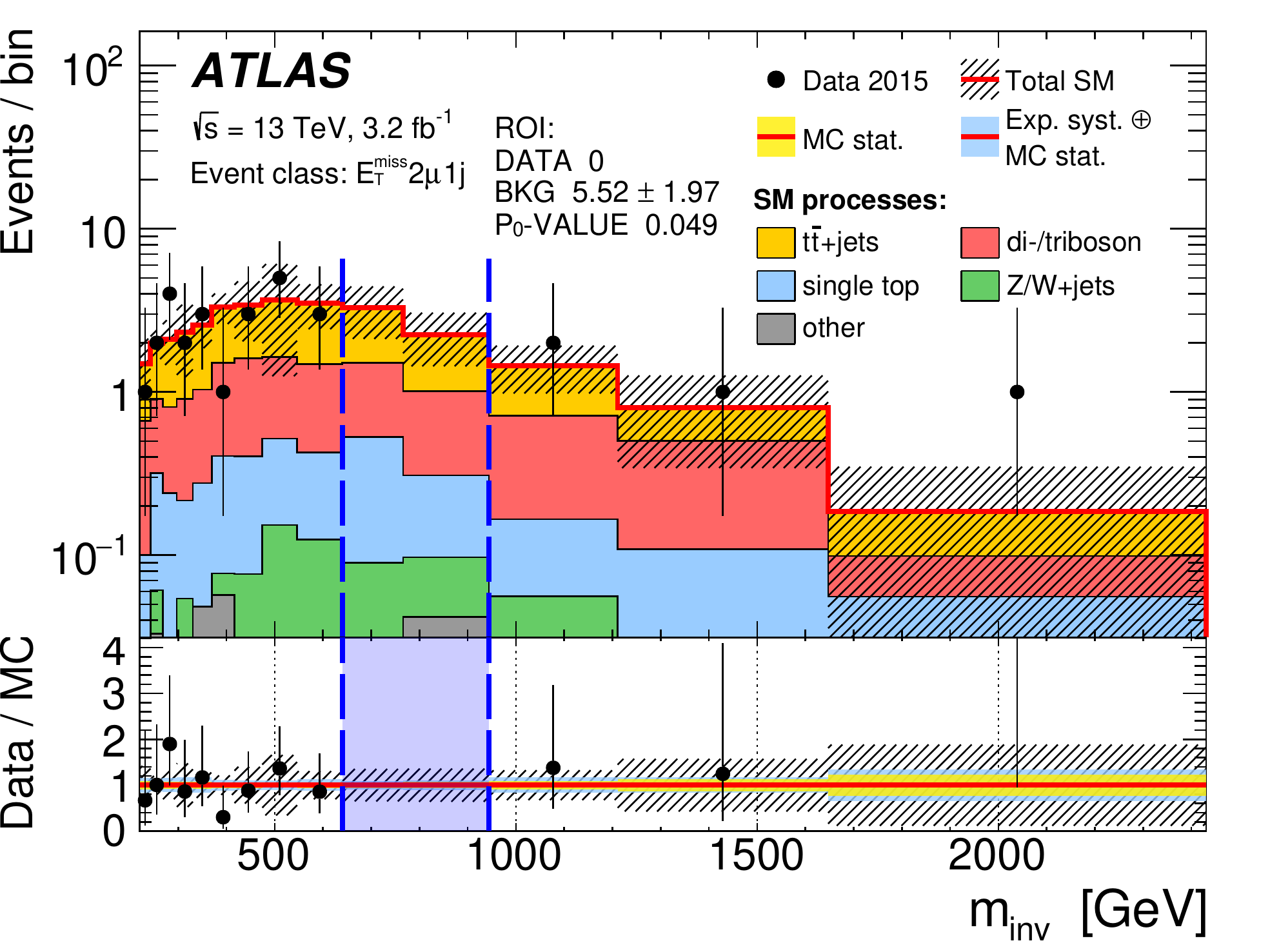}\label{fig:scan_example_e}}
\subfigure[]{\includegraphics[width=0.48\textwidth]{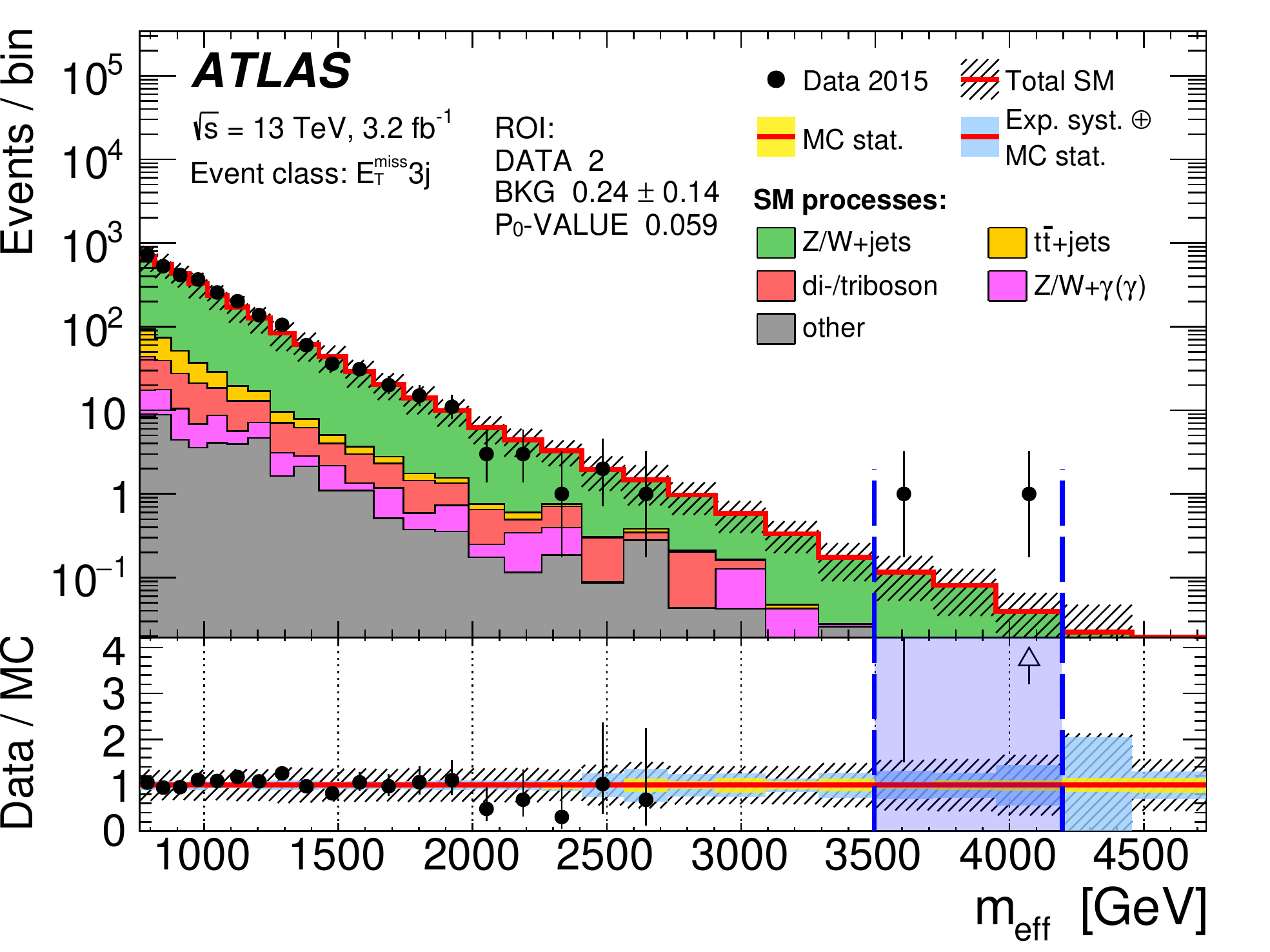}\label{fig:scan_example_f}}
\vspace*{-0.4cm} 
\caption{\label{fig:scan_example} 
Example distributions showing the region of interest (ROI), 
i.e.\ the region with the smallest \pvalue-value, 
between the vertical dashed 
lines.
(a) $\met{} 1\gamma 3j$ channel, 
which has the largest deviation in the \minv{} scan.
(b) $1\mu 1e 4b 2j$ channel, 
which has the largest deviation in the \meff{} scan.
(c) An upward fluctuation in the \minv{} distribution
of the $1e 1\gamma 2b 2j$ channel.
(d) A downward fluctuation in the \meff{} distribution
of the $6j$ channel.
(e) A downward fluctuation in the \minv{} distribution 
of the $\met{} 2\mu 1j$ channel.
(f) An upward fluctuation in the \meff{} distribution 
of the $\met 3j$ channel.
The hatched band includes all systematic and statistical uncertainties from MC simulations.
In the ratio plots the inner solid uncertainty band shows the statistical uncertainty from MC simulations, 
the middle solid band includes the experimental systematic uncertainty, and the hatched band includes the theoretical systematic uncertainty.
}
\end{figure}

To illustrate 
the operation of the algorithm, 
six example distributions are presented. 
\Fig{\ref{fig:scan_example_a}} shows the invariant mass distribution
of the event class with one photon, three light jets and large missing transverse momentum
($\met{} 1\gamma 3j$), which has the smallest \pchannel{}-value in the \minv{} scan.
\Fig{\ref{fig:scan_example_b}} shows the effective mass distribution
of the event class with one muon, one electron, four $b$-jets and two light jets
($1\mu 1e 4b 2j$), which has the smallest \pchannel{}-value in the \meff{} scan.
\Fig{\ref{fig:scan_example_c}} shows the invariant mass distribution
of the event class with one electron, one photon, two $b$-jets and two light jets ($1e 1\gamma 2b 2j$). \Fig{\ref{fig:scan_example_d}} shows the effective mass distribution
of the event class with six light jets ($6j$).
\Fig{\ref{fig:scan_example_e}} shows the invariant mass distribution
of the event class with two muons, a light jet and large missing transverse momentum ($\met{} 2\mu 1j$) and 
\Fig{\ref{fig:scan_example_f}} shows the effective mass distribution
of the event class with three light jets and large missing transverse momentum ($\met 3j$).
The regions with the largest deviation found by the search algorithm in these distributions,
an excess in \Figs{\ref{fig:scan_example_a}, \ref{fig:scan_example_b}, \ref{fig:scan_example_c}}{\ref{fig:scan_example_f}}, 
and a deficit in \Figs{\ref{fig:scan_example_d}}{\ref{fig:scan_example_e}},
are indicated by vertical 
dashed lines.

To minimize the impact of few MC events, 
$213992$
regions where the background prediction has a total relative uncertainty of
over 100\% are discarded by the algorithm. 
Discarding a region forces the algorithm to consider a different or larger region in the event class, or if no region in the event class satisfies the condition, to discard the entire event class.\footnote{
  In the \minv{} and \meff{} scan respectively, 72 and 87 event classes are discarded since they have no ROI.
}
For all discarded regions with $\Nobs>3$ a \pvalue-value is calculated.
If the \pvalue-value is smaller than the \pchannel-value
(or if there is no ROI and hence no \pchannel-value), it is evaluated manually by comparing it with the distribution of \pchannel-values from the scan.
This is done for 27 event classes among which the smallest \pvalue-value observed in a discarded region is $0.01$.
To model 
the analysis of discarded regions 
in pseudo-experiments, regions are allowed to have larger  uncertainties if they fulfil the $\Nobs>3$ criterion.

In addition to monitoring 
regions discarded due to a total uncertainty in excess of 100\%, 
regions discarded due to $\Nsm<0.01$ 
but with 
$\Nobs>3$ 
would also be monitored individually;
however, no such region has been observed.

\Tabs{\ref{tab:minv_deviations}}{\ref{tab:meff_deviations}} list the three event classes with the largest deviations in the \minv{} and \meff{} scans respectively.
The largest deviation reported by a dedicated search using the same dataset was observed in an inclusive diphoton data selection at a diphoton mass of around 750~\GeV{} 
with a local significance of $3.9\sigma$~\cite{HIGG-2016-08}.
Due to the different event
selections and background 
estimates the excess
has a lower significance 
in this analysis.
The excess was not confirmed
in a dedicated analysis with 2016 data~\cite{HIGG-2016-17}.

\begin{table}[htbp]
  \footnotesize
   \caption{\label{tab:minv_deviations} List of the three channels with the smallest \pchannel-values in the scan of the \minv{} distributions.}
   \begin{center}
      \begin{tabular}{l c r c c}
         \toprule
         \multicolumn{5}{c}{\textbf{Largest deviations in \minv{} scan}} \\
         Channel & \pchannel $(\cdot10^{-3})$ & \Nobs & $\Nsm \pm \delta \Nsm$ & Region $[\GeV]$ \\
         \midrule
         $\met\, 1\gamma\, 3j$      & $2.81$    & $9$    & $2.15 \pm 0.66$    & 670--732       \\
         $1\mu\, 1e\, 4b\, 2j$      & $2.91$    & $2$    & $0.042\pm 0.037$   & 1227--1569     \\
         $1e\, 1b\, 4j$             & $3.44$    & $160$  & $105  \pm 14$      & 726--809       \\
         \bottomrule
      \end{tabular}
   \end{center}
\end{table}

\begin{table}[ht!]
  \footnotesize
   \caption{\label{tab:meff_deviations} List of the three channels with the smallest \pchannel-values in the scan of the \meff{} distributions.}
   \begin{center}
      \begin{tabular}{l c r c c}
         \toprule
         \multicolumn{5}{c}{\textbf{Largest deviations in \meff{} scan}}                                             \\
         Channel  & \pchannel $(\cdot10^{-3})$  & \Nobs     & $\Nsm \pm \delta \Nsm$ & Region $[\GeV]$ \\
         \midrule
         $1\mu\, 1e\, 4b\, 2j$      & $2.66$   &  $2$    & $0.040 \pm 0.036$  & 992--1227         \\
         $1\mu\, 1\gamma\, 5j$      & $3.98$   &  $4$    & $0.45  \pm 0.18$   & 750--895          \\
         $3b\, 1j$                  & $4.87$   &  $4$    & $0.42  \pm 0.24$   & 3401--3923        \\
         \bottomrule
      \end{tabular}
   \end{center}
\end{table}

   \subsection{Step 4: Generation of pseudo-experiments}
      As described in \Sect{\ref{sec:pseudoexperiments}}, 
pseudo-experiments 
are generated to derive the probability 
of finding
a \pvalue-value of a given size,
for a given 
observable and algorithm.
The \pchannel-value distributions of the pseudo-experiments and their statistical properties
can be compared with the \pchannel-value distribution obtained from data.
Correlations in the uncertainties of the SM expectation affect this probability  
and their effect is taken into account in the generation of pseudo-data as 
outlined in the following.

For the experimental uncertainties, 
each of the 35 sources of uncertainty is
varied independently by drawing a value at random from a Gaussian pdf.
This value is assumed to be 100\% correlated across all bins and event classes.
The uncertainty in the normalization of the various backgrounds 
is also considered as 100\% correlated.
Likewise, 
theoretical shape uncertainties, 
including those estimated from scale variations or
the differences with alternative generators, 
are assumed to be 100\% correlated,
with the exception of 
the uncertainties which are used for some SM processes with small cross-sections.
The latter uncertainties are assumed to be uncorrelated, 
both between event classes and between bins of the same event class.
Scale variations are applied in the generation of pseudo-experiments by varying the renormalization, factorization, resummation and merging scales independently. The values for each scale of a given pseudo-experiment are 100\% correlated between all bins and event classes.
The scales are correlated between processes of the same type which are generated with a similar generator set-up, i.e.\ scales are correlated among the $W/Z/\gamma+\text{jets}$ processes, among all the diboson processes, among the $t\bar{t}+W/Z$ processes, and among the single-top processes.

Changing the size of the theoretical uncertainties by a factor of two leads to a change of less than 5\% in the $-\log_{10} (p_{\text{min}})$ thresholds at which a dedicated analysis is triggered. The correlation assumptions in the theoretical uncertainties were also tested. 
\Fig{\ref{fig:toys_svcorr}} shows the effect of changing the correlation assumption for 
all theoretical shape uncertainties that are nominally taken as 100\% correlated.
This test decorrelates the bin-by-bin variations due to the theoretical shape uncertainties in the pseudo-data while retaining the correlation when summing over selected bins in the scan, thus testing the impact of an incorrect assumption in the correlation model.
By comparing the nominal assumption of 100\% correlation with a 50\% correlated component, and a fully uncorrelated assumption, the 
threshold at which a dedicated analysis is triggered 
is changed by a negligible amount.

\begin{figure}[htbp]
  \subfigure[center][]{
  	\includegraphics[width=0.495\textwidth,page=3]{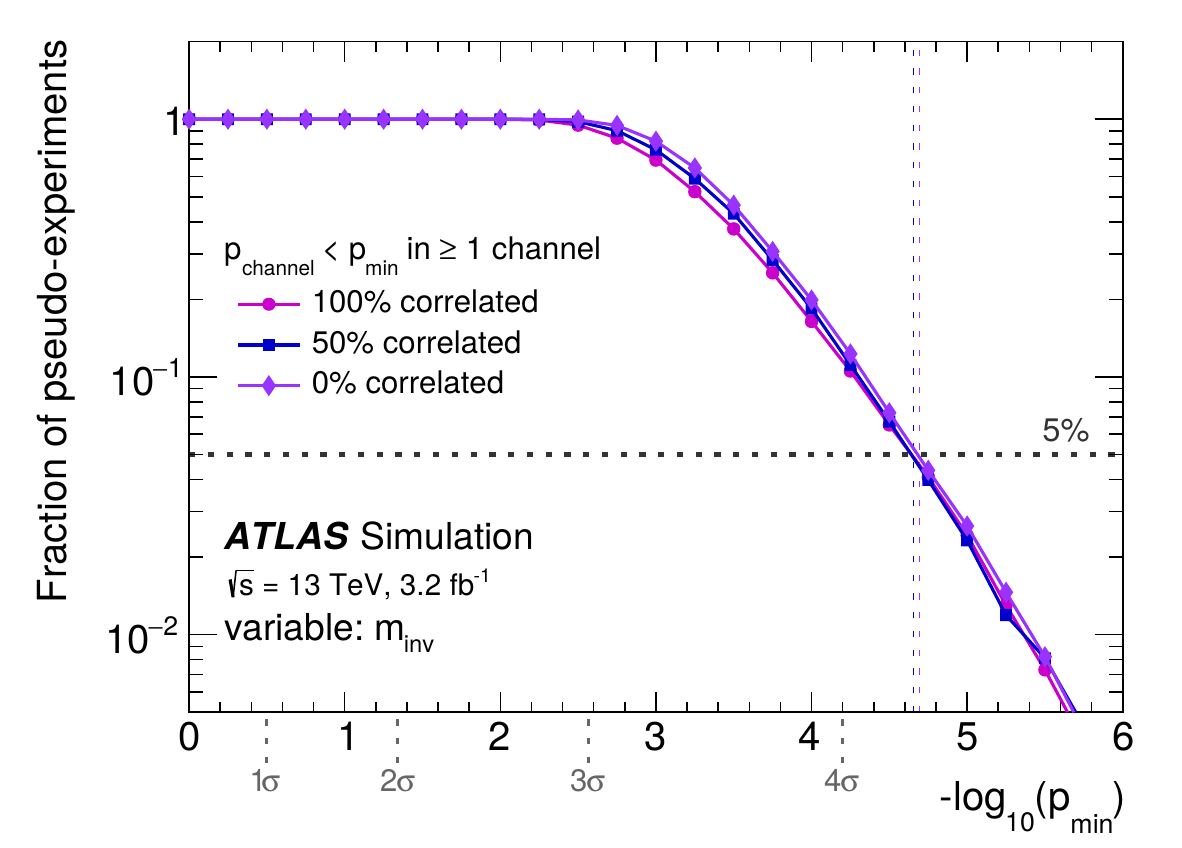}
  }
  \subfigure[center][]{
  	\includegraphics[width=0.495\textwidth,page=2]{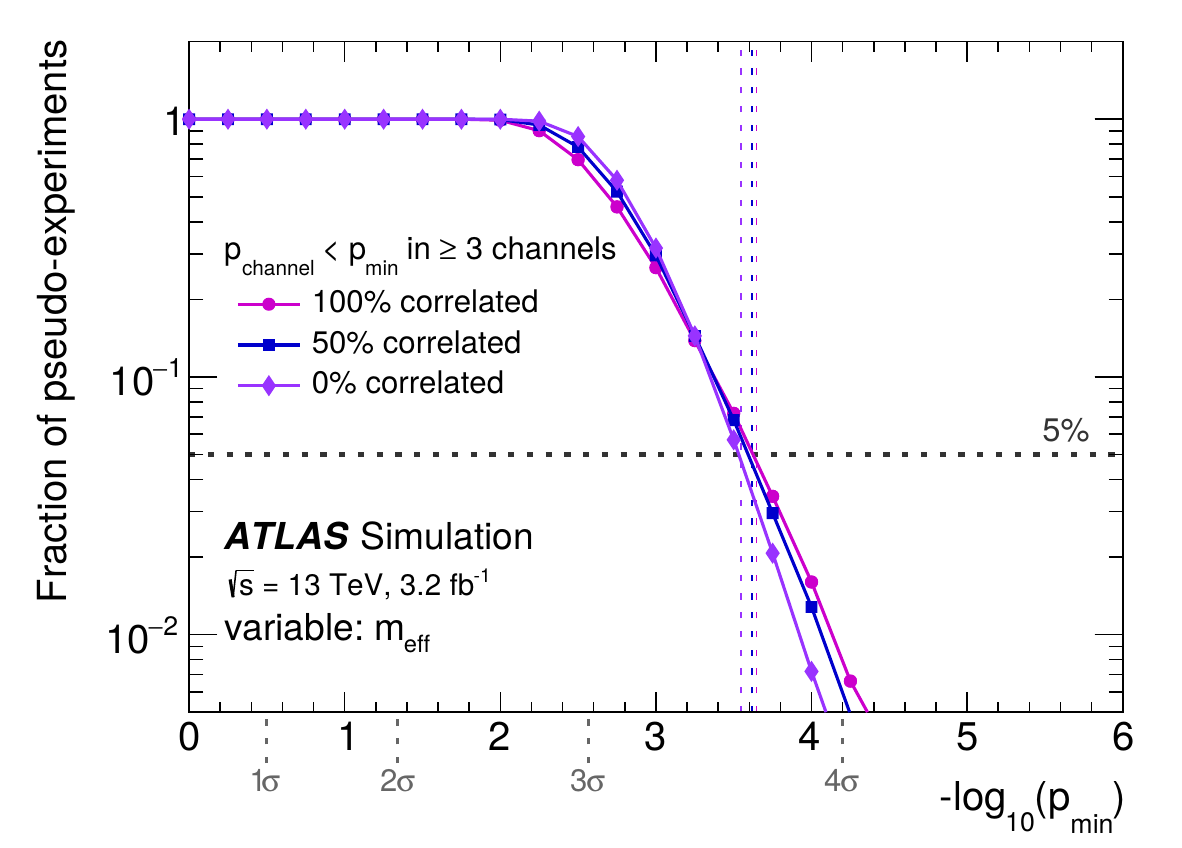}
  } \\
    \vspace*{-0.3cm}
  \caption{A comparison of different correlation assumptions for scale variations: 100\% correlated; 50\% correlated and 50\% uncorrelated; and 100\% uncorrelated. The fractions of pseudo-experiments in the scan of the \minv{} distribution having at least one \pchannel-value smaller than $p_{\text{min}}$ are shown on the left (a), 
  while the fractions in the scan of the \meff{} distribution having at least three \pchannel-values smaller than $p_{\text{min}}$ are shown on the right (b).
  \label{fig:toys_svcorr}
  }
\end{figure}

   \subsection{Step 5: Evaluation of the sensitivity of the strategy}
      \label{sec:sensitivity}
      \subsubsection{Sensitivity to Standard Model processes}
The sensitivity of the procedure 
is evaluated with two different methods
that 
either use a modified background estimation through the removal of SM processes 
or 
in which signal contributions are added to the pseudo-data sample.
As a figure of merit, the fraction of
`signal' pseudo-experiments 
with $\pexp < 5\%$ for $i=1,2,3$ is computed.

\begin{figure}[htbp]
  \subfigure[center][]{
  	\includegraphics[page=5,width=0.495\textwidth]{figures/PAPER/toyplots.pdf}\label{fig:pvalues1:a}
  }
  \subfigure[center][]{
  	\includegraphics[page=4,width=0.495\textwidth]{figures/PAPER/toyplots.pdf}\label{fig:pvalues1:b}
  } \\
  \subfigure[center][]{
  	\includegraphics[width=0.495\textwidth]{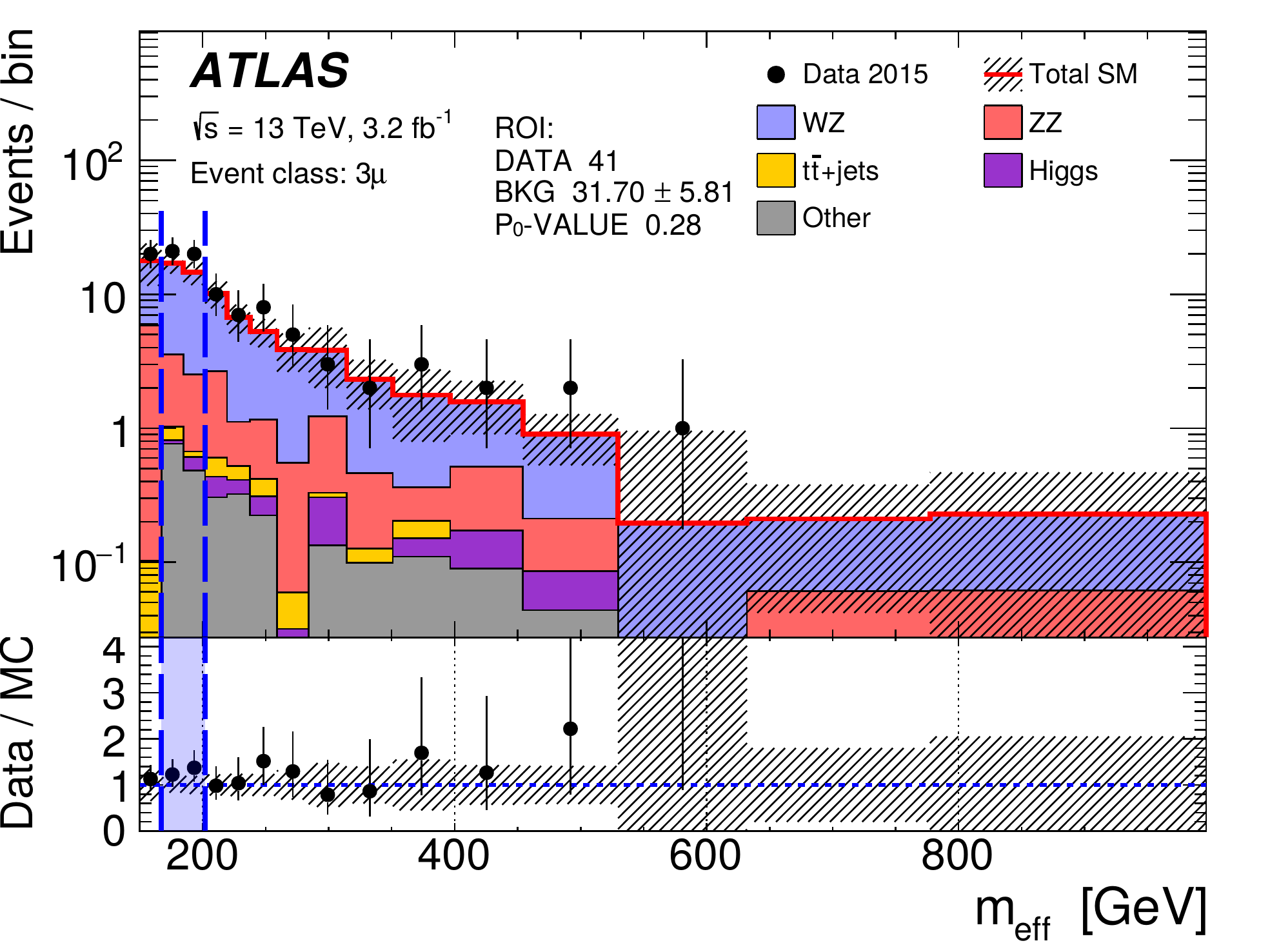}\label{fig:pvalues1:c}
  }
  \subfigure[center][]{
  	\includegraphics[width=0.495\textwidth]{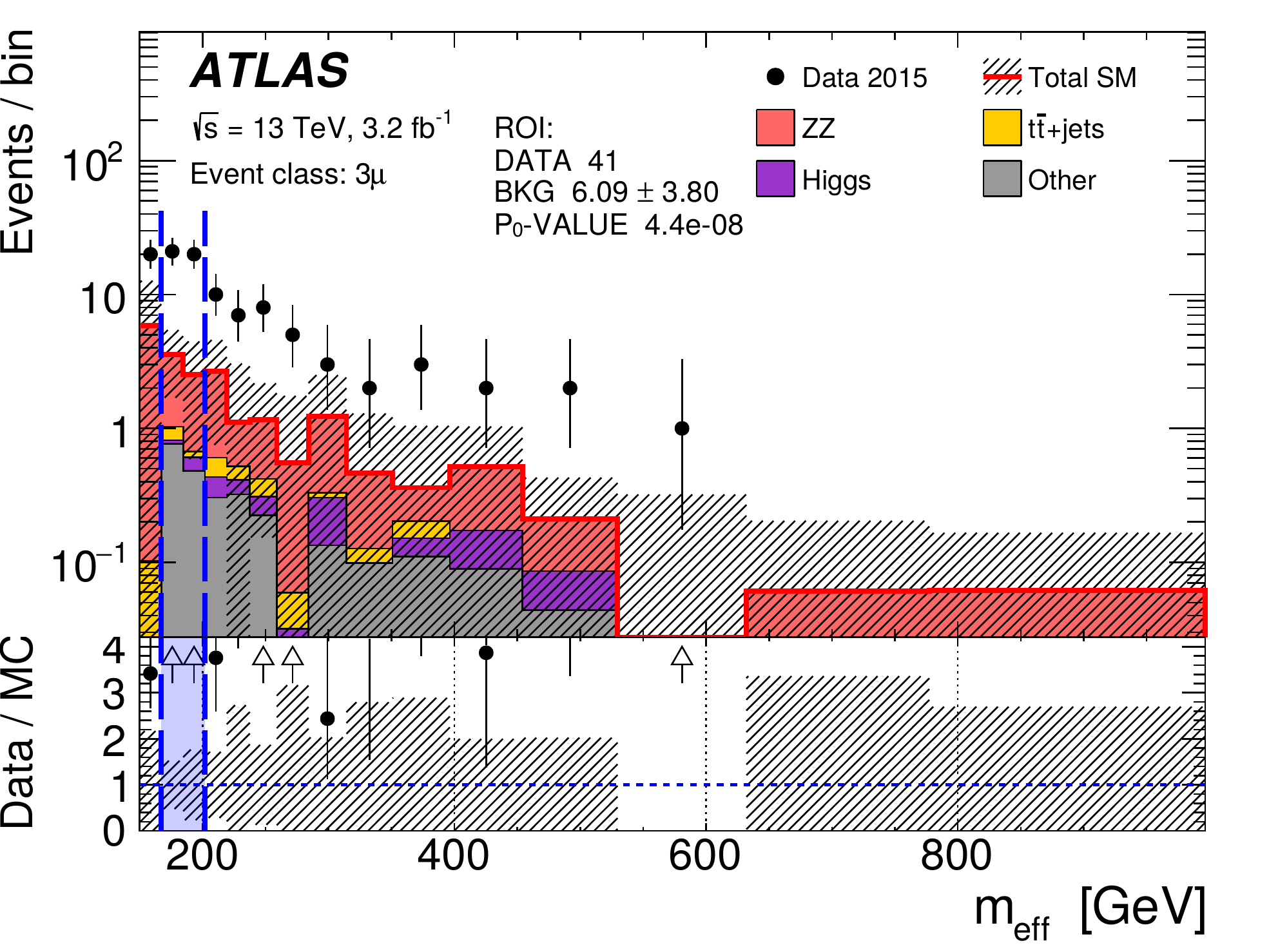}\label{fig:pvalues1:d}
  }
  \vspace*{-0.3cm}
  \caption{The fraction of pseudo-experiments which have at least one, two and three \pchannel-values below a given $p_{\min}$, 
    given for both the pseudo-experiments generated from the nominal SM expectation and tested against the nominal expectation (dashed) 
    and for those tested against the modified expectation (`SM, $WZ$ removed') in which the $WZ$ diboson process is removed (solid). 
  	The \minv{} scan is shown in (a) and the \meff{} scan in (b).
    The horizontal dotted lines show the fractions of pseudo-experiments yielding $\pexp < 5\%$ when tested against the modified background prediction.
    The scan results of the data tested against the modified background prediction
    are indicated with solid arrows. For reference the scan results under the SM hypothesis are
    plotted as dashed arrows.
    The largest deviation 
    after removing the $WZ$ process from the background expectation
    is found in the \meff{} distribution of the $3\mu$ event class. The distributions of the data and the expectation with both $WZ$ included
    and $WZ$ removed are shown in (c) and (d) respectively.  
    \label{fig:pvalues1}}
\end{figure}

\Fig{\ref{fig:pvalues1}} shows how removing the $WZ$ process from the background prediction affects the three smallest expected \pchannel-values. 
In \Figs{\ref{fig:pvalues1:a}}{\ref{fig:pvalues1:b}}, the dashed 
curves show the nominal expected \pchannel distribution obtained from pseudo-experiments. These define the $p_{\text{min}}$ thresholds for which $\pexp < 5\%$ 
and vertical dotted lines are drawn at the threshold values. 
The solid lines show the \pchannel distributions obtained by testing pseudo-experiments generated from the SM prediction against the modified background prediction 
which has the $WZ$ diboson process removed.
It can be observed that in this case the \meff{} scan is more sensitive; 
the fraction of `signal' pseudo-experiments 
with $\pexp < 5\%$ 
is about 80\% in all three cases $i=1,2,3$.

Additionally, in \Figs{\ref{fig:pvalues1:a}}{\ref{fig:pvalues1:b}}, the three smallest \pchannel-values observed in the data are shown by arrows, 
both when tested against the full SM prediction (dashed) and when tested against the modified prediction (solid).
For all three cases ($i=1,2,3$), $P_{\textrm{exp},i}<5\%$ is found again.
This means that a dedicated analysis would be performed for the three
event classes in which the \pchannel-values are observed, i.e.\ $3\mu$, $1\mu2e1j$, and $2\mu1e1j$, likely resulting
in the discovery of 
an unexpected signal
due to $WZ$ production.
\Figs{\ref{fig:pvalues1:c}}{\ref{fig:pvalues1:d}} shows the \meff{} distributions of the data with the full SM prediction and the modified prediction respectively.
This test uses the conclusion from \Sect{\ref{sec:results}} and is performed in retrospect.
In the case of a significant deviation, this test would be performed with pseudo-data to assess the sensitivity of the search to a missing background.

\begin{figure}[htbp]
  \subfigure[center][]{
  	\includegraphics[page=7,width=0.495\textwidth]{figures/PAPER/toyplots.pdf}
  }
  \subfigure[center][]{
  	\includegraphics[page=6,width=0.495\textwidth]{figures/PAPER/toyplots.pdf}
  } \\
  \subfigure[center][]{
  	\includegraphics[width=0.495\textwidth]{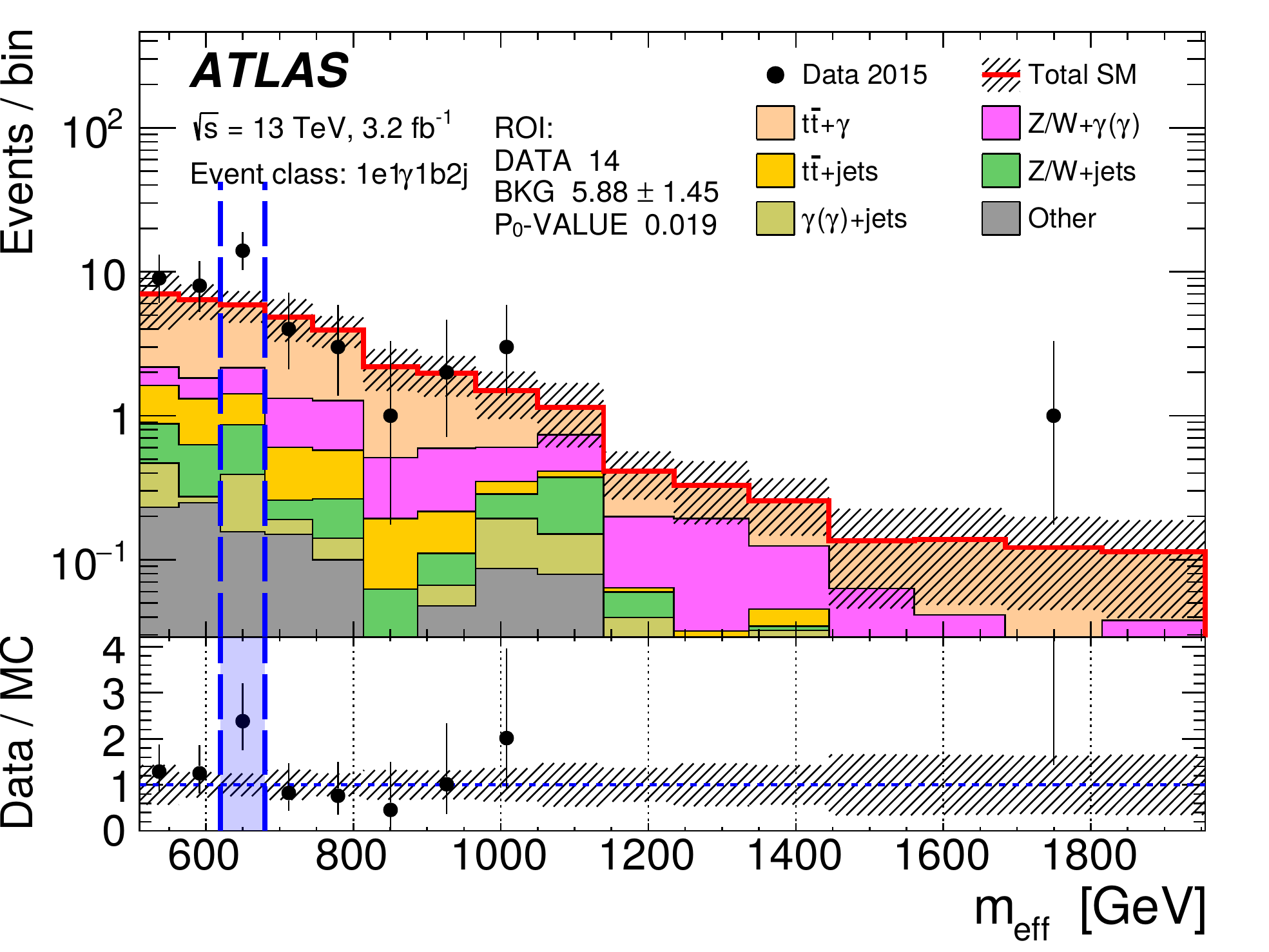}
  }
  \subfigure[center][]{
  	\includegraphics[width=0.495\textwidth]{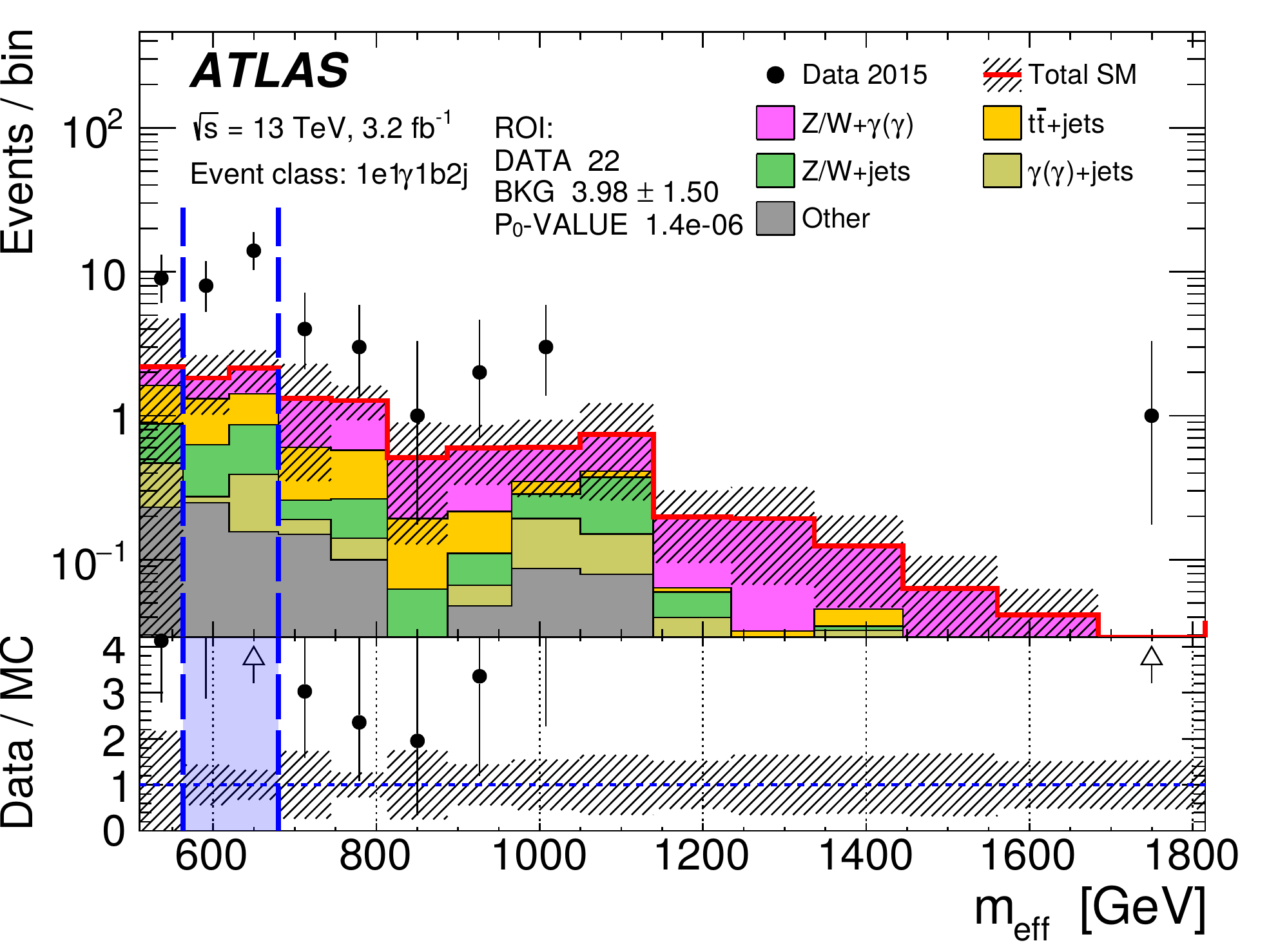}
  }
  \vspace*{-0.3cm}
  \caption{
  	The fraction of pseudo-experiments which have at least one, two and three \pchannel-values below a given $p_{\text{min}}$, 
  	given for both the pseudo-experiments generated from the nominal SM expectation 
    and tested against the nominal expectation (dashed) and for those tested against the modified expectation (`SM, $t\bar{t}\gamma$ removed') 
    in which the $t\bar{t}\gamma$ process is removed (solid). 
  	The \minv{} scan is shown in (a) and the \meff{} scan in (b).
    The horizontal dotted lines show the fractions of pseudo-experiments yielding $\pexp < 5\%$ when tested against the modified background prediction.
    The scan results of the data tested against the modified background prediction
    are indicated with solid arrows. For reference the scan results under the SM hypothesis are
    plotted as dashed arrows.
    The largest deviation 
    after removing the $t\bar{t}\gamma$ process from the background expectation
    is found in the \meff{} distribution of the $1e1\gamma1b2j$ event class. 
    The distributions of the data and the expectation with both $t\bar{t}\gamma$ included
    and $t\bar{t}\gamma$ removed are shown in (c) and (d) respectively.  
    \label{fig:pvalues2}}
\end{figure}

\Fig{\ref{fig:pvalues2}} shows the
effect of removing the $t \bar{t} + \gamma$ process. 
Again the \meff{} scan is slightly more sensitive, 
and about 70\%
of `signal' pseudo-experiments 
have $\pexp < 5\%$ 
in all three cases $i=1,2,3$. 
In the data, $\pexp < 5\%$ is found again for all three cases ($i=1,2,3$).
A dedicated analysis would be performed for the three classes $1\mu1\gamma2b1j$, $1e1\gamma2b2j$, and $1\mu1\gamma1b3j$, 
likely
resulting  in the discovery
of 
an unexpected signal due to
$t \bar{t} + \gamma$ production.

It is interesting to note that these discoveries would have been made
without a priori knowledge of the existence of these processes.

\subsubsection{Sensitivity to new-physics signals}
\begin{figure}[htbp]
  \subfigure[center][]{
  	\includegraphics[page=9,width=0.495\textwidth]{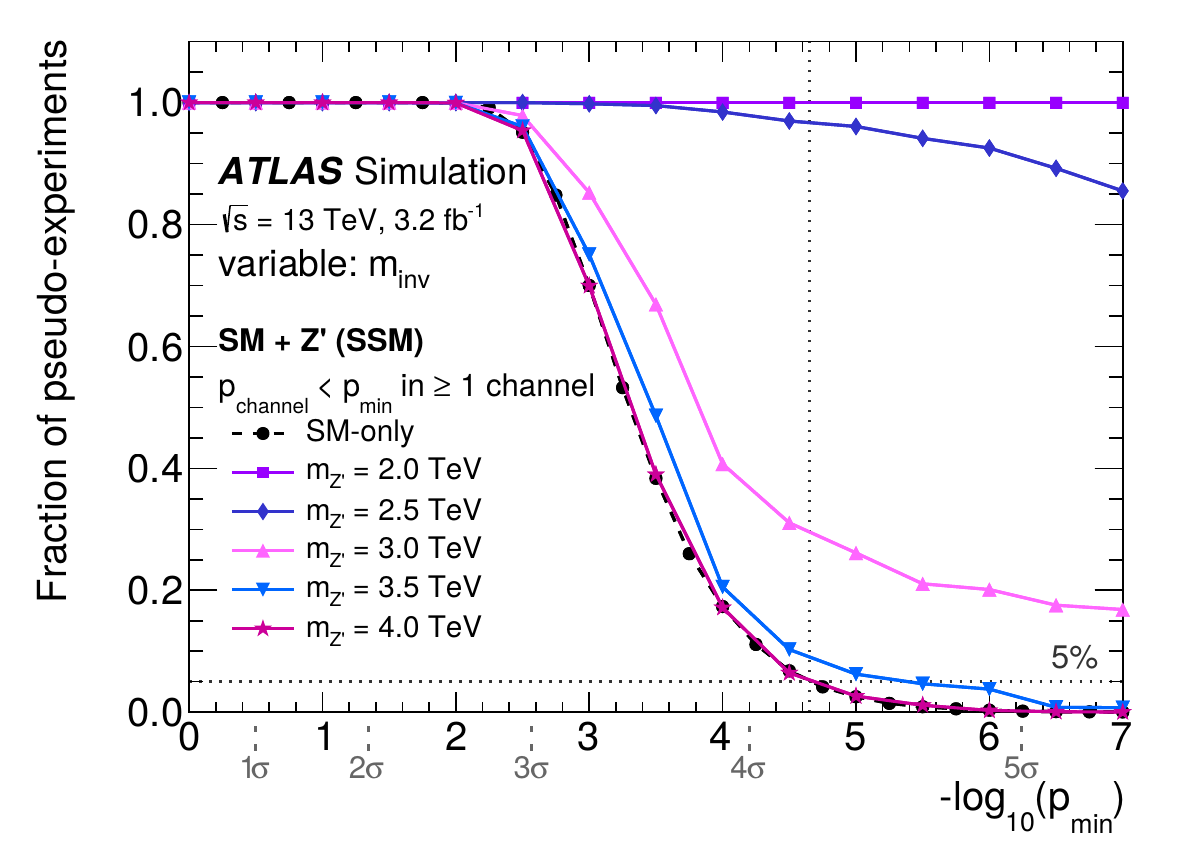}
  }
  \subfigure[center][]{
  	\includegraphics[page=8,width=0.495\textwidth]{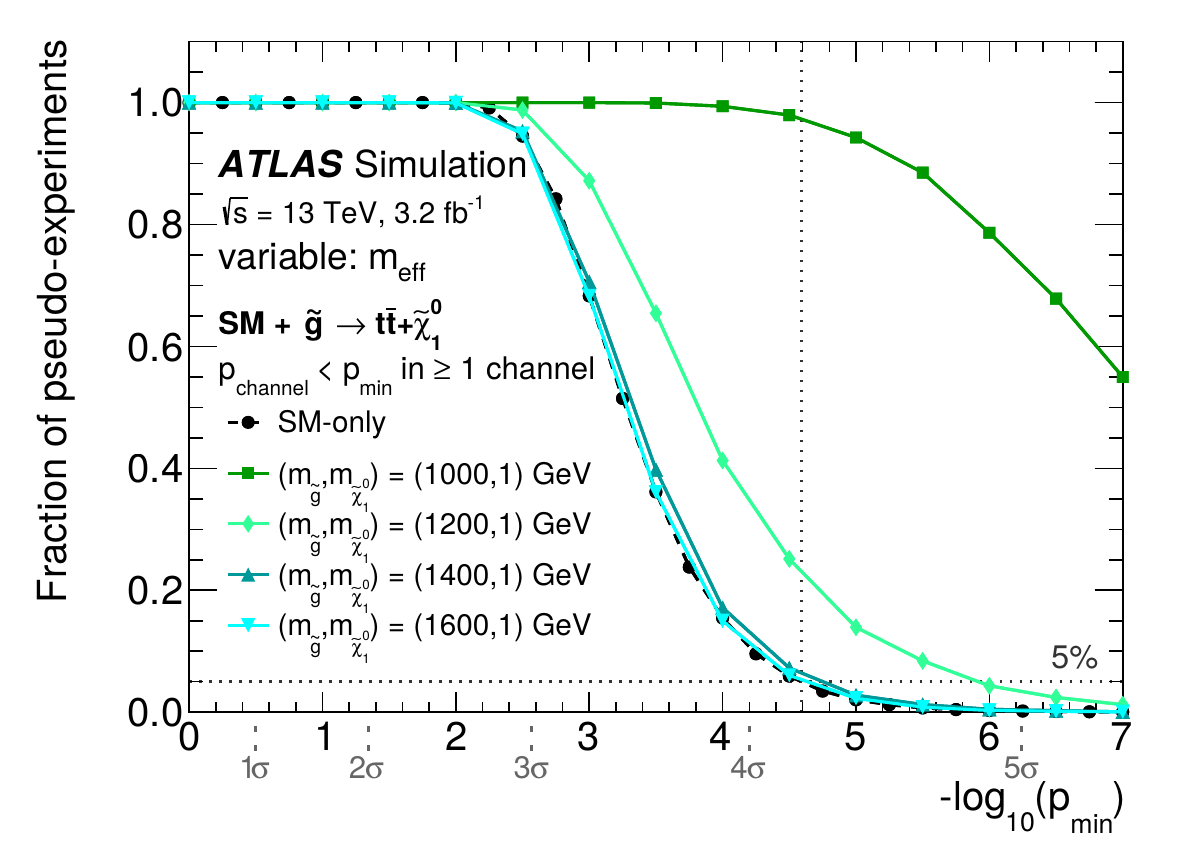}
  }
  \vspace*{-0.3cm}
  \caption{\label{fig:sensitivities} 
    The fraction of pseudo-experiments in which a deviation is found with a \pchannel-value smaller than a given $p_{\text{min}}$.
    Distributions are shown for pseudo-experiments generated from the SM expectation (circular markers),
    and after injecting signals of (a) inclusive $Z'$ decays or (b) gluino pairs with $\tilde{g} \to t\bar{t}\tilde{\chi}_1^0$ decays and various masses.
    The line corresponding to the injection of a $Z'$ boson with a mass of 4~\TeV{} and a gluino with a mass of 1600~\GeV{} overlap with the line obtained from the SM-only pseudo-experiments due to the small signal cross-section.
  }
\end{figure}

\Fig{\ref{fig:sensitivities}} shows the sensitivity for the two benchmark signals considered 
as a function of the mass of the produced particle.
For the $Z'$ model, where the mass of the resonance can be reconstructed from its decay products,
the sensitivity to the signal is found to be the largest in the scan of the \minv{} distribution.
Gluinos undergo a cascade decay process to the lightest neutralino, which is undetected and leads to missing transverse momentum.
It is not possible to fully reconstruct an event from gluino pair production due to the presence of neutralinos in the final state.
The sensitivity to the gluino signal is 
therefore 
found to be the largest in the \meff{} scan,
where a broad excess at large values of this quantity is 
expected.

Exclusion and discovery sensitivity have to be carefully distinguished
when the results of this search are compared with model-based searches.
An exclusion sensitivity at the 95\%~CL in a dedicated search 
roughly corresponds to a single class having a \pvalue-value for a discovery test smaller than 0.05.
Consequently, the sensitivity to a benchmark signal
corresponding to a given particle mass
should be compared with the discovery sensitivity of other searches for a $Z'$ boson or gluino.

As previously described, a deviation for which $\pexp<5\%$ 
promotes the selection to a signal region for a dedicated analysis.
By applying this sensitivity criterion, it can be seen in \Fig{\ref{fig:sensitivities}}
that this search is sensitive to a $Z'$ boson with a mass of about 2.5~\TeV{} 
as more than 90\% of the signal-injected pseudo-experiments show a deviation for which $P_{\text{exp},1}<5\%$. 
Similar sensitivity is expected for a gluino with a mass of about 1~\TeV{}. 
The probability of discovering a new-physics signal in a new dataset with a dedicated search in the selected event classes is estimated in the next section.

\subsubsection{Sensitivity of a second independent dataset }
In step 7 a dedicated analysis of a deviation is performed on an independent dataset. 
The sensitivity of step 7
is evaluated with  
pseudo-experiments.

A first pseudo-experiment 
emulates the original dataset 
on which this analysis is performed. 
The scan algorithm is applied 
after which eight different cases can be distinguished where $P_{\text{exp},i}$ is either larger or smaller than 5\% for $i=1,2,3$.
In seven cases at least one $P_{\text{exp},i}<5\%$ and 
a new independent pseudo-experiment is generated
to emulate a new independent dataset with the same
integrated luminosity.
The one, two or three data selections for which
$\pexp<5\%$ are applied
to the second pseudo-experiment to obtain the \pvalue-values for these selections. 
Although the systematic uncertainties may be reduced by applying data-driven estimates of the background,  
they were assumed to have the same size in the second pseudo-experiment to make a conservative estimate.
The systematic uncertainties are also expected to be partially correlated between two datasets but here they were assumed to be uncorrelated.

In four of the seven cases
$P_{\text{exp},1}<5\%$ and these cases are
grouped together into a `one signal region' class.
This class shows the sensitivity when there would be only a single data-derived signal region.
The case where only $P_{\text{exp},3}<5\%$
is called the `three signal region' class.
The two remaining cases where 
$P_{\text{exp},2}<5\%$ define the
`two signal region' class.
These classes show
cases when a data-derived signal region
is found only by a combination
of two and three regions.

\begin{figure}[htbp]
  \subfigure[center][]{
  	 \includegraphics[page=12,width=0.495\textwidth]{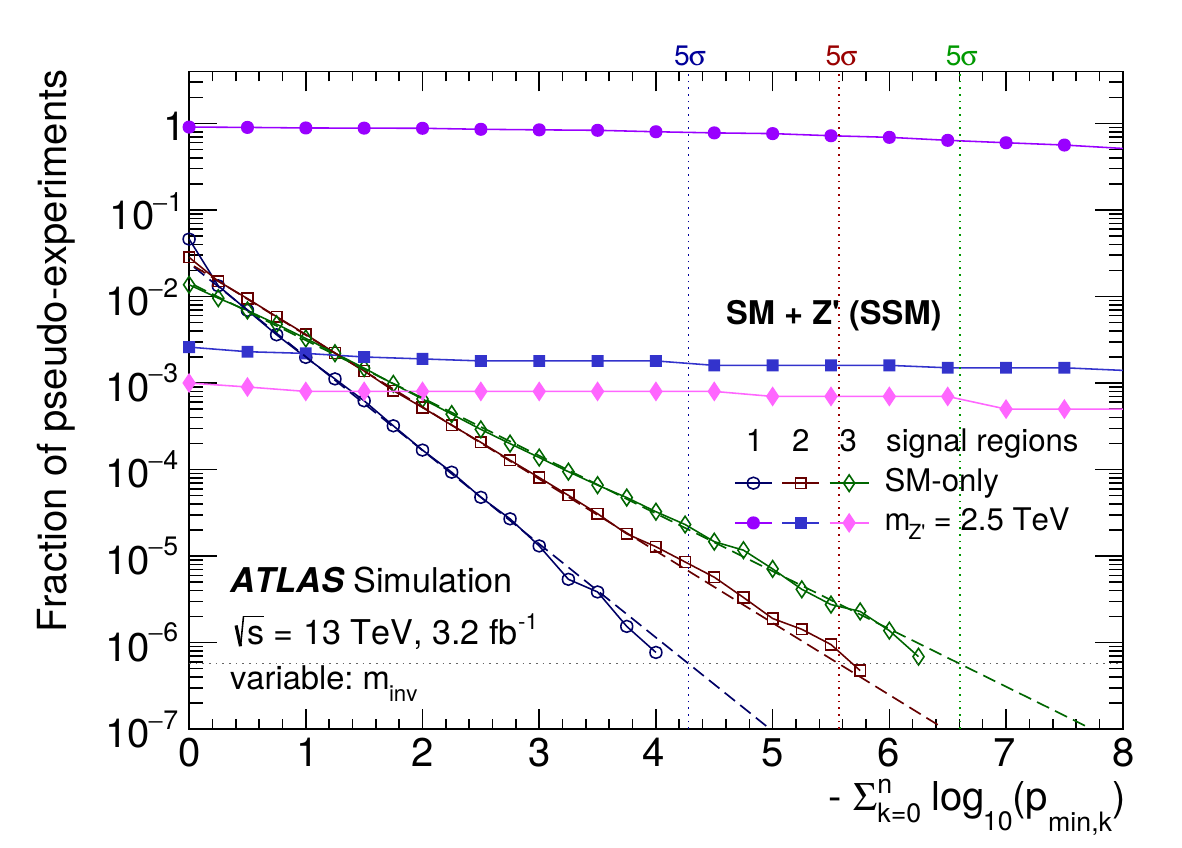}
  }
  \subfigure[center][]{
     \includegraphics[page=10,width=0.495\textwidth]{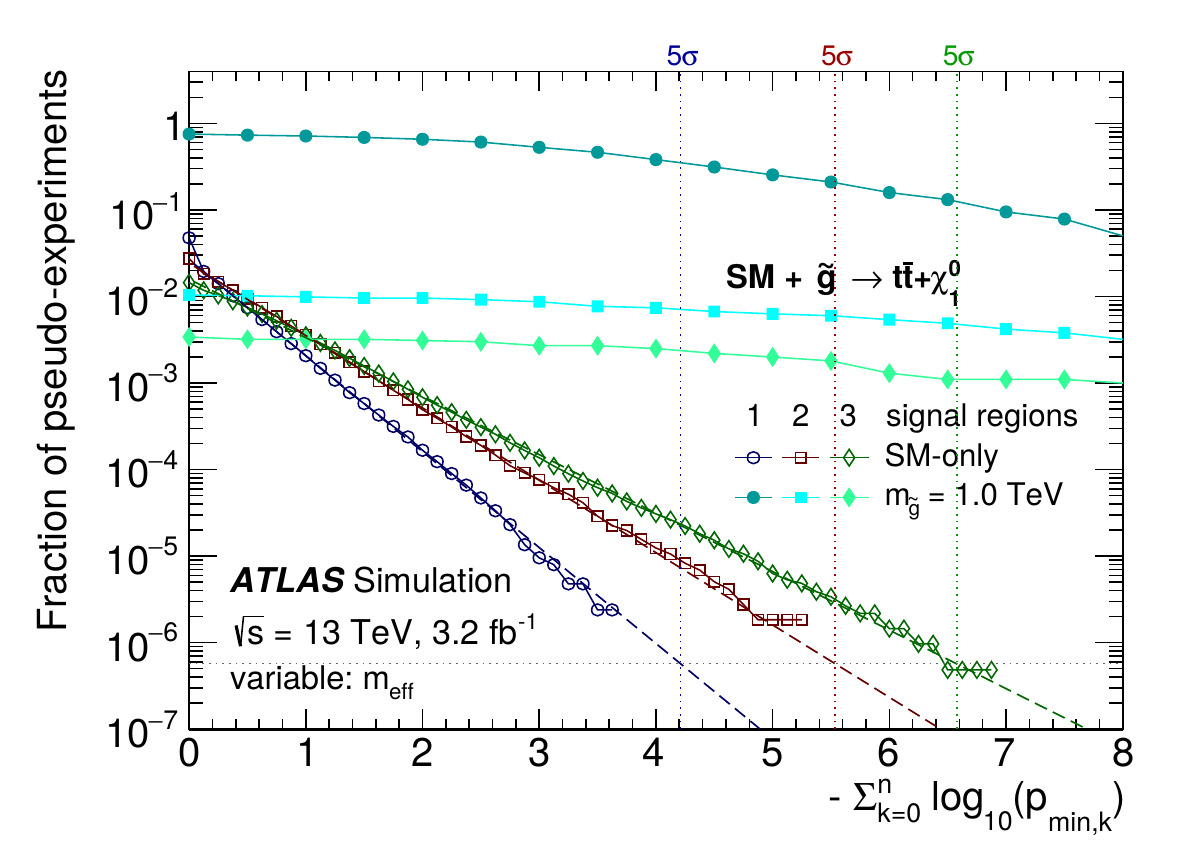}
  }
  \vspace*{-0.3cm}
  \caption{\label{fig:sensitivities2} 
    The fraction of cases in which the first pseudo-experiment has $\pexp<5\%$ 
    and triggers a second pseudo-experiment which yields a value 
    for $-\sum_{k=1}^{n}\log_{10}({\pvalue}_k)$ smaller than a given 
    value ($-\sum_{k=1}^{n}\log_{10}(p_{\text{min},k})$).
    Here $n$ denotes 
    the minimum number  of event classes (1, 2 or 3) for which $P_{\text{exp},i}<0.05$.
    The distribution is shown for pseudo-experiments generated from the SM expectation,
    and from an SM-plus-signal expectation of (a) inclusive decays of a $Z'$ boson with mass $m_{Z'} = 2.5$~\TeV{} 
    or (b) gluino pairs with mass $m_{\tilde{g}} = 1.0$~\TeV{} and $g\to t\bar{t}\tilde{\chi}_1^0$ decays.
    The signal region tested in the follow-up pseudo-experiment is defined by the preceding pseudo-experiment.
    The $5\sigma$ thresholds
    are obtained by extrapolating the SM-only fractions to $5.7 \cdot 10^{-7}$ and 
    are indicated at the top of the figure 
    for $n=1$ (left), $n=2$ (middle) or $n=3$ (right) event classes.
}
\end{figure}

For each of the three classes ($n=1,2,3$) the statistical estimator
$\prod_{k=1}^{n} {\pvalue}_k$ is computed.  
\Fig{\ref{fig:sensitivities2}} shows,
as a function of the estimator 
in the logarithmic form $-\sum_{k=1}^{n}\log_{10}({\pvalue}_k)$
the fraction of cases in which the first pseudo-experiment has $\pexp<5\%$ and triggers a second pseudo-experiment 
which yields a value for $-\sum_{k=1}^{n}\log_{10}({\pvalue}_k)$ above 
a threshold given by the value on the horizontal axis.
This is done for pseudo-experiments generated from the SM expectation (SM-only) and for pseudo-experiments generated from the SM expectation plus $Z'$ or gluino signal contributions.
The $5\sigma$ lines are derived from the fractions given by the SM-only lines, as these correspond to the probability of false positives which defines the level of significance.
It should be noted that the 
SM-only lines with circular markers 
start at a fraction of $0.05$ by construction of the $\pexp<5\%$ definition.
The $n=2$ and $n=3$ lines  show the gain in sensitivity when a deviation in one or two channels, respectively, is not large enough to define a data-derived signal region.
It does not show the gain in sensitivity from considering multiple channels when a single channel defines a signal region.
Signals which produce one or more large deviations therefore lower the number of cases in the $n=2$ and 3 categories, 
while signals producing deviations close to the $\pexp<5\%$ threshold (e.g.\ for higher $Z'$ masses) would raise the number of cases in the $n=2$ and 3 categories.
A $Z'$ boson with a mass of 2.5~\TeV{}
would yield a 
discovery
in almost all cases.

In the
case of a $1.0$~\TeV{} gluino  the sensitivity is about $5\sigma$. 
The sensitivity increases to about $1.1$~\TeV{} if the integrated luminosity of the two datasets combined is increased to about 10~\ifb{} 
by doubling the size of the second dataset to 6.4~\ifb{}.
ATLAS has determined the discovery sensitivity of the dedicated searches 
for 
gluinos 
decaying to quarks, a \Wboson boson and a neutralino.
This dedicated search estimates a local significance (i.e. not corrected for  trial factors of the dedicated searches) of $5\sigma$
with a luminosity of 10~\ifb{}
for gluinos with a mass of $1.35$~\TeV{}
assuming a systematic uncertainty of 25\%~\cite{ATL-PHYS-PUB-2015-005}.

It should be noted that, with this strategy, these signals are found without
any 
a priori 
assumptions about the model, including the mass
and the decay chain of the 
gluinos  or the  $Z'$ boson.
It can therefore be concluded that this 
procedure could also 
be sensitive to possible
unexpected signals for new physics.

   \subsection{Step 6: Results}
      \label{sec:results}
      In step 6 the \pchannel-values found in the 
analysis of the 2015 ATLAS data are interpreted
by comparing them with the \pchannel-values
found in the pseudo-experiments.

\begin{figure}[htbp]
  \subfigure[center][]{
  	 \includegraphics[page=15,width=0.495\textwidth]{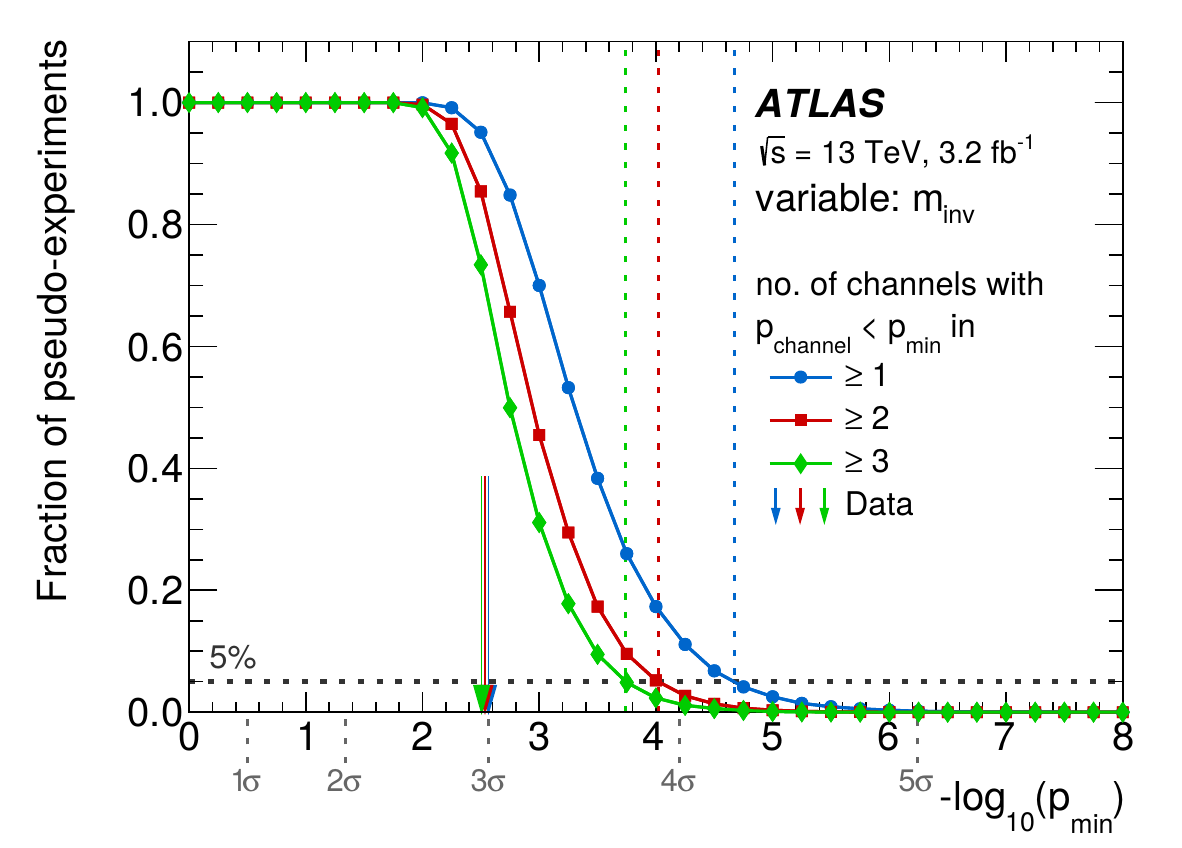}
  }
  \subfigure[center][]{
     \includegraphics[page=14,width=0.495\textwidth]{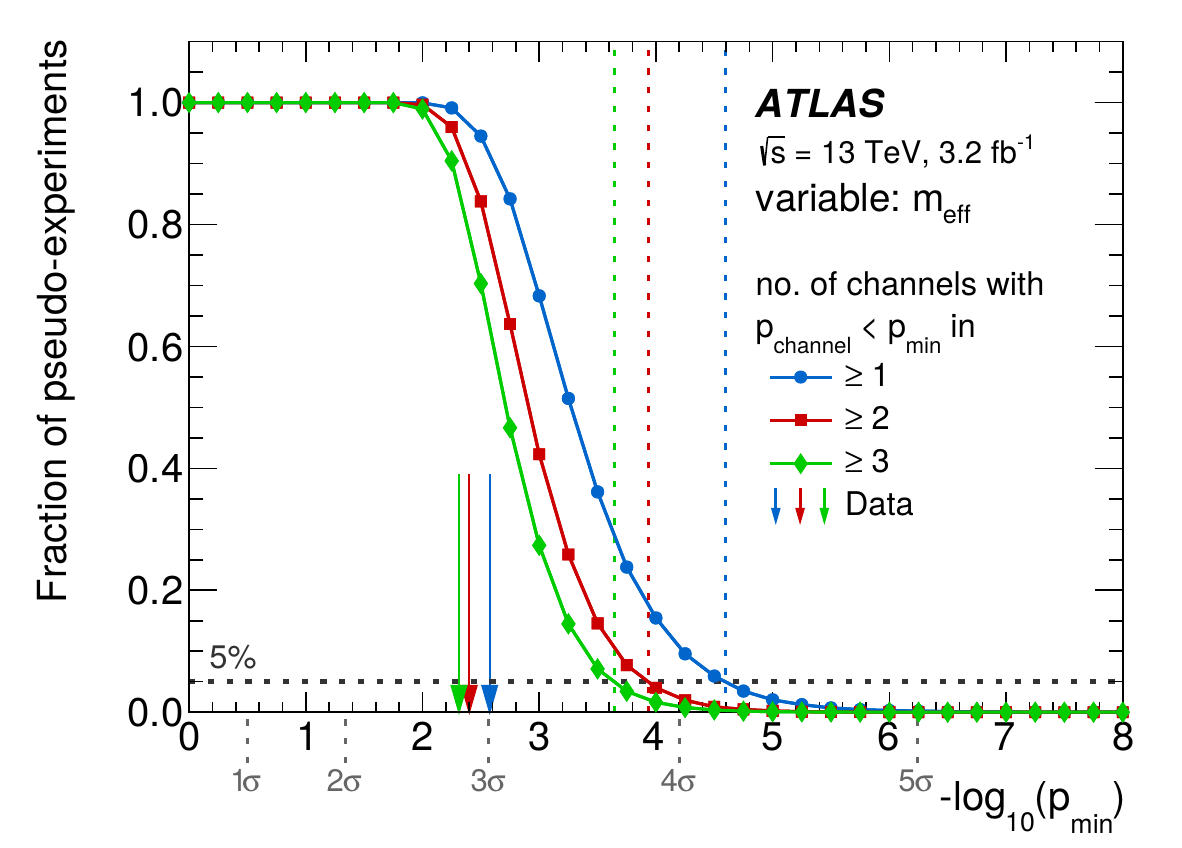}
  }
  \vspace*{-0.3cm}
  \caption{\label{fig:pvalues} 
The fractions of pseudo-experiments ($P_{\text{exp},i}(p_{\text{min}})$) 
which have at least one, two or three \pchannel-values 
(circular, square, and rhombic markers respectively) 
smaller than a given threshold ($p_{\text{min}}$)
in the scans of (a) the \minv{} distributions, and of (b) the \meff{} distributions.
The coloured arrows represent the three smallest \pchannel-values observed in data. 
Dashed lines are drawn at $P_{\textrm{exp},i}=5\%$ and at the $p_{\text{min}}$-values corresponding to 
$1\sigma$, $2\sigma$, $3\sigma$, $4\sigma$ and $5\sigma$ local significances. Details of the deviations can be found in \Tabs{\ref{tab:minv_deviations}}{\ref{tab:meff_deviations}}.}
\end{figure}

\Fig{\ref{fig:pvalues}} shows the fractions of pseudo-experiments
that 
have at least one, two or three \pchannel-values below a given threshold ($p_{\text{min}}$)
in the scans of the \minv{} and the \meff{} distributions.
The statistical tests in both distributions for 
the three leading \pchannel-values 
are all consistent at the $\pexp>50\%$ level with the SM expectation of \pchannel-values 
obtained from pseudo-experiments.
Changing the size of the theoretical shape uncertainties by a factor of two leads to a change in the three smallest \pchannel-values of a factor of two.
It therefore does not lead to an appreciable change in the result.

In conclusion, 
no significant deviations are found in the 2015 dataset and consequently no dedicated analysis using data-derived signal regions (step 7) is initiated.

   \section{Conclusions}
      \label{sec:Conclusions}
      A strategy for a model-independent general search to find potential 
indications of new physics is presented.
Events are classified according to their final state into many event classes.
For each event class an automated search algorithm tests whether the data is compatible with the 
Monte Carlo simulated Standard Model expectation in several distributions sensitive to the effects of new physics. 
For each distribution the search algorithm is repeated on many pseudo-experiments 
to make a frequentist estimate of the statistical
significance of the three largest deviations. 
A data selection in which a significant deviation is observed
defines a data-derived signal region which will be tested on a new dataset in a dedicated analysis with an improved background model.

The strategy has been applied to the data collected by the ATLAS experiment at the LHC during 2015, corresponding 
to a total of 3.2~\ifb{} of 13~\TeV{} \pp collisions.
In this dataset, exclusive event classes containing electrons, muons, photons, $b$-tagged jets, non-$b$-tagged jets and missing transverse momentum 
have been scanned for deviations from the MC-based SM prediction in the distributions
of the effective mass and the   invariant mass. 
Sensitivity studies 
with various toy signals
($t\bar{t} + \gamma$, $WZ$, gluino, and $Z'$ production)
have shown that the strategy could discover signals for new physics without an a priori knowledge of the existence of the processes.

No significant deviations are found in the 2015 dataset and consequently no dedicated analysis using data-derived signal regions is performed.
The strategy discussed in this paper will be useful to search for signals of unknown particles and interactions in the subsequent Run 2 datasets.

\clearpage

\section*{Acknowledgements}

We thank CERN for the very successful operation of the LHC, as well as the
support staff from our institutions without whom ATLAS could not be
operated efficiently.

We acknowledge the support of ANPCyT, Argentina; YerPhI, Armenia; ARC, Australia; BMWFW and FWF, Austria; ANAS, Azerbaijan; SSTC, Belarus; CNPq and FAPESP, Brazil; NSERC, NRC and CFI, Canada; CERN; CONICYT, Chile; CAS, MOST and NSFC, China; COLCIENCIAS, Colombia; MSMT CR, MPO CR and VSC CR, Czech Republic; DNRF and DNSRC, Denmark; IN2P3-CNRS, CEA-DRF/IRFU, France; SRNSFG, Georgia; BMBF, HGF, and MPG, Germany; GSRT, Greece; RGC, Hong Kong SAR, China; ISF and Benoziyo Center, Israel; INFN, Italy; MEXT and JSPS, Japan; CNRST, Morocco; NWO, Netherlands; RCN, Norway; MNiSW and NCN, Poland; FCT, Portugal; MNE/IFA, Romania; MES of Russia and NRC KI, Russian Federation; JINR; MESTD, Serbia; MSSR, Slovakia; ARRS and MIZ\v{S}, Slovenia; DST/NRF, South Africa; MINECO, Spain; SRC and Wallenberg Foundation, Sweden; SERI, SNSF and Cantons of Bern and Geneva, Switzerland; MOST, Taiwan; TAEK, Turkey; STFC, United Kingdom; DOE and NSF, United States of America. In addition, individual groups and members have received support from BCKDF, CANARIE, CRC and Compute Canada, Canada; COST, ERC, ERDF, Horizon 2020, and Marie Sk{\l}odowska-Curie Actions, European Union; Investissements d' Avenir Labex and Idex, ANR, France; DFG and AvH Foundation, Germany; Herakleitos, Thales and Aristeia programmes co-financed by EU-ESF and the Greek NSRF, Greece; BSF-NSF and GIF, Israel; CERCA Programme Generalitat de Catalunya, Spain; The Royal Society and Leverhulme Trust, United Kingdom. 

The crucial computing support from all WLCG partners is acknowledged gratefully, in particular from CERN, the ATLAS Tier-1 facilities at TRIUMF (Canada), NDGF (Denmark, Norway, Sweden), CC-IN2P3 (France), KIT/GridKA (Germany), INFN-CNAF (Italy), NL-T1 (Netherlands), PIC (Spain), ASGC (Taiwan), RAL (UK) and BNL (USA), the Tier-2 facilities worldwide and large non-WLCG resource providers. Major contributors of computing resources are listed in Ref.~\cite{ATL-GEN-PUB-2016-002}.

\clearpage

\appendix
\part*{Appendix}
\addcontentsline{toc}{part}{Appendix}
\section{Details of the Monte Carlo samples}
\label{appendixA}
\subsection{Monte Carlo programs and settings}
\label{appendixA1}
Samples of multijet production were simulated with  $2\to2$ matrix elements (ME) 
at leading order (LO) using the \PYTHIAV{8.186} generator~\cite{Sjostrand:2007gs}.
The A14~\cite{ATL-PHYS-PUB-2014-021} set of shower and multiple parton interactions parameters (tune) was used
together with the NNPDF2.3LO PDF set~\cite{Ball:2012cx}.
Alternative multijet samples with  $2\to2$ ME at LO were generated with \textsc{Herwig++} 2.7.1~\cite{Corcella:2000bw}
with the UEEE5 underlying-event 
tune 
and the CTEQ6L1~\cite{Pumplin:2002vw} PDF set,
and with \textsc{Sherpa} 2.1.1~\cite{Gleisberg:2008ta} with ME for $2\to2$ and $2\to3$ partons at LO merged using the ME+PS\@LO prescription.
All \textsc{Sherpa} samples use   the  CT10~\cite{Lai:2010vv} PDF set
and  the \textsc{Sherpa} parton shower~\cite{Schumann:2007mg}
with a dedicated shower tuning developed by the \textsc{Sherpa} authors.

Events containing leptonic decays of a $W$ or $Z$ bosons with associated jets ($W$/$Z+\text{jets}$) 
were simulated using the \textsc{Sherpa}~2.1.1 generator~\cite{ATL-PHYS-PUB-2016-003}.
Matrix elements were calculated using the Comix~\cite{Gleisberg:2008fv} and Open-Loops~\cite{Cascioli:2011va} generators.
They include up to two partons at NLO and four partons at leading order (LO), 
merged using the ME+PS@NLO prescription~\cite{Hoeche:2012yf}.
Samples with $W$ and $Z$ decaying hadronically
are also generated with \textsc{Sherpa}~2.1.1
including up to four partons at LO.
The $W$/$Z+\text{jets}$ events were normalized to their inclusive next-to-next-to-leading-order (NNLO) cross-sections~\cite{Gavin:2010az,Gavin:2012sy}.
Simulated samples of massive vector bosons produced in association with one or two real photons 
were also generated with \textsc{Sherpa} 2.1.1 with a ME calculated at LO for up to three partons.
They are scaled to their NLO cross-sections computed with MCFM~\cite{Campbell:1999ah,Campbell:2011bn}.

Samples of prompt photon production in association with jets ($\gamma+\text{jets}$) were generated using \textsc{Sherpa} 2.1.1.
For these samples up to four real parton emissions are included at LO.
Events containing two prompt photons ($\gamma\gamma+\text{jets}$) were also generated with \textsc{Sherpa} 2.1.1.
Matrix elements were calculated with up to two partons at LO. The gluon-induced box process is also included. 
These samples were scaled to data following the procedure described in \Sect{\ref{sec:reweighting}}.

Top-quark pair production events, and single top quarks in the $W t$- and s-channels,
were simulated using the \textsc{Powheg-Box} v2~\cite{Alioli:2010xd} generator
with the CT10 PDF set, as detailed in~\Ref{\cite{ATL-PHYS-PUB-2016-004}}. 
The top quark mass was set to 172.5~\GeV.
The $h_{\textrm{damp}}$ parameter, which regulates the transverse momentum  of the first extra emission beyond the Born configuration
and thus controls the \pt{} of the $t\bar{t}$ system, was set to the mass of the top quark.
Electroweak t-channel single-top-quark events were generated using the \textsc{Powheg-Box} v1 generator.
This generator uses the four-flavour scheme for the NLO matrix-element calculations
together with the fixed four-flavour PDF set CT10f4.
For all top-quark processes, top-quark spin correlations are preserved
(for the single-top t-channel, top quarks were decayed using MadSpin~\cite{Artoisenet:2012st}).
An alternative sample of $t\bar{t}$ was generated with the \textsc{Sherpa} 2.1.1 generator,
including up to one additional parton at NLO and up to four additional partons at LO accuracy,
interfaced to the parton shower using the ME+PS@NLO prescription.
The parton shower (PS), fragmentation, and the underlying event of the \textsc{Powheg-Box} samples 
were simulated using \PYTHIAV{6.428}~\cite{Sjostrand:2006za}
with the CTEQ6L1 PDF set and the corresponding Perugia 2012 tune (P2012)~\cite{Skands:2010ak}.
The  $t\bar{t}$ and single-top-quark events were normalized to the NNLO  cross-section
including the resummation of soft gluon emission at next-to-next-to-logarithm accuracy~\cite{Kidonakis:2010ux,Kidonakis:2010tc,Kidonakis:2011wy} 
using \verb|Top++2.0|~\cite{Czakon:2011xx}.
Both the default and the alternative $t\bar{t}$ samples were corrected to reproduce 
the NNLO prediction~\cite{Czakon:2015owf,Czakon:2016dgf}  of the top quark \pt{} and the \pt{} of the $t\bar{t}$ system. 
The contribution of $t\bar{t}+b\bar{b}$ was generated separately with \textsc{Sherpa}
2.1.1 at NLO;
the calculation was performed in the four-flavour scheme and with the CT10f4 PDF set.

Diboson samples were generated with the \textsc{Sherpa} 2.1.1 generator,
and are described in~\Ref{\cite{ATL-PHYS-PUB-2016-002}}.
The matrix elements contain the 
$WW$, $WZ$ and $ZZ$ processes 
and all other diagrams with four or six electroweak vertices
(such as same-electric-charge $W$ boson production in association with two jets, $W^{\pm}W^{\pm}jj$).
Fully leptonic triboson processes ($WWW$, $WWZ$, $WZZ$ and $ZZZ$)
with on-shell bosons and up to six charged leptons were also simulated using \textsc{Sherpa} 2.1.1.
The ME for the $ZZ$ processes were calculated 
at NLO for up to one additional parton;
final states with two and three additional partons were calculated at LO.
The $WZ$ and $WW$ processes were calculated at NLO
with up to three extra partons at LO using the ME+PS@NLO prescription.
The $WW$ final states were generated without bottom quarks
in the hard-scattering process, to avoid contributions from top-quark-mediated processes.
The triboson processes were calculated with the same configuration 
and with up to two extra partons at LO.
The generator cross-sections were used for the normalization of these backgrounds.

Samples of top quark production in association with vector bosons~\cite{ATL-PHYS-PUB-2016-005}
($W$, $Z$, $\gamma$ and $WW$, including the non-resonant $\gamma^*/~Z$ contributions)
were generated at LO with MG5\_aMC@NLO~2.2.2~\cite{Alwall:2014hca} interfaced to \PYTHIAV{8.186},
with up to two ($t\bar{t}W$), one ($t\bar{t}Z$) or no ($t\bar{t}WW$, $t\bar{t}\gamma$) extra partons included in the matrix element.
The  A14 tune was used together with the NNPDF2.3LO PDF set.
The $t\bar{t}\gamma$ sample uses a fixed QCD renormalization and factorization scale of $2 m_{t}$,
and the top decay was performed in MG5\_aMC@NLO 
to account for hard photon radiation from the top decay products.
The same generator was also used to simulate the $tZ$,  3-top and 4-top quarks processes.
The $t\bar{t}W$, $t\bar{t}Z$,  $t\bar{t}WW$  and 4-top samples 
were normalized to their NLO cross-sections~\cite{Alwall:2014hca}
while the LO cross-section from the generator was used for $tZ$ and 3-top quarks.

The Higgs boson mass was set to 125~\GeV{} and all SM Higgs boson decay modes were considered.
The production of the SM Higgs boson in the gluon--gluon fusion (ggF) and vector-boson fusion (VBF) channels
was modelled using the \textsc{Powheg-Box} v2 generator with the CT10 PDF set.
It was interfaced to \PYTHIAV{8.186} with the CTEQ6L1 PDF set and the AZNLO tune~\cite{STDM-2012-23}.
Production of a Higgs boson in association with a pair of top quarks was simulated using MG5\_aMC@NLO~2.2.2
interfaced to \textsc{Herwig++}~2.7.1~\cite{Bahr:2008pv} for showering and hadronization. 
The UEEE5 underlying-event tune was used together with the CT10  
(matrix element) and CTEQ6L1 (parton shower) PDF sets.
Simulated samples of SM Higgs boson production in association with a $W$ or $Z$ boson were produced with \PYTHIAV{8.186},
using the A14 tune and the NNPDF2.3LO PDF set. Events were normalized to their most accurate cross-sections calculations (typically NNLO)~\cite{Dittmaier:2012vm}.

To avoid double counting, events with a hard photon from final-state radiation were removed
from the multijet, $t\bar{t}$ and $W$/$Z+\text{jets}$ samples.

\subsection{Theoretical uncertainties}
\label{appendixA2}
The inclusive $W$ and $Z$ cross-sections are known at NNLO, with an uncertainty of about 5\%~\cite{Gavin:2010az,Gavin:2012sy}.
Modelling uncertainties for $W+\text{jets}$ and $Z+\text{jets}$ are determined by varying the renormalization, 
factorization and resummation scales in the ME by factors $0.5$ and $2$,
together with a change of the merging scale from 20~\GeV{} to 15~\GeV{} or 30~\GeV.

For top quark pair or single-top production, processes known to NNLO+NNLL~\cite{Czakon:2011xx}
or approximate NNLO~\cite{Kidonakis:2010ux,Kidonakis:2010tc,Kidonakis:2011wy}, respectively, the cross-section uncertainty is 7\%. 
The modelling uncertainty for $t\bar{t}$ is determined by comparing the nominal
\textsc{Powheg+Pythia}
NLO+PS sample
with an alternative sample generated with \textsc{Sherpa} including up to two partons at NLO and four at LO accuracy in the ME.
The single-top quark uncertainty is estimated by varying the renormalization and factorization scales,
and by changing the $h_{\textrm{damp}}$ parameter and the shower tune 
of the \textsc{Powheg+Pythia} sample.
An uncertainty in the interference between the $Wt$ and $t\bar{t}$ production 
is estimated by comparing the nominal $Wt$ sample,
where all doubly resonant NLO $Wt$ diagrams are removed, 
with a sample where the cross-section contribution from Feynman diagrams containing two top quarks is subtracted~\cite{Frixione:2008yi}.

Diboson cross-sections ($WW$, $WZ$ and $ZZ$) are calculated at NLO, 
and a 6\% uncertainty, evaluated with the MCFM program~\cite{Campbell:1999ah,Campbell:2011bn}, is applied to their cross-sections.
Their modelling uncertainty is evaluated analogously to $V+\text{jets}$ by varying the scales used to perform the calculation.
For $W+\gamma$ and $Z+\gamma$  samples, which are computed at LO, a 20\% uncertainty in the cross-section is assumed,
with a further 20\%  modelling uncertainty assigned in accord with the measurement in~\Ref{\cite{STDM-2012-07}}.

The cross-sections for top-quark pair production in association with one or two vector bosons  
are calculated at NLO and an uncertainty of 15\% is used~\cite{Alwall:2014hca}.
Their modelling uncertainty is evaluated from variations of the renormalization and factorization scales,
together with a change in the merging scale.
For $t\bar{t}+\gamma$, an additional 12\% uncertainty in the normalization is assumed,
while a uniform 30\% uncertainty is assigned to the modelling~\cite{ATL-PHYS-PUB-2016-005}.

Multijet and $\gamma+\text{jets}$ processes are scaled to data
following the procedure described in \Sect{\ref{sec:Validation_procedures}}.
Therefore, no uncertainty is applied to their normalization.
For multijets the maximum bin-by-bin difference between the \textsc{Pythia} 8 nominal sample
and alternative samples generated with \textsc{Sherpa} and \textsc{Herwig++} 
is considered as a shape uncertainty.
In addition the 
standard deviation of the 100 replica sets of the NNPDF2.3LO PDF is used.
The modelling uncertainty for $\gamma+\text{jets}$  is estimated from scale variations
with the same methodology as for the $V+\text{jets}$ samples.
The uncertainty in the $\gamma\gamma+\text{jets}$ modelling is instead taken to be 30\%
from parton-level comparisons of samples with varied scales.

A conservative 
uncertainty of 20\%~\cite{Dittmaier:2012vm} is assumed for Higgs production in the ggF, VBF and $VH$ channels.
A further uncertainty of 20\% is assigned as a shape uncertainty.

For the subdominant triboson (including $V+\gamma\gamma$), $ttH$, 3-top, 4-top and $tZ$ production processes
a 50\% uncertainty is assigned to the event yields, similarly to~\Ref{\cite{SUSY-2015-09}}.
Uncertainties associated to PDFs are found to be small in all channels compared to the modelling uncertainties of the MC simulations.

\section{Details of the object reconstruction}
\label{appendixB}
Electron candidates are reconstructed from an isolated electromagnetic calorimeter energy deposit
matched to an ID track and are required to have $|\eta|<2.47$, a transverse momentum $\pt > 10$~\GeV, 
and to pass a loose likelihood-based identification requirement~\cite{PERF-2016-01, ATLAS-CONF-2016-024}.
The likelihood input variables include mea\-sure\-ments of calorimeter shower shapes
and measurements of track properties from the ID.
The candidate electrons are selected if the matched tracks 
have a transverse impact parameter significance 
relative to the reconstructed primary vertex of $|d_0|/\sigma(d_0) < 5$.
Candidates within the transition region between the barrel and endcap
electromagnetic calorimeters, $1.37<|\eta|< 1.52$, are removed.

Muon candidates are reconstructed in the region $|\eta|< 2.7$
from muon spectrometer tracks matched to ID tracks.
The muon candidates are selected if they have a transverse momentum above 10~\GeV{}
and pass the medium  identification requirements
defined in~\Ref{\cite{PERF-2015-10}}, based on selections on the number of hits
in the different ID and muon spectrometer subsystems, and the significance of the charge to momentum ratio $q/p$.

All candidate leptons (electrons and muons) are used for the object overlap removal,
as discussed in \Sect{\ref{sec:Objects}}. 
Tighter requirements on the lepton candidates are imposed, which are then referred to as `signal' electrons or muons and are used further in the analysis, 
i.e.\ to establish the accuracy of the background modelling processes or to classify the events.
Signal electrons must satisfy a tight likelihood-based identification requirement~\cite{PERF-2016-01, ATLAS-CONF-2016-024}. 
Signal muons must fulfil the requirement of $|d_0|/\sigma(d_0) < 3$.
The track associated with a signal lepton must have a longitudinal impact parameter
relative to the reconstructed primary vertex, $z_0$, satisfying $|z_0\cdot \sin\theta| < 0.5$ mm.
Isolation requirements are applied to both the signal electrons and muons. 
The calorimeter isolation is computed as  the sum of the energies of calorimeter energy clusters
in a cone of size $\Delta R =0.2$ around the lepton.
Track isolation is defined as the scalar sum of the \pt{} of tracks
within a variable-size cone around the lepton, 
in a cone of size $\Delta R = 0.2$ (0.3) for
electron (muon) transverse momenta $\pT < 50$~\GeV{} ($\pT < 33$~\GeV)
and of size $\Delta R = 10~\GeV/\pT$ for $\pT > 50$~\GeV{}  ($\pT > 33$~\GeV).
The efficiency of these criteria increases with
the lepton transverse momentum, reaching 95\% at 25~\GeV{} and 99\%
at 60~\GeV, as determined in a control sample of $Z$ decays into leptons
selected with a tag-and-probe technique~\cite{PERF-2015-10,ATLAS-CONF-2016-024}. 
Corrections are applied to the MC samples to match the leptons' trigger, reconstruction and isolation efficiencies in data.

Photon candidates 
are reconstructed from an isolated electromagnetic calorimeter energy deposit 
and are required to satisfy the tight identification criteria described in Refs.~\cite{ATL-PHYS-PUB-2016-014,PERF-2013-04}.
Furthermore, photons are required to have $\pt > 25$~\GeV{} and $|\eta| < 2.37$,
excluding the barrel--endcap calorimeter transition in the range $1.37 < |\eta| < 1.52$.
Photons must further satisfy isolation criteria based on both track and calorimeter information~\cite{ATL-PHYS-PUB-2016-014}.
After correcting for contributions from pile-up, 
the energy within a cone of $\Delta R = 0.4$ around the cluster barycentre 
is required to be less than $2.45~\GeV + 0.022 \times \pt^{\gamma}$, 
where $\pt^{\gamma}$ is the transverse momentum of the photon candidate.
The energy of tracks in a cone of $\Delta R = 0.2$ should be less than $0.05\times \pt^{\gamma}$.

Jet candidates are reconstructed with the anti-$k_t$ algorithm~\cite{Cacciari:2008gp} implemented in the FastJet package~\cite{Cacciari:2011ma} with a radius parameter of $R = 0.4$,
using as input three-dimensional energy clusters in the calorimeter~\cite{PERF-2014-07} calibrated to the electromagnetic scale.
The reconstructed jets are then calibrated to the jet energy scale (JES) 
derived from simulation and in situ corrections based on 13 TeV data~\cite{PERF-2016-04,ATLAS-CONF-2015-037}.
For all jets the expected average energy contribution 
from pile-up clusters is subtracted according to the jet area prescription~\cite{PERF-2014-03}.
Quality criteria are imposed to identify jets arising from non-collision sources
or detector noise and any event containing such a jet is removed~\cite{ATLAS-CONF-2015-029}.
All jet candidates are required to have $\pt>20~\GeV$ and $|\eta| < 2.8$.

Identification of jets containing $b$-hadrons ($b$-tagging)
is performed with a multivariate discriminant, \textsc{MV2c20},
making use of track impact parameters, the $b$- and $c$-hadron flight paths inside the jet and reconstructed secondary vertices~\cite{PERF-2012-04,ATL-PHYS-PUB-2015-022}.
The algorithm working point used corresponds
to a 77\% average efficiency obtained for $b$-jets in simulated $t\bar{t}$ events.
The rejection factors for light-quark jets, $c$-quark jets 
and hadronically decaying $\tau$-leptons in simulated $t\bar{t}$ events
are approximately 140, 4.5 and 10, respectively~\cite{ATL-PHYS-PUB-2015-022}. 
Jets with $|\eta|<2.5$ which satisfy this $b$-tagging requirement are identified as $b$-jet candidates.
To compensate for differences between data and MC simulation
in the $b$-tagging efficiencies and mistag rates,
correction factors are applied to the simulated samples~\cite{ATL-PHYS-PUB-2015-022}.

\clearpage
\section{Event yields for all event classes}
\label{appendixA3}
\begin{figure}[htbp]
  \begin{center}
 \includegraphics[width=\textwidth]{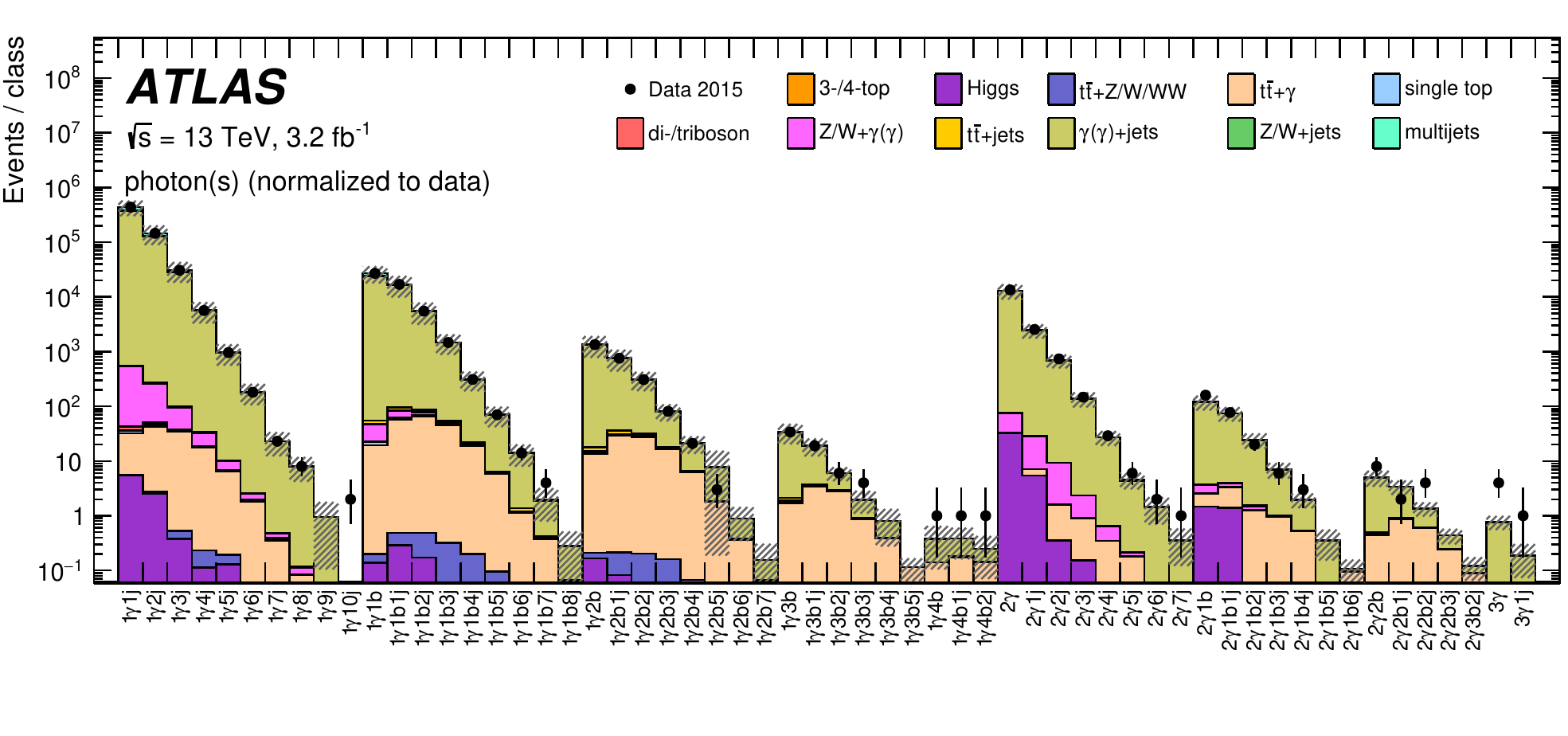}
  \end{center}
 \vspace*{-0.5cm}
  \caption{\label{fig:global2} The number of events in data, and for the different SM background predictions considered,
    for classes with at least one photon and ($b$-)jets (no leptons or \met{}).
    The classes are labelled according to the multiplicity and type
    ($e$, $\mu$, $\gamma$, $j$, $b$, $\met$) of the reconstructed objects for the given event class.
    In event classes with four or more data events, the $\gamma+\text{jets}$ and $\gamma\gamma+\text{jets}$ MC samples
    are scaled to data in the single-photon and diphoton event classes respectively.
    The hatched bands indicate the total uncertainty of the SM prediction. 
    }
\end{figure}
\begin{figure}[htbp]
  \begin{center}
 \includegraphics[height=16cm,width=16cm,keepaspectratio]{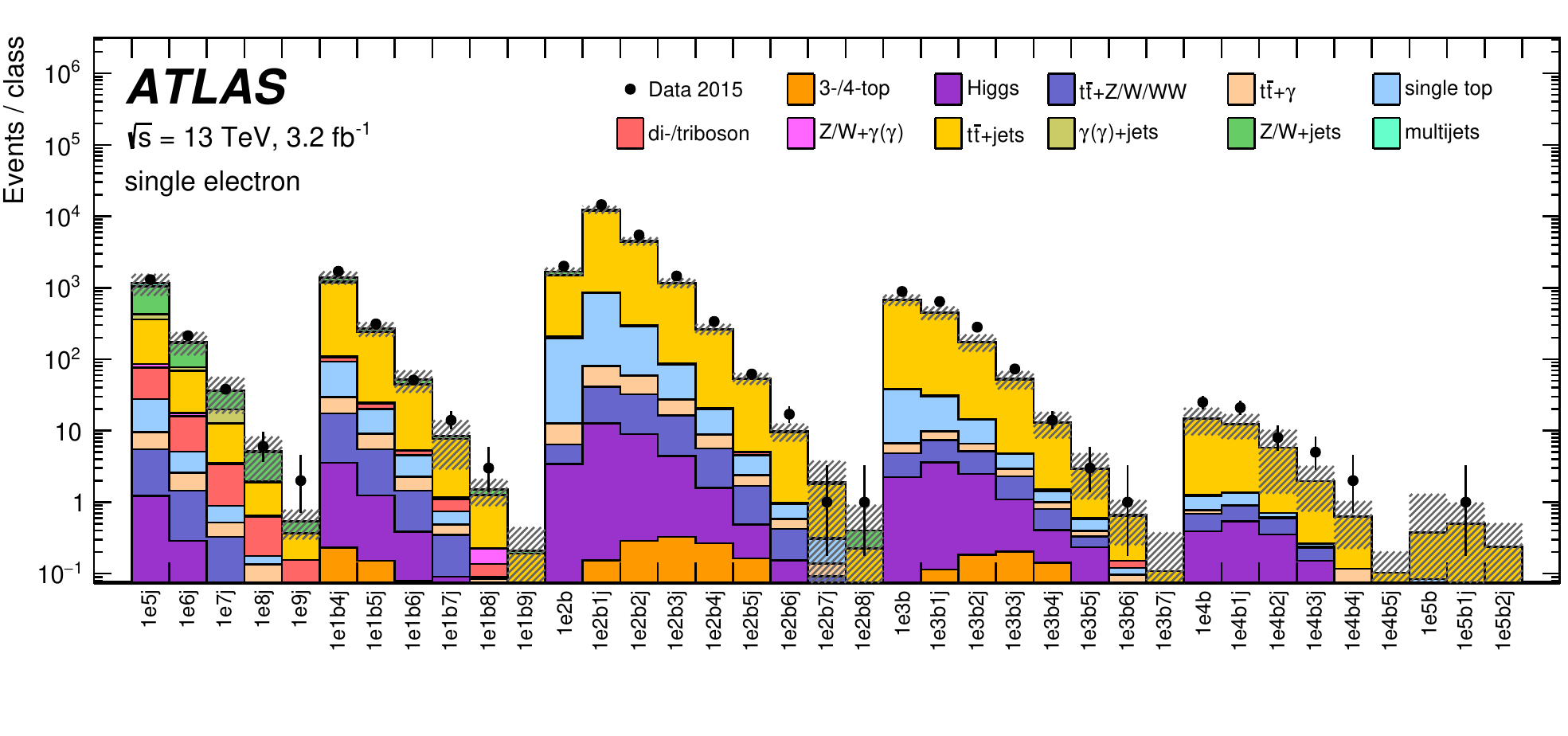}
  \end{center}
 \vspace*{-0.5cm}
  \caption{\label{fig:global3} The number of events in data, and for the different SM background predictions considered,
    for classes with one electron and ($b$-)jets (no muons, photons or \met{}).
    The classes are labelled according to the multiplicity and type
    ($e$, $\mu$, $\gamma$, $j$, $b$, $\met$) of the reconstructed objects for the given event class.
    The hatched bands indicate the total uncertainty of the SM prediction. 
    }
\end{figure}
\begin{figure}[htbp]
  \begin{center}
 \includegraphics[height=16cm,width=16cm,keepaspectratio]{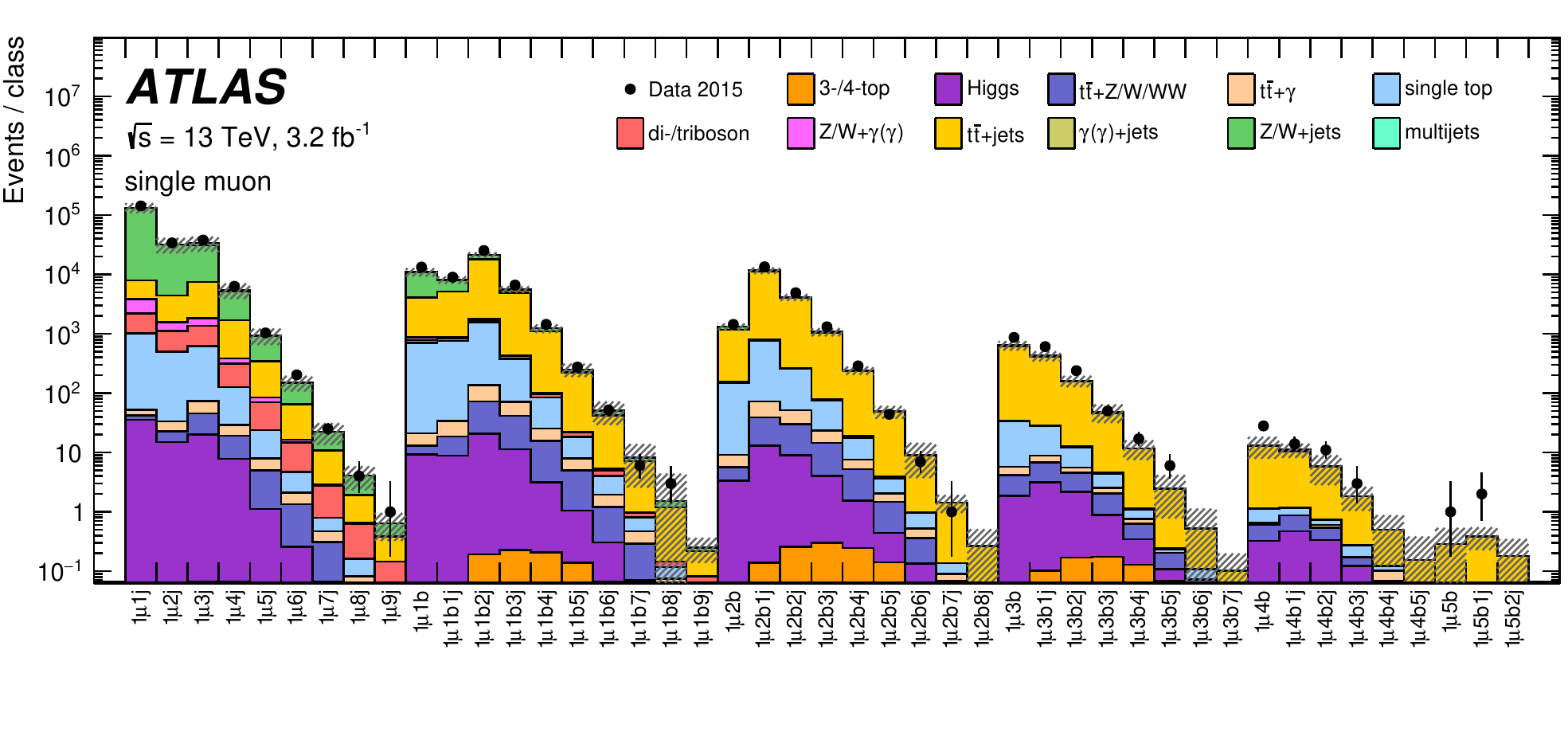}
  \end{center}
 \vspace*{-0.5cm}
  \caption{\label{fig:global4} The number of events in data, and for the different SM background predictions considered,
    for classes with one muon and ($b$-)jets (no electrons, photons or \met{}).
    The classes are labelled according to the multiplicity and type
    ($e$, $\mu$, $\gamma$, $j$, $b$, $\met$) of the reconstructed objects for the given event class.
    The hatched bands indicate the total uncertainty of the SM prediction. 
    }
\end{figure}
\begin{figure}[htbp]
  \begin{center}
 \includegraphics[height=16cm,width=16cm,keepaspectratio]{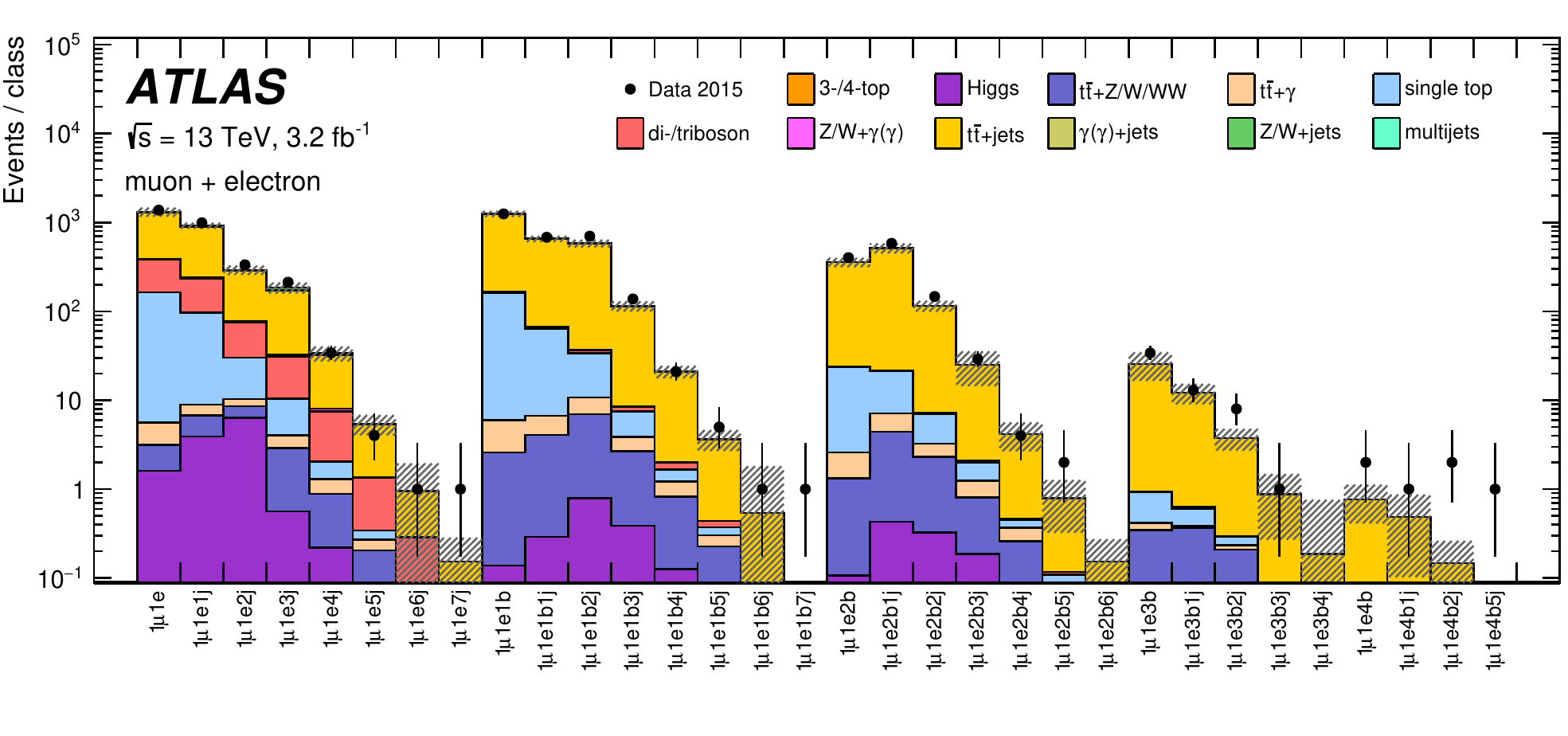}
  \end{center}
 \vspace*{-0.5cm}
  \caption{\label{fig:global5} The number of events in data, and for the different SM background predictions considered,
  	for classes with one muon, one electron and ($b$-)jets (no photons or \met{}).
    The classes are labelled according to the multiplicity and type
    ($e$, $\mu$, $\gamma$, $j$, $b$, $\met$) of the reconstructed objects for the given event class.
    The hatched bands indicate the total uncertainty of the SM prediction. 
    }
\end{figure}
\begin{figure}[htbp]
  \begin{center}
 \includegraphics[height=16cm,width=16cm,keepaspectratio]{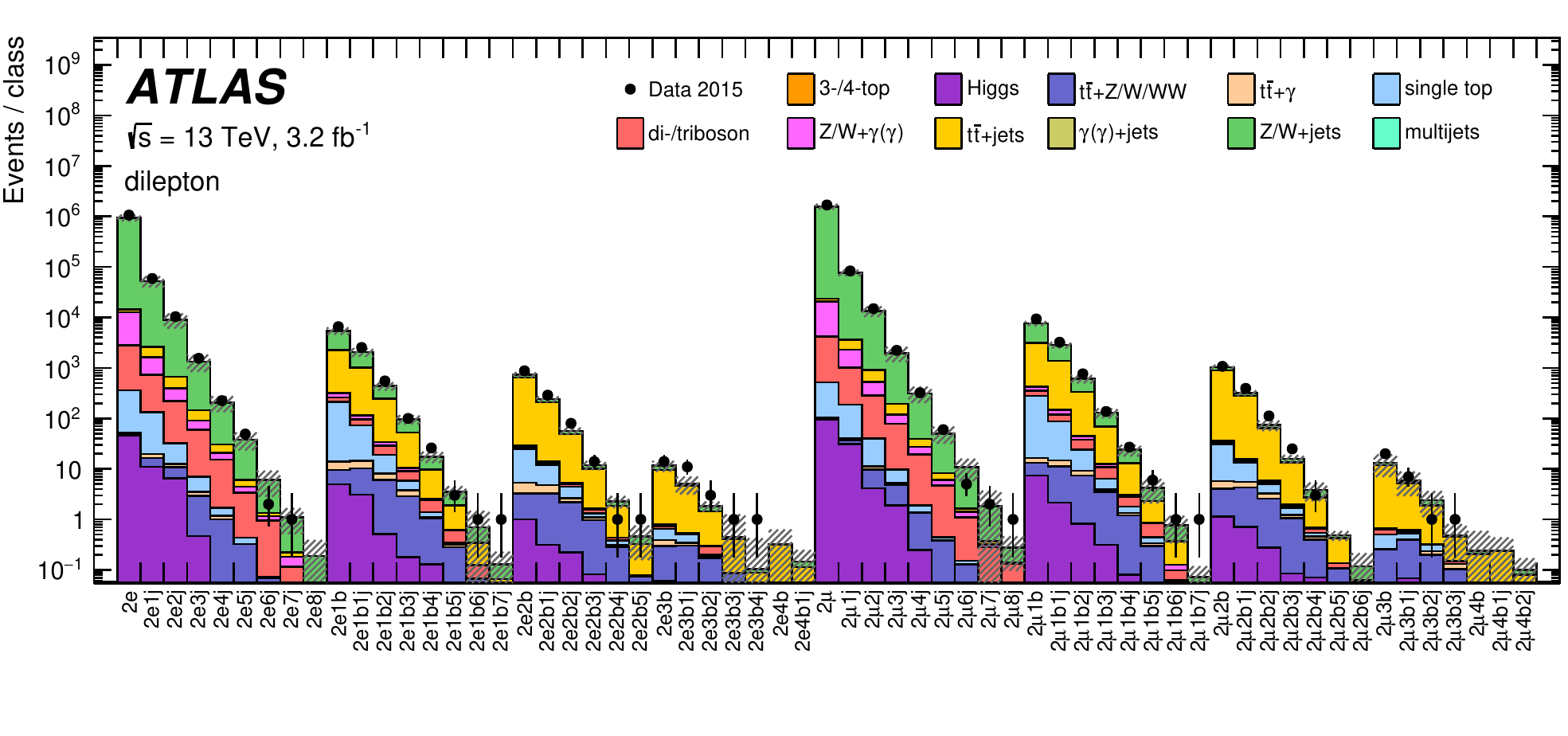}
  \end{center}
 \vspace*{-0.5cm}
  \caption{\label{fig:global6} The number of events in data, and for the different SM background predictions considered,
  	for classes with two same-flavour leptons and ($b$-)jets (no photons or \met{}).
    The classes are labelled according to the multiplicity and type
    ($e$, $\mu$, $\gamma$, $j$, $b$, $\met$) of the reconstructed objects for the given event class.
    The hatched bands indicate the total uncertainty of the SM prediction. 
    }
\end{figure}
\begin{figure}[htbp]
  \begin{center}
 \includegraphics[height=16cm,width=16cm,keepaspectratio]{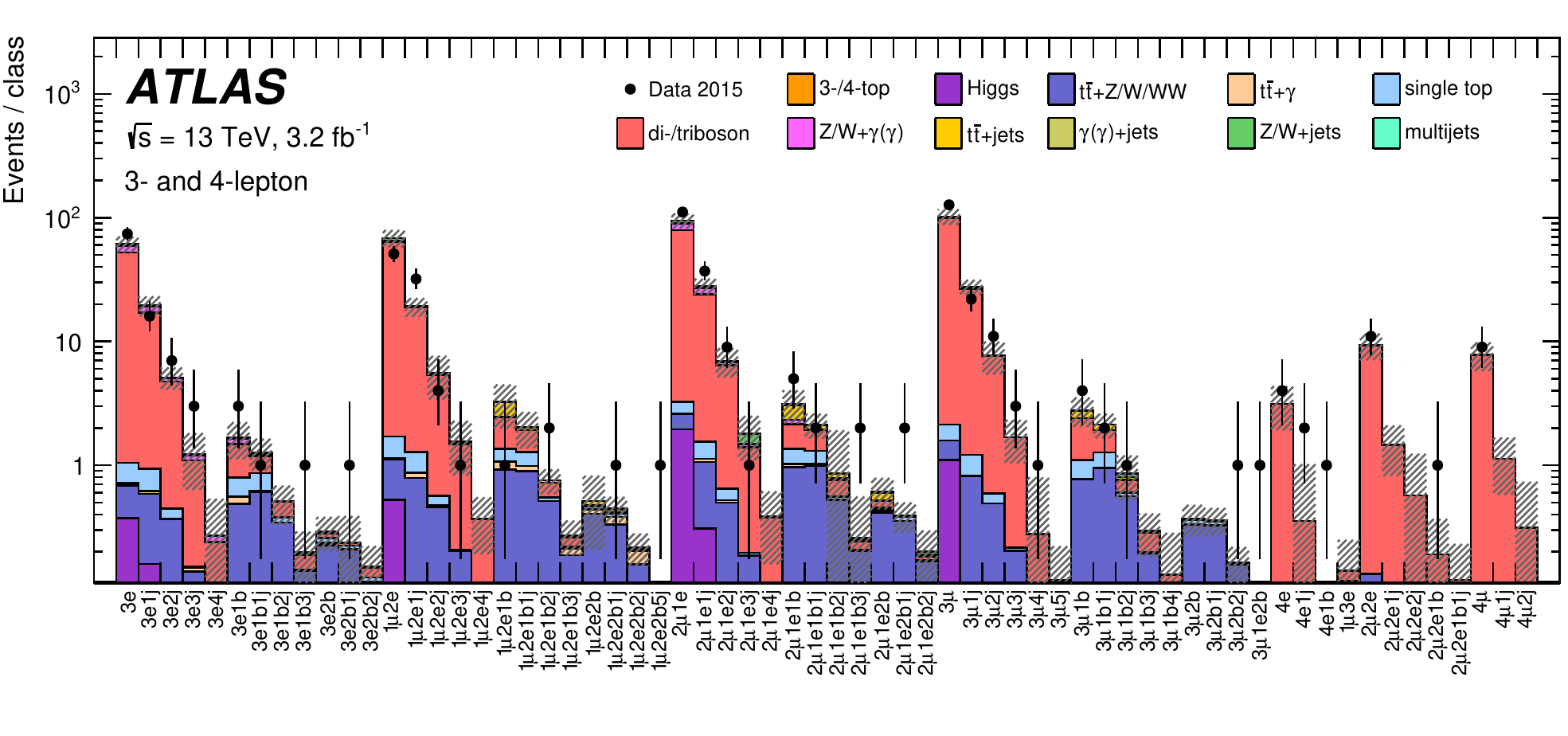}
  \end{center}
 \vspace*{-0.5cm}
  \caption{\label{fig:global7} The number of events in data, and for the different SM background predictions considered,
  	for classes with three or four leptons and ($b$-)jets (no photons or \met{}).
    The classes are labelled according to the multiplicity and type
    ($e$, $\mu$, $\gamma$, $j$, $b$, $\met$) of the reconstructed objects for the given event class.
    The hatched bands indicate the total uncertainty of the SM prediction. 
    }
\end{figure}
\begin{figure}[htbp]
  \begin{center}
 \includegraphics[height=16cm,width=16cm,keepaspectratio]{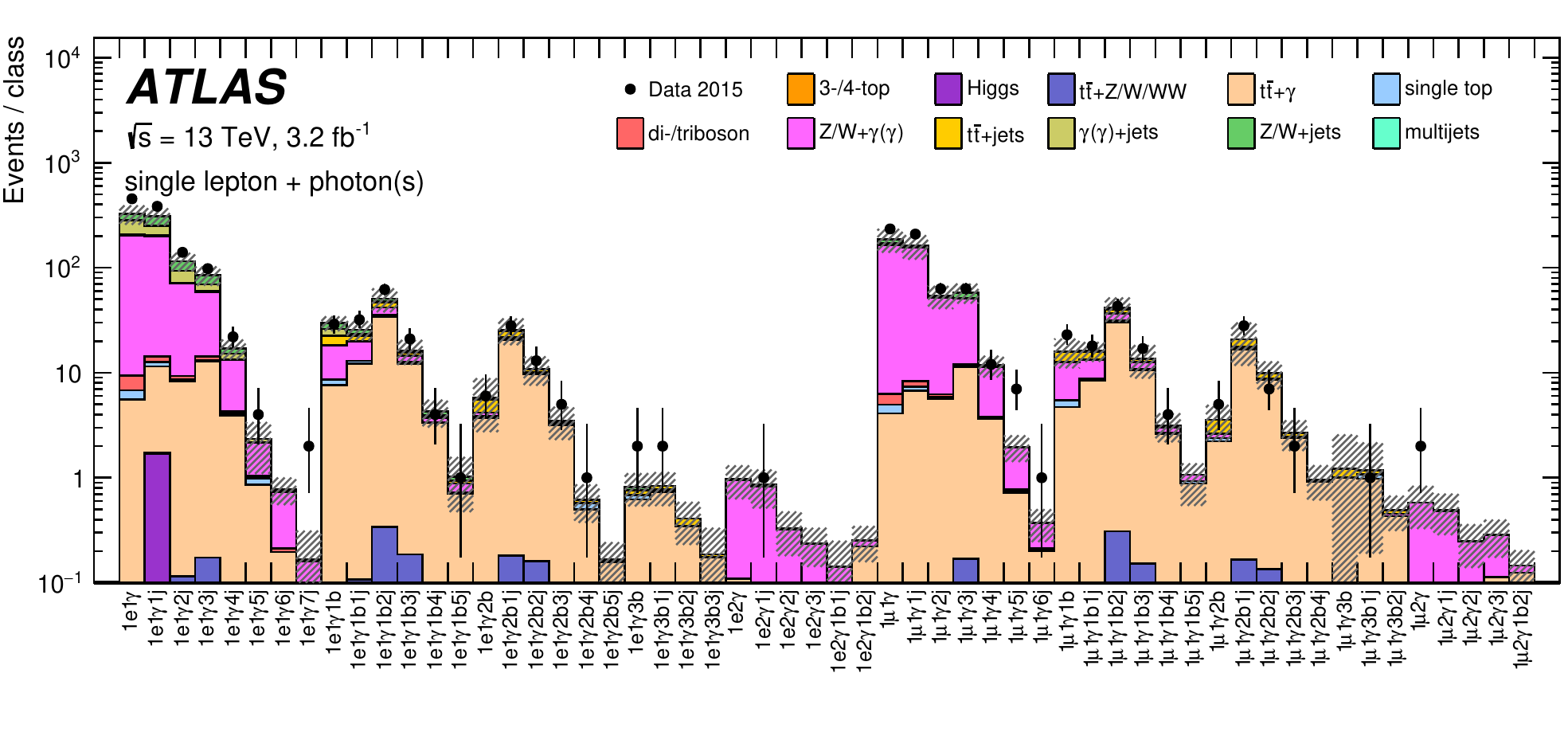}
  \end{center}
 \vspace*{-0.5cm}
  \caption{\label{fig:global8} The number of events in data, and for the different SM background predictions considered,
  	for classes with one lepton, at least one photon and ($b$-)jets (no \met{}).
    The classes are labelled according to the multiplicity and type
    ($e$, $\mu$, $\gamma$, $j$, $b$, $\met$) of the reconstructed objects for the given event class.
    The hatched bands indicate the total uncertainty of the SM prediction. 
    }
\end{figure}
\begin{figure}[htbp]
  \begin{center}
 \includegraphics[height=16cm,width=16cm,keepaspectratio]{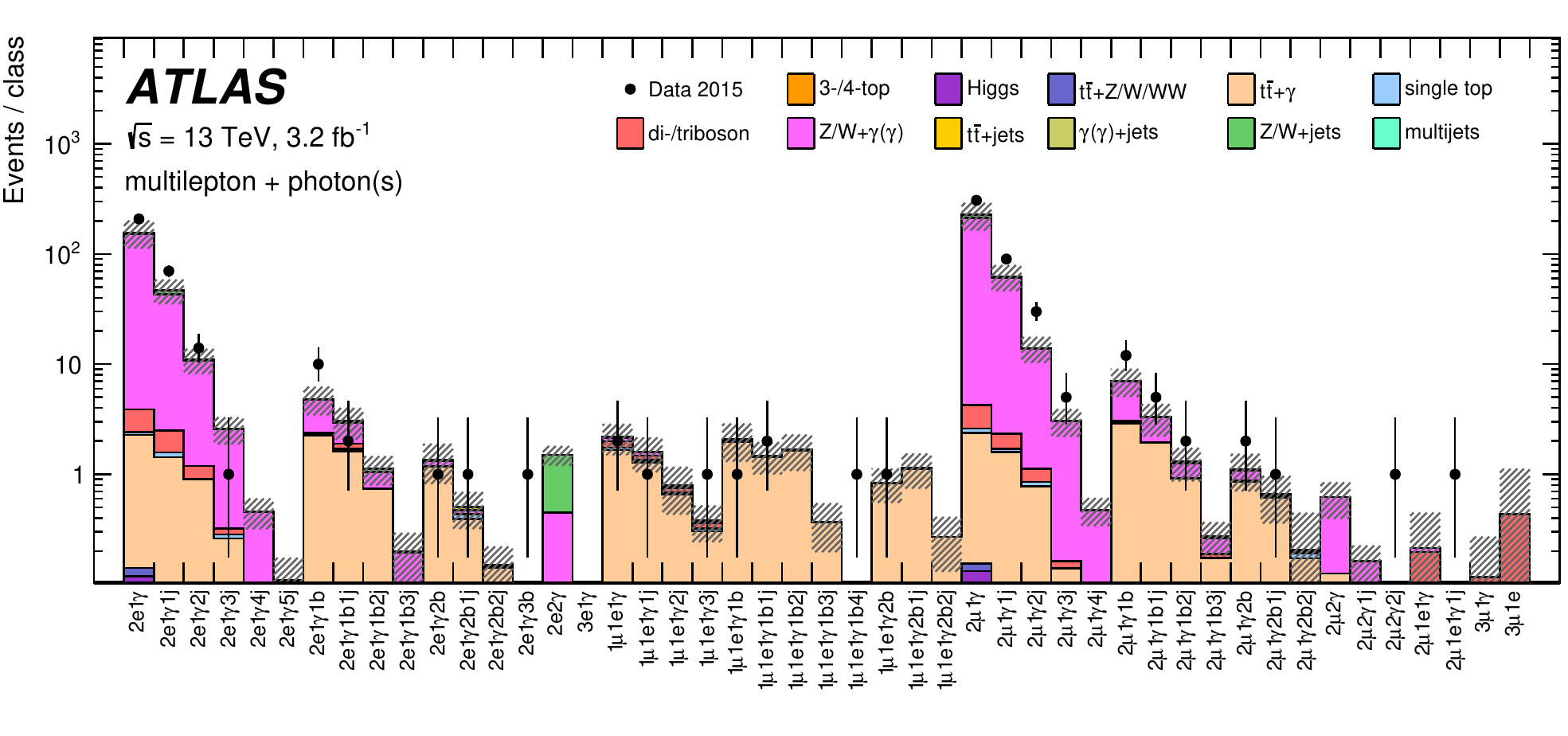}
  \end{center}
 \vspace*{-0.5cm}
  \caption{\label{fig:global9} The number of events in data, and for the different SM background predictions considered,
  	for classes with at least two leptons, at least one photon and ($b$-)jets (no \met{}).
    The classes are labelled according to the multiplicity and type
    ($e$, $\mu$, $\gamma$, $j$, $b$, $\met$) of the reconstructed objects for the given event class.
    The hatched bands indicate the total uncertainty of the SM prediction. 
    }
\end{figure}
\begin{figure}[htbp]
  \begin{center}
 \includegraphics[height=16cm,width=16cm,keepaspectratio]{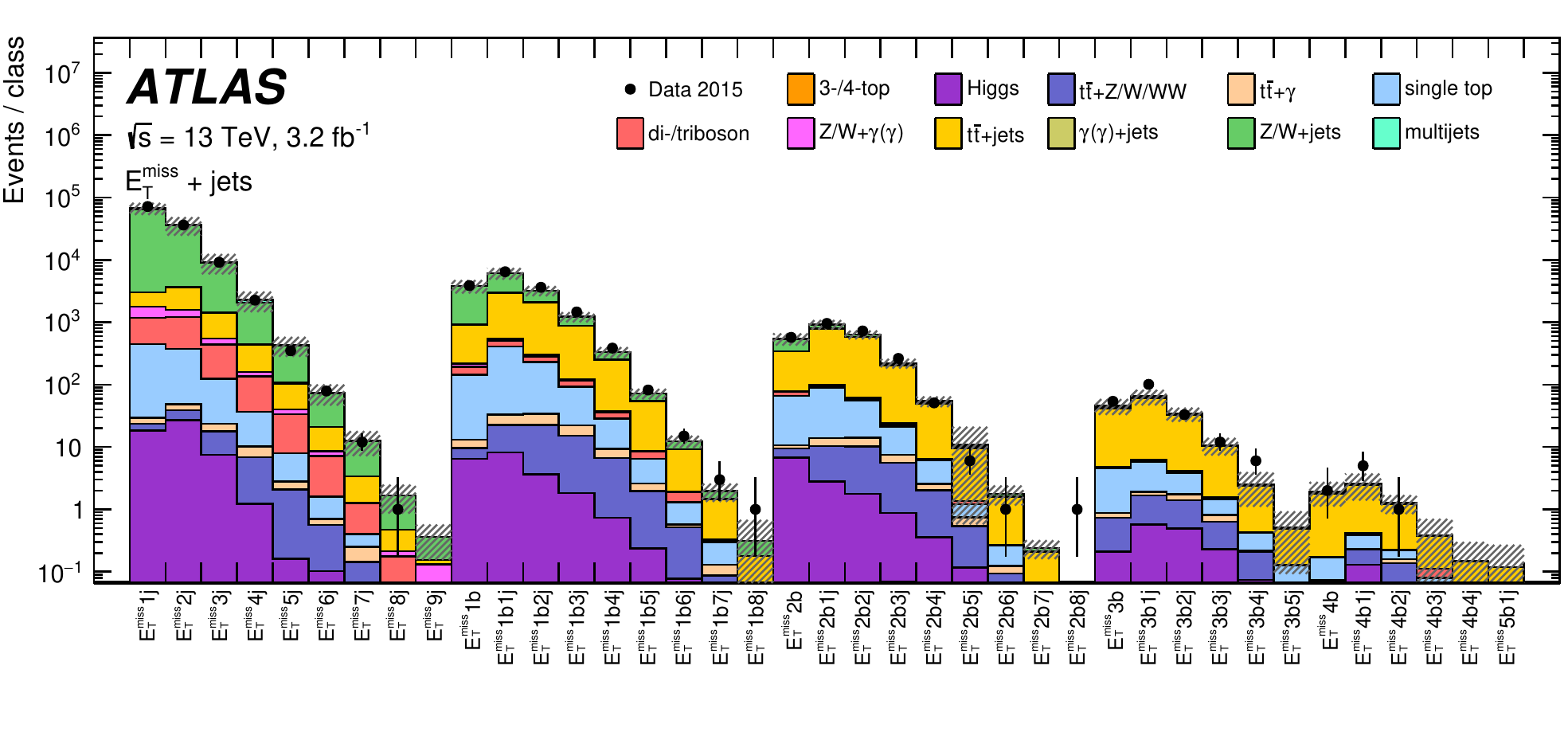}
  \end{center}
 \vspace*{-0.5cm}
  \caption{\label{fig:global10} The number of events in data, and for the different SM background predictions considered,
  	for classes with large \met{} and ($b$-)jets (no leptons or photons).
    The classes are labelled according to the multiplicity and type
    ($e$, $\mu$, $\gamma$, $j$, $b$, $\met$) of the reconstructed objects for the given event class.
    The hatched bands indicate the total uncertainty of the SM prediction. 
    }
\end{figure}
\begin{figure}[htbp]
  \begin{center}
 \includegraphics[height=16cm,width=16cm,keepaspectratio]{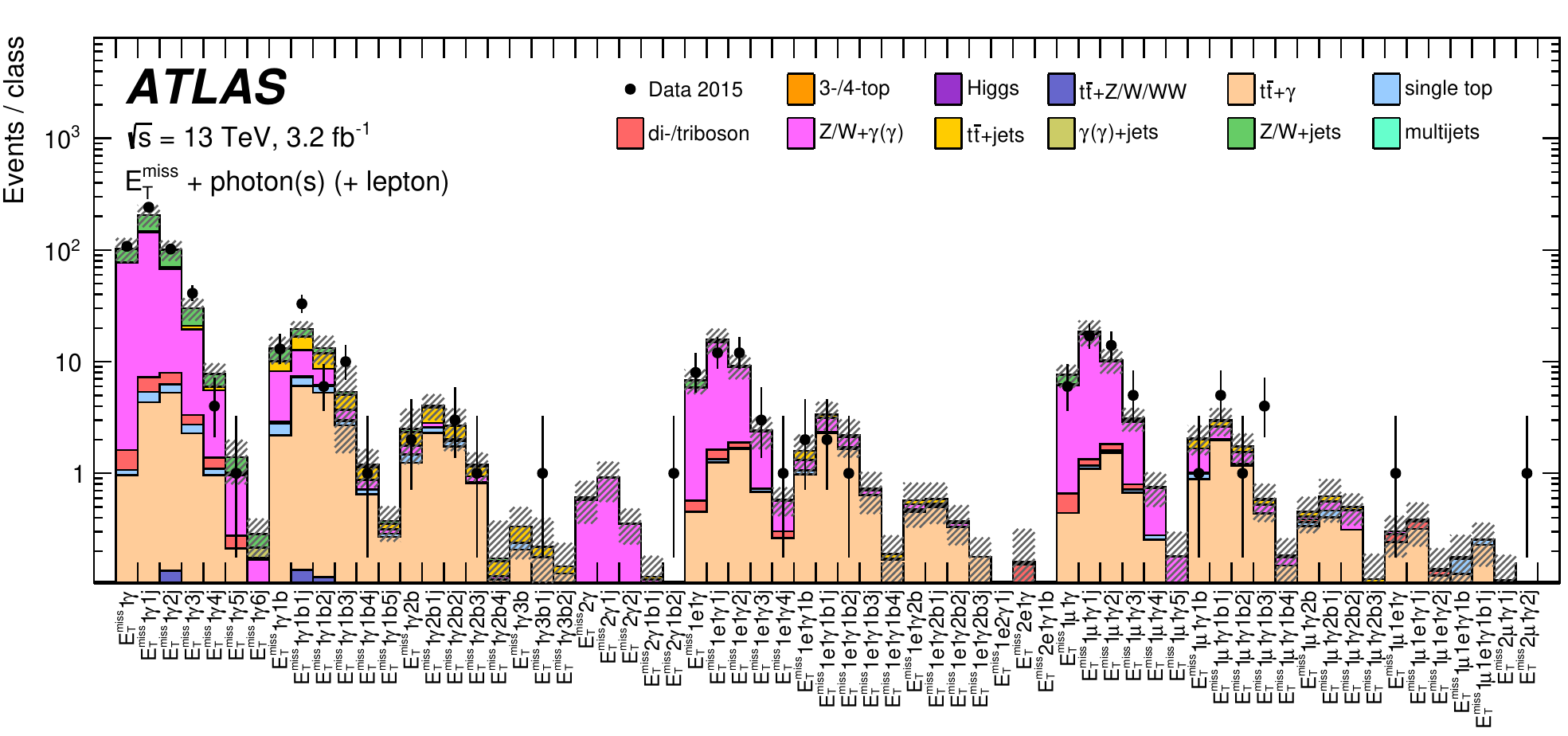}
  \end{center}
 \vspace*{-0.5cm}
  \caption{\label{fig:global11} The number of events in data, and for the different SM background predictions considered,
  	for classes with large \met{}, at least one photon, leptons and ($b$-)jets.
    The classes are labelled according to the multiplicity and type
    ($e$, $\mu$, $\gamma$, $j$, $b$, $\met$) of the reconstructed objects for the given event class.
    The hatched bands indicate the total uncertainty of the SM prediction. 
    }
\end{figure}
\begin{figure}[htbp]
  \begin{center}
 \includegraphics[height=16cm,width=16cm,keepaspectratio]{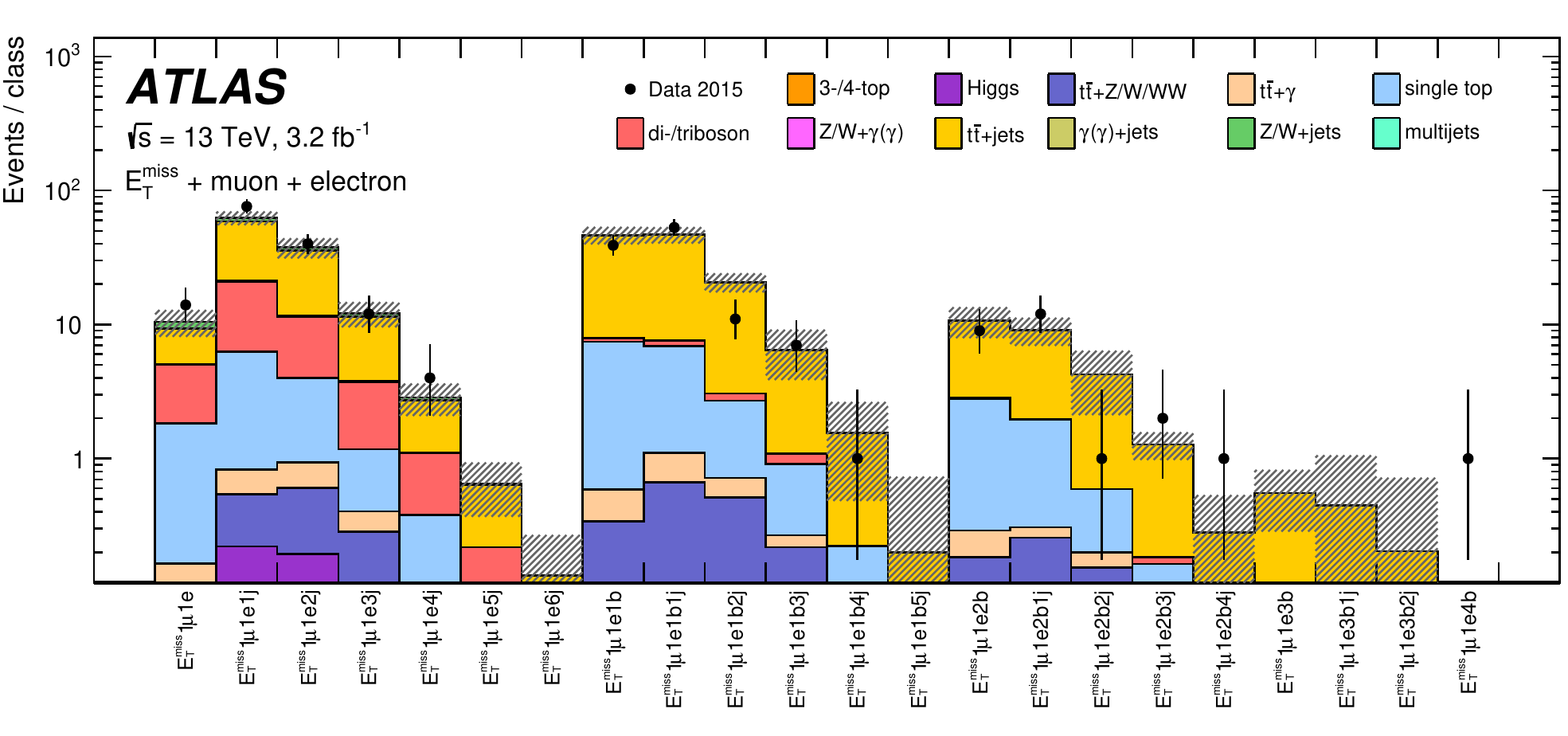}
  \end{center}
 \vspace*{-0.5cm}
  \caption{\label{fig:global12} The number of events in data, and for the different SM background predictions considered,
  	for classes with large \met{}, one muon, one electron and ($b$-)jets (no photons).
    The classes are labelled according to the multiplicity and type
    ($e$, $\mu$, $\gamma$, $j$, $b$, $\met$) of the reconstructed objects for the given event class.
    The hatched bands indicate the total uncertainty of the SM prediction. 
    }
\end{figure}
\begin{figure}[htbp]
  \begin{center}
 \includegraphics[height=16cm,width=16cm,keepaspectratio]{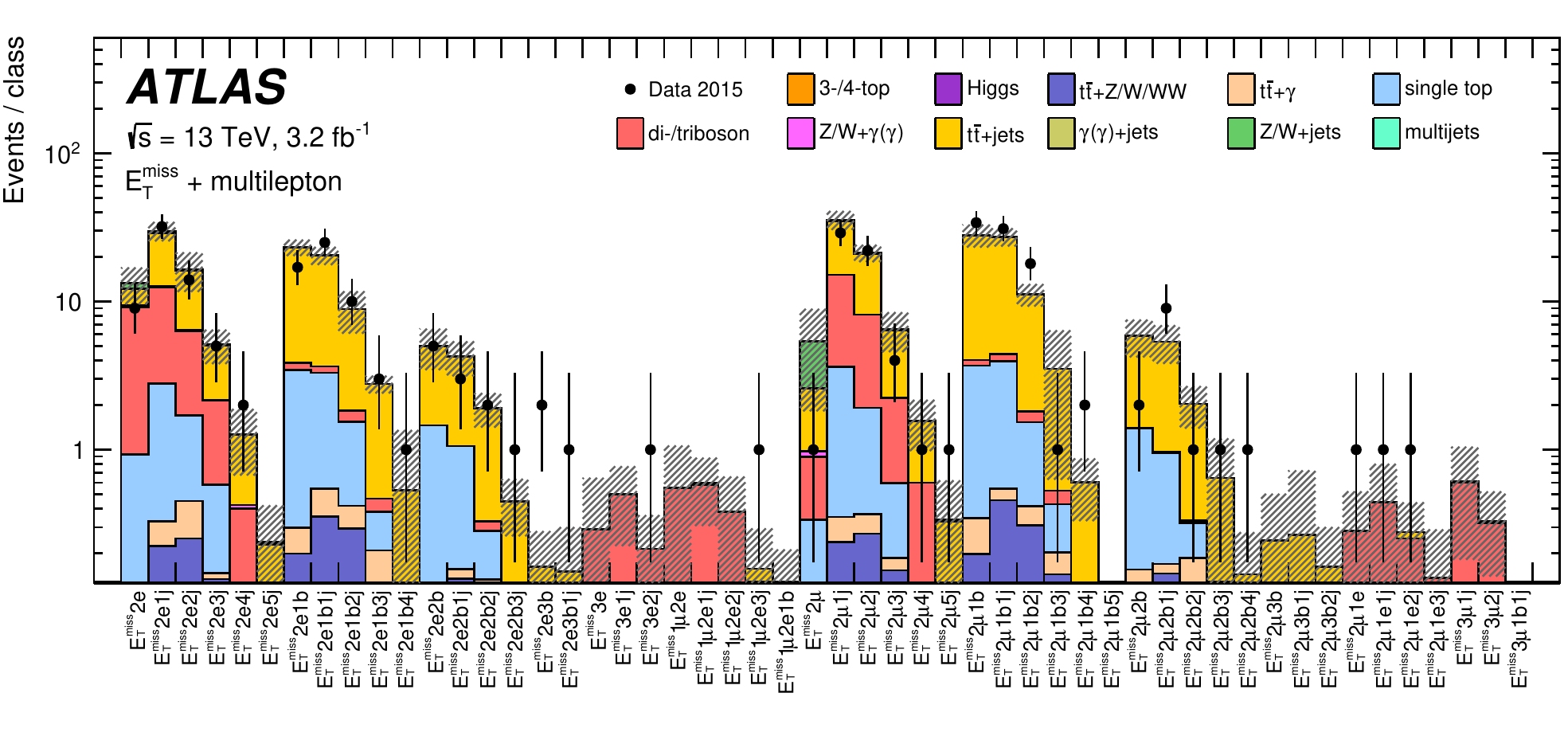}
  \end{center}
 \vspace*{-0.5cm}
  \caption{\label{fig:global13} The number of events in data, and for the different SM background predictions considered,
  	for classes with large \met{}, at least one pair of same flavour leptons and ($b$-)jets (no photons).
    The classes are labelled according to the multiplicity and type
    ($e$, $\mu$, $\gamma$, $j$, $b$, $\met$) of the reconstructed objects for the given event class.
    The hatched bands indicate the total uncertainty of the SM prediction. 
    }
\end{figure}
\begin{figure}[htbp]
  \begin{center}
 \includegraphics[height=16cm,width=16cm,keepaspectratio]{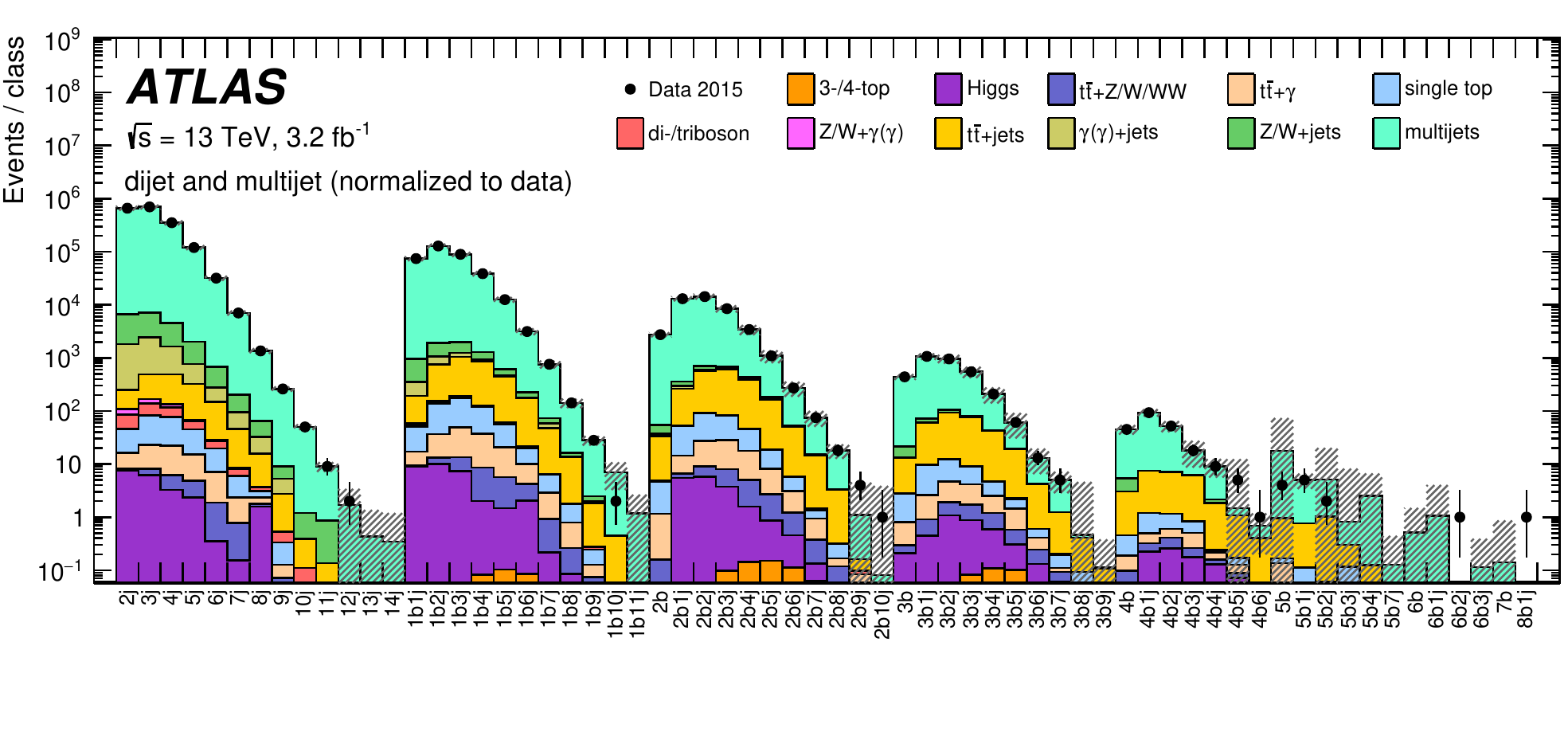}
  \end{center}
 \vspace*{-0.5cm}
  \caption{\label{fig:global14} The number of events in data, and for the different SM background predictions considered,
  	for classes with ($b$-)jets (no \met{}, leptons or photons).
    The classes are labelled according to the multiplicity and type
    ($e$, $\mu$, $\gamma$, $j$, $b$, $\met$) of the reconstructed objects for the given event class.
    In event classes with four or more data events, the multijet MC sample is scaled to data.
    The hatched bands indicate the total uncertainty of the SM prediction. 
    }
\end{figure}

\clearpage

\printbibliography
\clearpage

 
\begin{flushleft}
{\Large The ATLAS Collaboration}

\bigskip

M.~Aaboud$^\textrm{\scriptsize 34d}$,    
G.~Aad$^\textrm{\scriptsize 99}$,    
B.~Abbott$^\textrm{\scriptsize 124}$,    
O.~Abdinov$^\textrm{\scriptsize 13,*}$,    
B.~Abeloos$^\textrm{\scriptsize 128}$,    
S.H.~Abidi$^\textrm{\scriptsize 165}$,    
O.S.~AbouZeid$^\textrm{\scriptsize 143}$,    
N.L.~Abraham$^\textrm{\scriptsize 153}$,    
H.~Abramowicz$^\textrm{\scriptsize 159}$,    
H.~Abreu$^\textrm{\scriptsize 158}$
% ...
\bigskip
\\

% ...
$^{13}$Institute of Physics, Azerbaijan Academy of Sciences, Baku; Azerbaijan.\\
% ...
$^{34}$$^{(a)}$Facult\'e des Sciences Ain Chock, R\'eseau Universitaire de Physique des Hautes Energies - Universit\'e Hassan II, Casablanca;$^{(b)}$Centre National de l'Energie des Sciences Techniques Nucleaires (CNESTEN), Rabat;$^{(c)}$Facult\'e des Sciences Semlalia, Universit\'e Cadi Ayyad, LPHEA-Marrakech;$^{(d)}$Facult\'e des Sciences, Universit\'e Mohamed Premier and LPTPM, Oujda;$^{(e)}$Facult\'e des sciences, Universit\'e Mohammed V, Rabat; Morocco.\\
% ...
$^{99}$CPPM, Aix-Marseille Universit\'e, CNRS/IN2P3, Marseille; France.\\
% ...
$^{124}$Homer L. Dodge Department of Physics and Astronomy, University of Oklahoma, Norman OK; United States of America.\\
% ...
$^{128}$LAL, Universit\'e Paris-Sud, CNRS/IN2P3, Universit\'e Paris-Saclay, Orsay; France.\\
% ...
$^{143}$Santa Cruz Institute for Particle Physics, University of California Santa Cruz, Santa Cruz CA; United States of America.\\
% ...
$^{153}$Department of Physics and Astronomy, University of Sussex, Brighton; United Kingdom.\\
% ...
$^{158}$Department of Physics, Technion, Israel Institute of Technology, Haifa; Israel.\\
$^{159}$Raymond and Beverly Sackler School of Physics and Astronomy, Tel Aviv University, Tel Aviv; Israel.\\
% ...
$^{165}$Department of Physics, University of Toronto, Toronto ON; Canada.\\
% ...
$^{a}$ Also at  Department of Physics, University of Malaya, Kuala Lumpur; Malaysia.\\
$^{b}$ Also at Borough of Manhattan Community College, City University of New York, NY; United States of America.\\
$^{c}$ Also at Centre for High Performance Computing, CSIR Campus, Rosebank, Cape Town; South Africa.\\
$^{d}$ Also at CERN, Geneva; Switzerland.\\
$^{e}$ Also at CPPM, Aix-Marseille Universit\'e, CNRS/IN2P3, Marseille; France.\\
% ...
$^{*}$ Deceased

\end{flushleft}


\end{document}